\documentclass[%
 preprint,
 superscriptaddress,
 nofootinbib, 
 amsmath,amssymb,
 aps,
 longbibliography,
]{revtex4-2}

\usepackage{graphicx}
\usepackage{bm}
\usepackage{hyperref}

\usepackage{subcaption} 
\usepackage{float}
\usepackage{tikz}
\usepackage{tikz-feynman}
\usetikzlibrary{shapes.geometric,decorations.pathreplacing, positioning, calc} 
\usepackage{multirow}
\usepackage{ragged2e}

\begin{document}

\title{$b\to c \bar u q$ decay and $CP$ violating observables in the presence of new physics contributions}%

\author{Xuanning Guo}
\email{guoxn24@mails.jlu.edu.cn}
\affiliation{Center for Theoretical Physics and College of Physics,
Jilin University, Changchun, 130012, China}

\author{Albertus Hariwangsa Panuluh}
\email{panuluh@usd.ac.id}
\affiliation{Department of Physics Education, Faculty of Teacher Training and Education, Sanata Dharma University, Paingan, Maguwohardjo, Sleman, Yogyakarta 55282, Indonesia}

\author{Hiroyuki Umeeda}
\email{umeeda@jlu.edu.cn}
\affiliation{Center for Theoretical Physics and College of Physics,
Jilin University, Changchun, 130012, China}
\author{Jinglong Zhu}
\email{zhujl25@mails.jlu.edu.cn}
\affiliation{Center for Theoretical Physics and College of Physics,
Jilin University, Changchun, 130012, China}

\date{\today\\[1cm]}

\begin{abstract}
In this work, a comprehensive analysis for processes related to $b\to c\bar{u}q~(q=d, s)$ transitions
is carried out, including new physics contributions. In light of a recent tension between branching fractions for $B_{(s)}\to D_{(s)}^{(*)}M$ ($M=\pi, K$) decays in the QCD factorization approach and relevant experimental results, phenomenological constraints on complex-valued Wilson coefficients are discussed. Analyzed observables contain direct $CP$ asymmetry ($A_{\text{CP}}$) in $B^-\to D^0\pi^-$ decays and $\gamma/\phi_3$, one of the angles in the unitarity triangle, combined with others from $\tau_{B^+}/\tau_{B_d}$, $\Delta\Gamma_{q}/\Gamma_q$, and $A_{\rm SL}^{q}~(q=d, s)$. 
We constrain the complex Wilson coefficients at $1\sigma$ and $2\sigma$ levels under color-singlet and color-rearranged scenarios. These constraints yield correlated predictions for $\Delta\Gamma_d/\Gamma_d$, $A_{\rm SL}^d$ and $A_{\text{CP}}$.
\end{abstract}

\maketitle


\newpage
\section{Introduction}
Measurement of $B$-meson decays enables us to test various aspects of the standard model (SM), and possibly search for new physics (NP) contributions. Among a number of decay modes, nonleptonic channels require a dedicated theoretical discussion and can be compared with recent precision data in flavor factories. A theoretical framework to analyze, {\it e.g.}, $B_d\to D^-K^+$ and $B_s\to D_s^-\pi^+$ is given by the QCD factorization (QCDF) approach \cite{Beneke:2000ry}, where the NP contributions can be incorporated in its short-distance parts. The analysis within the SM has been performed \cite{Huber:2016xod} at the next-to-next-to leading order (NNLO) in QCD corrections.
\par
Recently, a significant tension between the branching ratios for color-allowed channels in the QCDF approach and the experimental results was pointed out in Ref.~\cite{Bordone:2020gao}: It has been claimed that power-suppressed corrections are not sufficiently large to explain the data.\footnote{A recent work \cite{Piscopo:2023opf} based on light-cone sum rules discussed a possibility that particular nonfactorizable contributions lead to sizable shift, albeit large theoretical uncertainties.} Another work \cite{Endo:2021ifc} discussed corrections from quasielastic rescattering \cite{Chua:2001br,Chua:2005dt,Chua:2007qw,Chua:2018ikx}, one formulation for representing final-state interactions (FSIs), revealing that branching ratios are not simultaneously accommodated within the SM. In light of the status, NP is considered a potential candidate to resolve the discrepancy \cite{Bordone:2020gao,Iguro:2020ndk,Cai:2021mlt,Fleischer:2021cct,Endo:2021ifc,Lenz:2022pgw,Panuluh:2024cxc,Meiser:2024zea}. Moreover, it is worth noting that the constraints on such NP scenarios can also be complemented by high-energy collider observables, particularly through top quark measurements and new scalar searches~\cite{Atkinson:2024hqp, Araz:2026zlu}.
\par
The explanation based on NP to the $B_{(s)}\to D_{(s)}^{(*)}M$ puzzle, where $M$ represents a meson, is supposed to confront phenomenological constraints from processes related to $b\to c\bar{u}q~(q=d, s)$ transitions. Specifically, $B$-meson lifetimes include such contributions, for which recent analyses in the SM were performed in Refs.~\cite{Cheng:2018rkz,Lenz:2022rbq,Egner:2024lay} (See Refs.~\cite{Lenz:2022pgw,Lang:2025ios} for discussions including NP contributions). In practice, for a ratio of the lifetimes, the theoretical precision is more controlled, which would play a suitable role in constraining the NP scenarios.
Furthermore, $\Delta\Gamma_q$ (see Ref.~\cite{Bobeth:2014rda} for new physics study) and $A_{\rm SL}^q$ in $B_q^0-\bar{B}_q^0~(q=d, s)$ mixing (see Ref.~\cite{Albrecht:2024oyn} for the SM analysis, and recent result in Ref.~\cite{Nierste:2025muk} at NNLO) can be also used to constrain the $b\bar{q} \to c\bar{u}$ transition and its $CP$ conjugate. For this process, future precise measurement is expected in the upgrade of the LHCb \cite{LHCb:2018roe,Cerri:2018ypt}, especially for $q=d$. The analyses beyond the SM including those types of the observables are carried out in Refs.~\cite{Lenz:2022pgw,Meiser:2024zea} and with FSIs in Ref.~\cite{Panuluh:2024cxc}.
\par
Another observable sensitive to $b\to c\bar{u}q~(q=d, s)$ transitions is given by $\gamma$ or $\phi_3$, one of the angles in the unitarity triangle (see Ref.~\cite{Brod:2014bfa} for a previous work, and Ref.~\cite{Lenz:2019lvd} for an updated analysis). In light of the $B_{(s)}\to D_{(s)}^{(*)}M$ puzzle, the analysis for NP was carried out in Ref.~\cite{Fleischer:2021cct} for $\bar{B}_s\to D^+_sK^-$ and $\bar{B}^0_s\to D^-_sK^+$ channels, based on the methodology \cite{Fleischer:2003yb} proposed primarily for the SM. In addition to Ref.~\cite{Fleischer:2003yb}, methods discussed by Gronau-London-Wyler (GLW) \cite{Gronau:1990ra,Gronau:1991dp} and Atwood-Dunietz-Soni (ADS) \cite{Atwood:1996ci, Atwood:2000ck} also provide possible channels to constrain $\gamma/\phi_3$.
\par
In the presence of NP contributions in $b \to c\bar{u}q$ transitions, additional $CP$ phases possibly give direct $CP$ asymmetries in tree-level $B\to DM$ decays \footnote{Within the SM, amplitudes of tree-level $B \to DM$ decays have a common Cabibbo-Kobayashi-Maskawa (CKM) factor, $V_{cb} V_{uq}^*$, and have no interfering weak phases, leading to vanishing $CP$ asymmetries. Hence, measurement of direct $CP$ violation for those channels is regarded as a signal of NP.}. Such $CP$ violating observables from NP have been discussed for $B_s\to D_s^- K^+$ and $B_s\to D_s^+ K^-$ in Refs.~\cite{Fleischer:2021cct, Fleischer:2021cwb}. Recently, a first measurement has been carried out in the LHCb experiment \cite{LHCb:2026gyt} for decay-time-integrated $CP$ asymmetry in $B_s\to D_s^-\pi^+$ decays (see Ref.~\cite{Gershon:2021pnc} for theory discussions in this channel). Furthermore, direct $CP$ asymmetry for $B^-\to D^0\pi^-$ was measured in the Belle \cite{Belle:2006cuz,Belle:2021nyg,Belle:2023yoe} and LHCb \cite{LHCb:2013jqb} experiments, although the mentioned $CP$ violating observables are both consistent with zero as expected in the SM.
\par
In this work, the phenomenological analysis including the NP contributions in Wilson coefficients for current-current operators is carried out. From the perspective of the $CP$ violating observables, complex-valued Wilson coefficients from NP are considered. In particular, $\gamma/\phi_3$ and direct $CP$ asymmetry in $B^-\to D^0\pi^-$ decays are added to the set of the analyzed observables, which consist of the branching ratio for $B_{(s)}\to D_{(s)}^{(*)}M$ decay, the lifetime ratio $(\tau_{B^+}/\tau_{B_d})$, the width difference normalized by the total width $(\Delta\Gamma_{q}/\Gamma_{B_q})$, and semileptonic $CP$ asymmetry $(A_{\rm SL}^{q})$.
\par
This paper is organized as follows: 
In Sec.~\ref{Sec:II}, we establish the theoretical foundation by formulating the effective Hamiltonian for $b \rightarrow c \bar{u} q$ transitions and addressing the relevant experimental constraints. 
Sec.~\ref{Sec:III} explores the implications of NP for the extraction of the Unitarity Triangle angle $\gamma / \phi_3$ through GLW observables. 
The focus then shifts in Sec.~\ref{Sec:IV} to a detailed investigation of direct $CP$ asymmetries and branching fractions within this framework. 
Quantitative evaluations, including phenomenological constraints on the NP Wilson coefficients and resulting predictions in light of future measurements are performed in Sec.~\ref{Sec:V}. 
Finally, we summarize our core findings and provide concluding remarks in Sec.~\ref{Sec:VI}.
\section{Effective Hamiltonian and $\Delta B=0,2$ observables}\label{Sec:II}

To analyze the $\Delta B=0$ and $\Delta B=2$ observables, we start by defining the relevant effective Hamiltonian for the underlying $\Delta B=1$ transitions. In our framework, we consider the nonleptonic decays $b \to c \bar u q$ (with $q=d,s$). Since these are charged-current transitions, penguin operators are absent. Assuming further that NP effects preserve the SM chirality and do not introduce new operator structures (such as scalar or tensor currents), but only modify the Wilson coefficients $C_1^q$ and $C_2^q$, the weak effective Hamiltonian is expressed as:
\begin{align}\label{eq:Hamiltonian}
        \mathcal{H}_{\text{eff}}^{\Delta B=1} &= \frac{G_F}{\sqrt{2}} V_{cb} V_{uq}^* \left[ C_1^q(\mu) \mathcal{Q}_1^{q} + C_2^q(\mu) \mathcal{Q}_2^{q} \right] + \text{h.c.},
\end{align}
where $G_F$ is the Fermi constant and $V_{ij}$ denotes the CKM matrix elements. The current-current four-quark operators $\mathcal{Q}_{1,2}^q$ are defined in the specific basis as:
\begin{align}
    \mathcal{Q}_1^{q} = (\bar{c}^\alpha b^\beta)_{\text{V-A}} (\bar{q}^\beta u^\alpha)_{\text{V-A}}, \qquad \mathcal{Q}_2^{q} = (\bar{c}^\alpha b^\alpha)_{\text{V-A}} (\bar{q}^\beta u^\beta)_{\text{V-A}}.
\end{align}
Here, $\alpha$ and $\beta$ are color indices with implicit summation. The subscript $(\dots)_{\text{V-A}}$ corresponds to the left-handed chiral current $\gamma^{\mu}(1-\gamma_5)$. The short-distance physics is encapsulated in the Wilson coefficients $C_i^q(\mu)$. Under our assumption, the total coefficients are given by the sum of the SM and NP contributions:
\begin{align}
    C_i^{q}(M_W) = C_i^{\text{SM}}(M_W) + C_i^{q,\text{NP}}(M_W), \quad (i=1, 2).
\end{align}
In our analysis, the NP contributions to the $b \to c\bar{u}d$ and $b \to c\bar{u}s$ transitions are treated as independent parameters. For the sake of brevity, we will hereafter suppress the flavor superscript $q$ on the coefficients $C_i^q$, unless an explicit distinction is required. The values of these coefficients at the low-energy scale $\mu \sim m_b$ are obtained by evolving them from the electroweak scale $M_W$ via the renormalization group equations (RGEs) \cite{Buchalla:1995vs}. For the SM predictions of the observables discussed in the following sections, we adopt values from the recent publications which incorporate contributions from tree-level as well as penguin and other loop diagrams. In contrast, the NP corrections to these observables are calculated assuming modifications solely to Wilson coefficients at the leading order (LO), where the detail is given later. The Hamiltonian in Eq.~(\ref{eq:Hamiltonian}) serves as the fundamental input for determining the observables discussed in the following sections.
\subsection{Lifetime Ratio}

In this section, we analyze the lifetime ratio $\tau_{B^+}/\tau_{B^0}$. Theoretically, the deviation of the lifetime ratio of heavy hadrons containing a heavy quark $Q$ from unity arises primarily from spectator effects, such as weak annihilation and Pauli interference. Combining these contributions, the lifetime ratio can be expressed as:
\begin{align}
    \frac{\tau_{B^+}}{\tau_{B^0}} = \frac{\Gamma_{B^0}}{\Gamma_{B^+}} = \frac{\Gamma_{\mathrm{dec}}+\Gamma_{\mathrm{ann}}(B^0)+\Gamma_{\mathrm{semi}}}{\Gamma_{\mathrm{dec}}+\Gamma_{\mathrm{int}}(B^+)+\Gamma_{\mathrm{semi}}}.
\end{align}
Since we restrict NP effects to the $b \to c \bar u q$ transitions, the semileptonic decay width $\Gamma_{\mathrm{semi}}$ remains unchanged from the SM prediction. The ratio of the lifetimes is determined by the inverse ratio of the total decay widths. Including both SM and NP contributions, this is given exactly by \footnote{For r.h.s. in Eq.~(\ref{eq:lifetime ratio}), we do not expand it by $1/m_b$, and instead adopt Eq. (\ref{eq:lifetime ratio}) since it can be used to constrain parametrically large NP contributions in the later numerical analysis.}:
\begin{equation}
    \begin{split}
        \frac{\tau_{B^+}}{\tau_{B^0}}\left/\frac{\tau^{\mathrm{SM}}(B^+)}{\tau^{\mathrm{SM}}(B^0)}\right.
    = \frac{1+\frac{\Gamma_{\mathrm{dec}}^{\mathrm{NP}}+\Gamma_{\mathrm{ann}}^{\mathrm{NP}}(B^0)}{\Gamma^{\mathrm{SM}}(B^0)}}{1+\frac{\Gamma_{\mathrm{dec}}^{\mathrm{NP}}+\Gamma_{\mathrm{int}}^{\mathrm{NP}}(B^+)}{\Gamma^{\mathrm{SM}}(B^+)}},\\
    \end{split}
    \label{eq:lifetime ratio}
\end{equation}
where the SM lifetimes are defined as:
\begin{equation}
    \begin{aligned}
        \frac{1}{\tau^{\mathrm{SM}}(B^0)}&=\Gamma^{\mathrm{SM}}(B^0)=\Gamma_{\mathrm{dec}}^{\mathrm{SM}}+\Gamma_{\mathrm{semi}}+\Gamma_{\mathrm{ann}}^{\mathrm{SM}}(B^0),\\
        \frac{1}{\tau^{\mathrm{SM}}(B^+)}&=\Gamma^{\mathrm{SM}}(B^+)=\Gamma_{\mathrm{dec}}^{\mathrm{SM}}+\Gamma_{\mathrm{semi}}+\Gamma_{\mathrm{int}}^{\mathrm{SM}}(B^+).
    \end{aligned}
\end{equation}
\indent The expressions for the decay widths ($\Gamma_{\text{dec}}$, $\Gamma_{\text{ann}}$, and $\Gamma_{\text{int}}$) presented later are general and depend quadratically on the $\Delta B = 1$ Wilson coefficients $C_1$ and $C_2$. For the SM, LO results can be analyzed by $C_i^{\text{SM}}$ while we employ the full coefficients ($C_i$) to include NP contributions. The formulae remain valid under this extension; one simply needs to use the full Wilson coefficients containing both SM and NP contributions. Specifically, to explicitly extract the NP contributions introduced in Eq.~(5), we define them as the difference between the decay widths evaluated with the full Wilson coefficients and those evaluated with purely SM coefficients as was considered in Ref.~\cite{Panuluh:2024cxc}. For instance, the spectator-independent NP decay width reads:
\begin{equation}
    \Gamma_{\text{dec}}^{\text{NP}} = \Gamma_{\text{dec}}(C_1, C_2) - \Gamma_{\text{dec}}(C_1^{\text{SM}}, C_2^{\text{SM}}).
    \label{eq:NP_def}
\end{equation}
Similar relations hold for the spectator-dependent annihilation and Pauli interference terms, $\Gamma_{\text{ann}}^{\text{NP}}$ and $\Gamma_{\text{int}}^{\text{NP}}$. Equation (\ref{eq:NP_def}) is used to analyze the NP contributions in Eq.~(\ref{eq:lifetime ratio}), while the recent theoretical value \cite{Egner:2024lay} is adopted for $\tau^{\text{SM}}(B^{+,0})$ (or equivalently $\Gamma^{\text{SM}}(B^{+,0})$).
\par
The spectator-independent decay width $\Gamma_{\mathrm{dec}}$ for $b \to c \bar u q$ ($q=d,s$) is given by \cite{Cheng:2018rkz}:
\begin{equation}
    \begin{split}
        \Gamma_{\mathrm{dec}}=&\frac{G_F^2m_Q^5}{192\pi^3} \xi \Biggl\{\left(N_c |C_1|^2+N_c |C_2|^2+2Re(C_1^*C_2)\right)\\
    &\times\left[\left((1-x^2)(1-8x+x^2)-12x^2\ln x\right)\left(1-\frac{\mu_{\pi}^2-\mu_G^2}{2m_Q^2}\right)-2(1-x)^4\frac{\mu_G^2}{m_Q^2}\right]\\
    &-16Re(C_1^*C_2)(1-x)^3\frac{\mu_G^2}{m_Q^2}\Biggr\},
    \end{split}
\end{equation}
where the CKM factor $\xi$ is defined as $|V_{cb}|^2 |V_{uq}|^2$ for the corresponding $b \to c \bar u q$ transition, and $\mu^2_{\pi,G}$ represents a matrix element of a power-suppressed operator. We note that our convention for $C_{1,2}$ is interchanged relative to Ref.~\cite{Cheng:2018rkz}. Additionally, all terms involving products of Wilson coefficients are generalized to $|C_i|^2$ and $\text{Re}(C_i^* C_j)$. As to the spectator dependent effects, the neutral meson $B^0$ receives contributions from weak annihilation, whereas the charged meson $B^+$ is affected only by Pauli interference. In the $b \to c \bar u d$ transition, the expressions read \cite{Cheng:2018rkz}:
\begin{equation}
    \begin{aligned}
        \Gamma_{\mathrm{ann}}(B^0) =& - \frac{G_F^2 m_b^2}{\pi} |V_{cb}V_{ud}|^2 |\psi_{b\bar{q}}^B(0)|^2 (1-x)^2 \\ &\times \Biggl\{ \left( \frac{1}{N_c}|C_2|^2 + 2Re(C_1^*C_2) + N_c |C_1|^2 \right)\Biggl[ \left(1 + \frac{x}{2} \right) B_1
         - (1+2x)B_2 \\
         &+ \left( \frac{1+x+x^2}{1-x}\rho_3^d + \frac{6x^2}{1-x}\rho_4^d 
         - \frac{1}{2}\left(1+\frac{x}{2}\right)\rho_5^d - \frac{1}{2}(1+2x)\rho_6^d \right) \left( \frac{m_B^2}{m_b^2} - 1 \right) \Biggl] \\
         &+ 2|C_2|^2 \Biggl[ \left(1+\frac{x}{2}\right)\epsilon_1 - (1+2x)\epsilon_2 + \left(\frac{1+x+x^2}{1-x}\sigma_3^d \right. 
        \left. + \frac{6x^2}{1-x}\sigma_4^d \right. \\
        &\left.- \frac{1}{2}\left(1+\frac{x}{2}\right)\sigma_5^d - \frac{1}{2}(1+2x)\sigma_6^d \right) \left(\frac{m_B^2}{m_b^2}-1\right) \Biggr] \Biggr\},
        \label{eq:Gamma_ann}
    \end{aligned}
\end{equation}    
\begin{equation}
    \begin{split}
        \Gamma_{\mathrm{int}}(B^+) =& \frac{G_F^2 m_b^2}{\pi} |V_{cb}V_{ud}|^2 |\psi_{b\bar{q}}^B(0)|^2 (1-x)^2\\
        &\times \Biggl\{ (2N_c Re(C_1^* C_2) + |C_1|^2 + |C_2|^2) \biggl[ B_1 - \left( \frac{1+x}{1-x}\rho_3^u + \frac{1}{2}\rho_5^u \right)\left( \frac{m_B^2}{m_b^2} - 1 \right) \biggr]\\
        &+ 6(|C_1|^2 + |C_2|^2) \biggl[ \epsilon_1 - \left( \frac{1+x}{1-x}\sigma_3^u + \frac{1}{2}\sigma_5^u \right) \left( \frac{m_B^2}{m_b^2} - 1 \right) \biggr] \Biggr\},
        \label{eq:Gamma_int}
    \end{split}
\end{equation} 
where $m_Q$ denotes the heavy quark mass ($Q=b$ for $B$ mesons), and $N_c=3$ is the number of colors. 
Other relevant hadronic parameters include $d_H=3$, the mass ratio $x=(\bar m_c/\bar m_b)^2$, the decay constant $f_B$ entering via $|\psi_{b\bar{q}}^B(0)|^2 = \frac{1}{12}f_B^2 m_B$, and the bag parameters, $B_1$, $\epsilon_1$, $\rho_1$ and $\sigma_1$. The strange quark contributions similar to Eqs.~(\ref{eq:Gamma_ann}), (\ref{eq:Gamma_int}) can be obtained by the replacement of flavors.
\subsection{$B-\bar B$ mixing}

In the neutral $B_q-\bar{B}_q$ mixing system ($q=d,s$), the ratio of the absorptive part to the dispersive part of the mass matrix, $|\Gamma_{21}^q/M_{21}^q|$, is of the order $\mathcal{O}(m_b^2/m_t^2) \sim 10^{-3}$. Consequently, we can safely neglect terms of $\mathcal{O}(m_b^4/m_t^4)$ in the mixing observables. 
Based on this hierarchy, the width difference $\Delta \Gamma_q$ and the semileptonic $CP$ asymmetry $A_{\mathrm{SL}}^q$ are given by:
\begin{equation}
    \begin{gathered}
        \Delta \Gamma_q = -2|M_{21}^q| \mathrm{Re}\left(\frac{\Gamma_{21}^q}{M_{21}^q}\right), \\
        A_{\mathrm{SL}}^q \approx \mathrm{Im}\left(\frac{\Gamma_{12}^q}{M_{12}^q}\right) = -\mathrm{Im}\left(\frac{\Gamma_{21}^{q}}{M_{21}^{q}}\right),
    \end{gathered}
\end{equation}
where the off-diagonal matrix elements are expressed as \cite{Ciuchini:2003ww}:
\begin{equation}\label{eq:mixing-c}
    \begin{split}
        \Gamma_{21}^q &= -\frac{G_F^2 m_b^2}{24\pi M_{B_q}} \left[ c_1^q \langle \bar{B}_q | \mathcal{O}_1^q | B_q \rangle + c_2^q \langle \bar{B}_q | \mathcal{O}_2^q | B_q \rangle + \delta_{1/m}^q \right],\\
        M_{21}^q &= \frac{G_F^2 M_W^2 \eta_B}{(4\pi)^2 (2M_{B_q})} (V_{tb}^*V_{tq})^2 S_0(x_t) \langle \bar{B}_q | \mathcal{O}_1^q | B_q \rangle,
    \end{split}
\end{equation}
where renormalization scales for $\Delta B=1$ and $\Delta B=2$ operators, not displayed explicitly for brevity, are set to a common one. $\eta_B$ is the perturbative QCD correction factor, and $S_0(x_t)$ is the Inami-Lim function \cite{Inami:1980fz} describing the short-distance top-quark loop contribution to $\Delta B=2$ processes, with $x_t = m_t^2/M_W^2$:
\begin{align}
    S_0(x) = \frac{4x - 11x^2 + x^3}{4(1-x)^2} - \frac{3x^3 \ln x}{2(1-x)^3}.
\end{align}
The coefficients $c_i^q$ and the $1/m_b$ correction term $\delta_{1/m}^q$ in Eq.~(\ref{eq:mixing-c}) encapsulate the contributions from the charm and up quark loops in the correlator calculation \cite{Ciuchini:2003ww}:
\begin{equation}\label{eq:c and delta in mixing}
    \begin{split}
        c_i^q &= (V_{tb}^* V_{tq})^2 D_i^{uu} + 2V_{cb}^* V_{cq} V_{tb}^* V_{tq} (D_i^{uu} - D_i^{cu}) + (V_{cb}^* V_{cq})^2 (D_i^{cc} + D_i^{uu} - 2D_i^{cu}),\\
        \delta_{1/m}^q &= (V_{tb}^* V_{tq})^2 \delta_{1/m}^{uu,q} + 2V_{cb}^* V_{cq} V_{tb}^* V_{tq} (\delta_{1/m}^{uu,q} - \delta_{1/m}^{cu,q}) + (V_{cb}^* V_{cq})^2 (\delta_{1/m}^{cc,q} + \delta_{1/m}^{uu,q} - 2\delta_{1/m}^{cu,q}).
    \end{split}
\end{equation}
The detailed definitions of the $\Delta B=2$ operator basis, along with their corresponding hadronic matrix elements, are listed in the Appendix.

To study NP effects, we incorporate the modified Wilson coefficients $C_1$ and $C_2$ (from the $\Delta B=1$ sector) into the calculation of $\Gamma_{21}^q$. The mass difference $M_{21}^q$ remains dominated by the SM top-quark loops and is unaffected by the NP considered here. Consequently, the constraints from $B_q-\bar B_q$ mixing observables can be derived as:\\
\begin{equation}
\begin{aligned}
    \left.\Delta \Gamma_{q}\right|_{\text{exp}}
        &=-2\left|M_{21}^q\right|\mathrm{Re}\left(\frac{\Gamma_{21,\text{SM}}^q+\Gamma_{21,\text{NP}}^q}{M_{21}^q}\right)
        =\Delta \Gamma_{q,\text{SM}}-2\left|M_{21}^q\right|\mathrm{Re}\left(\frac{\Gamma_{21,\text{NP}}^q}{M_{21}^q}\right),\\
        \left.A_{\text{SL}}^q\right|_{\text{exp}} 
        &= -\mathrm{Im}\left(\frac{\Gamma_{21,\text{SM}}^q+\Gamma_{21,\text{NP}}^q}{M_{21}^q}\right)
        = A_{\text{SL,SM}}^q - \mathrm{Im}\left(\frac{\Gamma_{21,\text{NP}}^q}{M_{21}^q}\right),
        \label{uncertainty:mixing}
\end{aligned}
\end{equation}
where $\Gamma_{21,\text{SM}}^q$ and $\Gamma_{21,\text{NP}}^q$ are defined in a way similar to Eq.~(\ref{eq:NP_def}), i.e., by evaluating the off-diagonal decay matrix element with purely SM coefficients and by taking the difference between the full result and the SM one, respectively ($\Gamma_{21,\text{NP}}^q = \Gamma_{21}^q(C_1, C_2) - \Gamma_{21}^q(C_1^{\text{SM}}, C_2^{\text{SM}})$). For $\Gamma_{q,\text{SM}}$ and $A^q_{\text{SL,SM}}$, the recent precise theoretical values \cite{Albrecht:2024oyn} are used in the later numerical analysis.
\section{Unitarity Triangle and New Physics}\label{Sec:III}
\subsection{Parameterization of Tree-Level Decays}

Two-body hadronic $B$-meson decays are classified according to their flavor topology into color-allowed tree ($T$), color-suppressed tree ($C$), exchange ($E$), and annihilation ($A$) diagrams. Among these, the $T$ diagram is the only component reliably calculable using the QCDF approach. For instance, the factorization amplitude for the process $\bar B^0_s \to D^+_s K^-$ is given by:
\begin{align}\label{eq:T-diagram}
    T(\bar B_s^0 \to D_s^+ K^-) = \frac{G_F}{\sqrt{2}} V_{us}^* V_{cb} f_{K^-} F_0^{B_s \to D_s}(m_{K^-}^2) (m_{B_s}^2 - m_{D_s}^2) a_1(D_s K),
\end{align}
where $G_F$ is the Fermi constant, $V_{us}$ and $V_{cb}$ are the relevant CKM matrix elements, $f_{K^-}$ is the decay constant, and $F_0^{B_s \to D_s}$ is the transition form factor evaluated at $q^2=m_{K^-}^2$. 

To account for $E$ diagram, an effective parameter $a_{1}^{\text{eff}}$ is often defined to replace the Wilson coefficient combination,
\begin{align}
    a_{1}^{\text{eff}}(D_s K) = a_1(D_s K) \left( 1 + \frac{E_{D_s K}}{T_{D_s K}} \right).
\end{align}
A similar parametric form that absorbs the hadronic complexity into the effective coefficient is used as a notation for amplitudes proportional to $T + C$. In the following sections, we will discuss the evaluation of these hadronic ratios.
\subsection{New Physics Phase}

Our analysis of NP effects in the unitarity triangle determination relies on modifying the effective parameter $a_1$ governing the nonleptonic tree-level decays. It is crucial to distinguish the quark-level transitions $\bar{b} \to \bar{c} u \bar{q}$ (where $q=d,s$) and their $CP$-conjugate counterparts $b \to c \bar{u} q$.

For the $b \to c \bar u q$ transitions (e.g., in $B^-$ or $\bar B^0$ decays), the NP Wilson coefficients enter directly without complex conjugation for the convention in Eq. (\ref{eq:Hamiltonian}). Thus, we generalize the parameter $a_1$ as:
\begin{align}\label{eq:def. of a1}
    a_1(b \to c) = a_1^{\text{SM}} + C_2^{\text{NP}} + \frac{C_1^{\text{NP}}}{3},
\end{align}
where $C_1^{\text{NP}}$ and $C_2^{\text{NP}}$ are the complex coefficients of the operators $\mathcal{Q}_1^q$ and $\mathcal{Q}_2^q$ in the NP effective Hamiltonian respectively. In contrast, for the $CP$-conjugate transitions $\bar b \to \bar c u \bar q$ (e.g., in $B^+$ or $B^0$ decays), the Wilson coefficients must be complex conjugated ($C_i^{\text{NP}} \to C_i^{\text{NP}*}$). 

To parameterize this effect, we define a magnitude modification and a new weak phase $\rho$ based on the unconjugated $b$ transition:
\begin{align}
    a_1(b \to c) &= a_1^{\text{SM}} \left(1 + \frac{C_2^{\text{NP}} + C_1^{\text{NP}}/3}{a_1^{\text{SM}}} \right) 
    \equiv a_1^{\text{SM}} |1+\eta| e^{i\rho},
\end{align}
where we set $a_1^{SM}$ to its absolute value (since the strong phase of this quantity is rather tiny \cite{Beneke:2000ry, Huber:2016xod, Cai:2021mlt} in the context of our present work), and the auxiliary parameters are defined as:
\begin{align}
    \eta = \frac{C_2^{\text{NP}} + C_1^{\text{NP}}/3}{a_1^{\text{SM}}}, \qquad \qquad 
    \rho = Arg(1+\eta).
\end{align}
Consequently, the effective parameter for the conjugate $\bar b \to \bar c$ transitions is given by $a_1^{\text{SM}} |1+\eta| e^{-i\rho}$.
\subsection{Modification of Observables under New Physics}

To evaluate the impact of the NP phase $\rho$ defined in the previous section, we examine the ratio of the suppressed to favored amplitudes. In the SM, this ratio is parameterized as
\begin{align}
    \frac{A_{\text{sup}}}{A_{\text{fav}}} = r_B e^{i(\delta_B \pm \gamma)},
\end{align}
where $r_B$ is the magnitude ratio, $\delta_B$ is the strong phase difference, and the $\pm$ sign corresponds to $B^\mp$ decays.

We assume that the NP contribution enters only through the effective parameter $a_1$ in the Cabibbo-favored process ($b \to c$), as defined in Eq.~(\ref{eq:def. of a1}), without altering Cabibbo-suppressed decay ($b \to u$). Consequently, the amplitudes for the process $B^+ \to D K^+$ are modified as follows:
\begin{align}
    A(B^+ \to D^0K^+)&=|A_{\text{sup}}^{\text{SM}}|e^{i\gamma}e^{i\delta},\\
    A(B^+ \to \bar D^0K^+)&=|A_{\text{fav}}^{\text{SM}}|e^{i\bar \delta}|1+\eta| e^{-i\rho}.
    \label{eq:Amp. of B to DK}
\end{align}
Note that Eq.~(\ref{eq:Amp. of B to DK}) provides a compact parameterization for the amplitude, and remains fundamentally equivalent to the factorization framework used later for evaluating the branching fractions. In Sec.~\ref{Sec:Branching Ratio}, a model-independent treatment is further adopted to extract the color-suppressed hadronic contributions from experimental data to avoid large theoretical uncertainties. Substituting these into the ratio, we obtain the effective amplitude ratio:
\begin{align}
    \left. \frac{A_{\text{sup}}}{A_{\text{fav}}} \right|_{\text{eff}} 
    = \frac{A_{\text{sup}}^{\text{SM}}}{A_{\text{fav}}^{\text{SM}} (1+\eta^*)} 
    = \frac{r_B}{|1+\eta|} e^{i(\delta_B + \gamma + \rho)},
\end{align}
where $r_B = |A_{\text{sup}}^{\text{SM}}|/|A_{\text{fav}}^{\text{SM}}|$ and $\delta_B=\delta-\bar \delta$. Crucially, we observe that the functional form of the ratio remains invariant. The NP effects can be absorbed by redefining the experimental parameters:
\begin{align}
    r_B^{\text{exp}} = \frac{r_B}{|1+\eta|}, \qquad \gamma^{\text{exp}} = \gamma_i + \rho,
    \label{eq:new def. of rb and gamma}
\end{align}
where $i = \text{indirect or direct}$, and specific values of $\gamma_i$ are introduced later. The GLW observables are experimentally defined via the ratios of partial decay widths involving $CP$ eigenstates $D_{\text{CP}\pm}$ \cite{ParticleDataGroup:2024cfk}:
\begin{align}
    R_{\text{CP}\pm} &= \frac{\Gamma(B^{-} \to D_{\text{CP}\pm}K^{-}) + \Gamma(B^{+} \to D_{\text{CP}\pm}K^{+})}{\Gamma(B^{-} \to D^0K^{-}) + \Gamma(B^{+} \to \bar{D}^0K^{+})}, \\[5pt]
    A_{\text{CP}\pm} &= \frac{\Gamma(B^{-} \to D_{\text{CP}\pm}K^{-}) - \Gamma(B^{+} \to D_{\text{CP}\pm}K^{+})}{\Gamma(B^{-} \to D_{\text{CP}\pm}K^{-}) + \Gamma(B^{+} \to D_{\text{CP}\pm}K^{+})}.
\end{align}
Therefore, the standard expressions for the GLW observables $R_{\text{CP}\pm}$ and $A_{\text{CP}\pm}$ remain valid under the replacements $r_B \to r_B^{\text{exp}}$ and $\gamma \to \gamma^{\text{exp}}$:
\begin{align}
    R_{\text{CP}\pm} &= 1 + (r_B^{\text{exp}})^2 \pm 2r_B^{\text{exp}} \cos\delta_B \cos\gamma^{\text{exp}}, \\[10pt]
    A_{\text{CP}\pm} &= \frac{\pm 2r_B^{\text{exp}} \sin\delta_B \sin\gamma^{\text{exp}}}{R_{\text{CP}\pm}}.
\end{align}
This confirms that a NP phase in the tree-level Wilson coefficient manifests simply as a shift in the extracted CKM angle $\gamma$. In the later numerical analysis, the second relation in Eq. (\ref{eq:new def. of rb and gamma}) is used to constrain the Wilson coefficients.
\section{Direct $CP$ and Branching Ratio}\label{Sec:IV}

\subsection{Direct $CP$ violation}
\label{Sec:IVA}
Having established the phenomenological impact of NP on the GLW observables, we now turn to the theoretical calculation of the underlying partial widths and direct $CP$ asymmetries within the SM.
Based on the underlying mechanisms, $CP$ violation is usually classified into three types: direct violation, indirect violation, and $CP$ violation arising from the interference between decay and mixing. In charged B meson decays, there is no mixing phenomenon, and direct $CP$ violation is the only observable effect. Specifically, the corresponding asymmetry is defined as:
\begin{align}
    A_{\text{CP}}(B^+ \to f^+) \equiv \frac{\Gamma(B^{-} \to f^{-}) - \Gamma(B^{+} \to f^{+})}{\Gamma(B^{-} \to f^{-}) + \Gamma(B^{+} \to f^{+})} \, .
\end{align}

It is important to emphasize that the equation above represents the generic definition of direct $CP$ asymmetry for a specific final state $f$. This should be distinguished from the GLW observable $A_{\text{CP}\pm}$ discussed in the previous section, which specifically refers to the asymmetry involving $CP$ eigenstate modes ($D_{\text{CP}\pm}$) and involves a different normalization factor (the denominator is normalized to the flavor-specific modes).
The partial width for a specific decay, for instance $B^- \to D^0\pi^-$, can be calculated within the framework of perturbative QCD (PQCD). It is proportional to the squared magnitude of the decay amplitude \cite{Keum:2003js},
\begin{align}
    \Gamma(B^- \to D^0 \pi^-) = \frac{1}{128\pi} G_F^2 |V_{cb}|^2 |V_{ud}|^2 \frac{m_B^3}{r} |A_{D^0 \pi^-}|^2 \, ,
\end{align}
with $r$ defined in Eq.~(\ref{def: r,r0,[dx]}). 
In the PQCD approach, the total decay amplitude $A_{D^0 \pi^-}$ is systematically decomposed into contributions from different Feynman diagram topologies \cite{Keum:2003js}:
\begin{align}
    \label{eq:PQCD}
    A_{D^0 \pi^-} = f_{\pi}\xi_{\text{ext}} + f_D\xi_{\text{int}} + \mathcal{M}_{\text{ext}} + \mathcal{M}_{\text{int}} \, .
\end{align}
Here, $\xi_{\rm ext}$ and $\xi_{\rm int}$ denote the factorizable external $W$-emission (color-allowed) and internal $W$-emission (color-suppressed) contributions, respectively, while $\mathcal M_{\rm ext}$ and $\mathcal M_{\rm int}$ denote the corresponding nonfactorizable contributions. For this charged mode, the $W$-exchange topology does not enter the class-\uppercase\expandafter{\romannumeral 3}  amplitude. The explicit PQCD expressions for these building blocks, including the hard kernels, evolution factors and meson distribution amplitudes, are collected in Appendix~\ref{Appendix:PQCD}. 
\subsection{Branching Ratio}
\label{Sec:Branching Ratio}
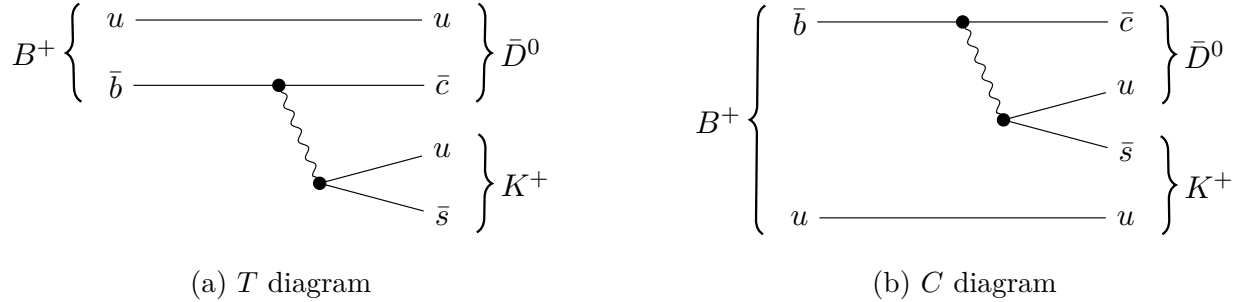
\begin{figure}[htb]
    \centering
    \begin{subfigure}[b]{0.45\textwidth}
        \centering
        \resizebox{\linewidth}{!}{%
            \begin{tikzpicture}
            \begin{feynman}
                \vertex (A) at (0, 2.4) {$u$}; 
                \vertex (B) at (0, 1.6) {$\bar b$};
                \vertex [dot] (V3) at (2, 1.6) {};
                \vertex [dot] (V4) at (2.5, 0.4) {};
                \vertex (C) at (4, 2.4) {$u$};
                \vertex (D) at (4, 1.6) {$\bar c$};
                \vertex (E) at (4, 0.8) {$u$};
                \vertex (F) at (4, 0) {$\bar s$};
                \diagram* {
                    (A) -- [plain] (C),
                    (B) -- [plain] (D),
                    (V4) -- [plain] (E),
                    (V4) -- [plain] (F),
                    (V3) -- [photon] (V4),
                };
            \end{feynman}
            \draw [decorate, decoration={brace, amplitude=5pt}, thick] 
                  ($(B.west)+(-0.2, -0.2)$) -- ($(A.west)+(-0.2, 0.2)$) 
                  node [midway, left=4pt] {$B^+$}; 
            \draw [decorate, decoration={brace, amplitude=5pt}, thick] 
                  ($(C.east)+(0.2, 0.2)$) -- ($(D.east)+(0.2, -0.2)$) 
                  node [midway, right=4pt] {$\bar D^0$};
            \draw [decorate, decoration={brace, amplitude=5pt}, thick] 
                  ($(E.east)+(0.2, 0.2)$) -- ($(F.east)+(0.2, -0.2)$) 
                  node [midway, right=4pt] {$K^+$};
            \end{tikzpicture}%
        }
        \caption{$T$ diagram}
        \label{fig:diagram_T}
    \end{subfigure}
    \hfill
    \begin{subfigure}[b]{0.45\textwidth}
        \centering
        \resizebox{\linewidth}{!}{%
            \begin{tikzpicture}
            \begin{feynman}
                \vertex (A) at (0, -0.8) {$u$}; 
                \vertex (B) at (0, 1.6) {$\bar b$};
                \vertex [dot] (V3) at (2, 1.6) {};
                \vertex [dot] (V4) at (2.5, 0.4) {};
                \vertex (C) at (4, -0.8) {$u$};
                \vertex (D) at (4, 1.6) {$\bar c$};
                \vertex (E) at (4, 0.8) {$u$};
                \vertex (F) at (4, 0) {$\bar s$};
                \diagram* {
                    (A) -- [plain] (C),
                    (B) -- [plain] (D),
                    (V4) -- [plain] (E),
                    (V4) -- [plain] (F),
                    (V3) -- [photon] (V4),
                };
            \end{feynman}
            \draw [decorate, decoration={brace, amplitude=5pt}, thick] 
                  ($(A.west)+(-0.2, -0.2)$) -- ($(B.west)+(-0.2, 0.2)$) 
                  node [midway, left=4pt] {$B^+$}; 
            \draw [decorate, decoration={brace, amplitude=5pt}, thick] 
                  ($(D.east)+(0.2, 0.2)$) -- ($(E.east)+(0.2, -0.2)$) 
                  node [midway, right=4pt] {$\bar D^0$};
            \draw [decorate, decoration={brace, amplitude=5pt}, thick] 
                  ($(F.east)+(0.2, 0.2)$) -- ($(C.east)+(0.2, -0.2)$) 
                  node [midway, right=4pt] {$K^+$};
            \end{tikzpicture}%
        }
        \caption{$C$ diagram}
        \label{fig:diagram_C}
    \end{subfigure}
    \caption{Topological diagrams contributing to $B^+ \to \bar D^0K^+$.}
    \label{fig:parallel_feynman_diagrams}
\end{figure}

For a general two-body decay $B \to M_1 M_2$, the branching ratio is given by
\begin{align}
    \mathcal{B}\left(B \to M_1 M_2\right) = \frac{\Gamma\left(B \to M_1 M_2\right)}{\Gamma_{\text{tot}}} = \frac{|\mathbf{p}|}{8\pi m_B^2} |\mathcal{M}|^2\tau_{B} ,
\end{align}
where $\tau_{B}$ is the lifetime of the $B$ meson. $|\mathbf{p}|$ denotes the magnitude of the 3-momentum of the final state particles in the rest frame of the decaying meson, defined as:
\begin{align}
    |\mathbf{p}| = \frac{\Phi(m_B, m_{M_1}, m_{M_2})}{2m_B}, \quad \text{with} \quad \Phi(x, y, z) \equiv \sqrt{[x^2 - (y+z)^2][x^2 - (y-z)^2]}.
\end{align}
Here, $\mathcal{M}$ represents the total amplitude, which is generally a linear combination of different topological amplitudes, such as $T$, $C$, $E$, and $A$ diagrams.
For the specific decay $B^+ \to \bar D^0 K^+$, the amplitude receives contributions from both $T$ and $C$ topologies, as illustrated in Fig.~\ref{fig:parallel_feynman_diagrams}. While the $T$ diagram can be reliably calculated using, e.g., the QCDF approach, the $C$ diagram suffers from large hadronic uncertainties and cannot be calculated directly with high precision.

\begin{figure}[h]
    \centering
    \begin{tikzpicture}
        \begin{feynman}
            \vertex (A) at (0, 2.4) {$d$}; 
            \vertex (B) at (0, 1.6) {$\bar b$};
            \vertex [dot] (V3) at (2, 1.6) {};
            \vertex [dot] (V4) at (2.5, 0.4) {};
            \vertex (C) at (4, 2.4) {$d$};
            \vertex (D) at (4, 1.6) {$\bar c$};
            \vertex (E) at (4, 0.8) {$u$};
            \vertex (F) at (4, 0) {$\bar s$};
            \diagram* {
                (A) -- [plain] (C),
                (B) -- [plain] (D),
                (V4) -- [plain] (E),
                (V4) -- [plain] (F),
                (V3) -- [photon] (V4),
            };
        \end{feynman}
        \draw [decorate, decoration={brace, amplitude=5pt}, thick] 
              ($(B.west)+(-0.2, -0.2)$) -- ($(A.west)+(-0.2, 0.2)$) 
              node [midway, left=4pt] {$B^0$}; 
        \draw [decorate, decoration={brace, amplitude=5pt}, thick] 
              ($(C.east)+(0.2, 0.2)$) -- ($(D.east)+(0.2, -0.2)$) 
              node [midway, right=4pt] {$D^-$};
        \draw [decorate, decoration={brace, amplitude=5pt}, thick] 
              ($(E.east)+(0.2, 0.2)$) -- ($(F.east)+(0.2, -0.2)$) 
              node [midway, right=4pt] {$K^+$};
    \end{tikzpicture}
    \caption{The reference decay channel $B^0 \to D^-K^+$, which proceeds via the $T$ diagram.}
    \label{fig:diagram_Bd}
\end{figure}
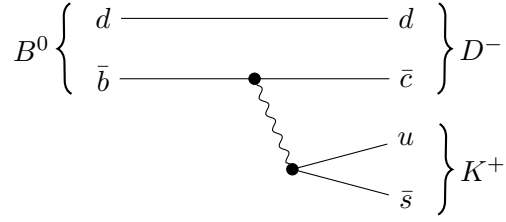

To evaluate the total amplitude $\mathcal{M}(B^+ \to \bar D^0 K^+) = T + C$, we adopt a phenomenological approach inspired by the method proposed in \cite{Fleischer:2021cct} (originally for $T+E$ topologies). 
We utilize $SU(2)$ isospin symmetry to relate the process of interest to the reference decay $B^0 \to D^- K^+$, which is dominated by the color-allowed tree diagram (see Fig.~\ref{fig:diagram_Bd}).

Using the experimental branching ratios and the theoretical expression for the $T$ diagram, we can extract the contribution of the $C$ diagram. The ratio of the squared amplitudes is related to the observables by:
\begin{align}
    \left|\frac{T_{\bar D^0K^+}}{T_{D^-K^+}}\right|^2 \left|1+\frac{C_{\bar D^0K^+}}{T_{\bar D^0K^+}}\right|^2
    = \frac{\tau_{B^0}}{\tau_{B^+}} \left(\frac{m_{B^+}}{m_{B^0}}\right)^3 
    \left[ \frac{\Phi(m_{B^0}, m_{D^-}, m_{K^+})}{\Phi(m_{B^+}, m_{\bar D^0}, m_{K^+})} \right]
    \left[ \frac{\mathcal{B}(B^+ \to \bar D^0 K^+)}{\mathcal{B}(B^0 \to D^- K^+)} \right].
    \label{eq:ratio of total amp. similar decay}
\end{align}
The theoretical ratio of the $T$ amplitudes is given by factorization as:
\begin{align}
    \left|\frac{T_{\bar D^0K^+}}{T_{D^-K^+}}\right| = \left[ \frac{F_0^{B^+ \to \bar D^0}(m_{K^+}^2)}{F_0^{B^0 \to D^-}(m_{K^+}^2)} \right] 
    \left[ \frac{m_{B^+}^2 - m_{\bar D^0}^2}{m_{B^0}^2 - m_{D^-}^2} \right] 
    \left| \frac{a_1(\bar D^0K^+)}{a_1(D^-K^+)} \right|.
    \label{eq:ratio of similar T amp.}
\end{align}
Finally, the total amplitude for the $B^+ \to \bar D^0K^+$ decay is evaluated as:
\begin{align}
    \left|\mathcal{M}(B^+ \to \bar D^0K^+)\right|^2 = \left|T_{\bar D^0K^+}\right|^2 \left|1+\frac{C_{\bar D^0K^+}}{T_{\bar D^0K^+}}\right|^2.
    \label{eq:total Amp. of T+C}
\end{align}
With the total amplitude determined, we can subsequently introduce NP corrections in $T$ diagrams. $|1 + C/T|^2$ in Eq.~(\ref{eq:total Amp. of T+C}) is model independently extracted from the data by using Eqs.~(\ref{eq:ratio of total amp. similar decay}), (\ref{eq:ratio of similar T amp.}).

\section{Numerical Results }\label{Sec:V}
\subsection{Experimental Inputs and SM Predictions}
For the $\gamma$ inputs, we consider the following three values:
\begin{align}
    \gamma_{\text{indirect}} &= 73.27^\circ, \label{eq:gamma_indirect} \\
    \gamma_{\text{direct}} &= (66.4^{+2.7}_{-2.8})^\circ, \label{eq:gamma_direct} \\
    \gamma_{\text{CKMfitter}} &= 66.33^\circ. \label{eq:gamma_CKMfitter}
\end{align}
In Eq.~(\ref{eq:gamma_indirect}), the indirect central value is obtained by using the unitarity relation $\alpha + \beta + \gamma = 180^\circ$ with $\alpha$ and $\beta$ from HFLAV \cite{HeavyFlavorAveragingGroupHFLAV:2024ctg,HFLAV:Online}. Because this indirect determination is almost free from tree-level NP effects in $b \to c\bar{u}q$ transitions, it is used as the primary input in our subsequent numerical analysis. As for Eq.~(\ref{eq:gamma_direct}), it is given by direct measurement~\cite{HeavyFlavorAveragingGroupHFLAV:2024ctg,HFLAV:Online}. For Eq.~(\ref{eq:gamma_CKMfitter}), $\gamma_{\text{CKMfitter}}$ is obtained from the definition $\gamma = -\arg\left(V_{ud}V_{ub}^*/V_{cd}V_{cb}^*\right)$ using the Wolfenstein parameters from CKMfitter~\cite{Charles:2015gya,CKMfitterSummer2025}.
\\
\indent Unlike the indirect value, $\gamma_{\text{direct}}$ is sensitive to NP contributions, while $\gamma_{\text{CKMfitter}}$ is also affected to a lesser extent. Direct measurements of $\gamma$ and the branching ratios used in our analysis are summarized in Tables I and II, respectively.
\par
For the lifetime ratio $\tau(B^+)/\tau(B^0)$, the comparison between the experimental world average~\cite{HeavyFlavorAveragingGroupHFLAV:2024ctg} and the SM prediction~\cite{Egner:2024lay} yields:
\begin{align}
    \left(\frac{\tau_{B^+}}{\tau_{B^0}}\right)_{\text{Exp}} = 1.076 \pm 0.004, \qquad & \qquad \left(\frac{\tau_{B^+}}{\tau_{B^0}}\right)_{\text{SM}} = 1.081 \pm 0.016.
\end{align}

Regarding the neutral meson mixing observables, the experimental values~\cite{HeavyFlavorAveragingGroupHFLAV:2024ctg} for the width difference $\Delta \Gamma_q/\Gamma_{B^0_q}$ and the semileptonic asymmetry $A_{\text{SL}}^q$ (with $q=d,s$) are:
\begin{equation}
    \begin{aligned}
    \left.\frac{\Delta \Gamma_d}{\Gamma_{B_d^0}}\right|_{\text{Exp}} &= 0.001 \pm 0.010, & A_{\text{SL,Exp}}^d &= -0.0021 \pm 0.0017, \\
   \left.\frac{\Delta \Gamma_s}{\Gamma_{B_s^0}}\right|_{\text{Exp}} &= 0.124 \pm 0.007, & A_{\text{SL,Exp}}^s &= -0.0006 \pm 0.0028.
    \end{aligned}
\end{equation}
\indent For the future experimental projection, the uncertainties are expected to be significantly reduced at LHCb \cite{LHCb:2018roe,Cerri:2018ypt}. These projections read:
\begin{equation}
    \delta \left( \frac{\Delta \Gamma_d}{\Gamma_d} \right)_{\text{future}} = 1 \times 10^{-3}, \quad \delta \left( \mathcal{A}_{\text{SL}}^d \right)_{\text{future}} = 2 \times 10^{-4}.
\end{equation}
\indent The corresponding SM predictions have been presented in Ref.~\cite{Albrecht:2024oyn}:
\begin{equation}
    \begin{aligned}
    \Delta \Gamma_{d,\text{SM}} &= 0.0027 \pm 0.0004 \text{ ps}^{-1}, & A_{\text{SL,SM}}^d &= -0.00051 \pm 0.00005, \\
    \Delta \Gamma_{s,\text{SM}} &= 0.091 \pm 0.015 \text{ ps}^{-1}, & A_{\text{SL,SM}}^s &= 0.000022 \pm 0.000002.
    \end{aligned}
\end{equation}
In the analysis, while contributions of NP and interference between NP and the SM are evaluated at LO in QCD, the SM contributions are extracted from recent publications. The SM prediction for $\tau(B^+)/\tau(B^0)$ is taken from the NLO-QCD HQE result~\cite{Egner:2024lay}, and the SM predictions for the neutral meson mixing observables are adopted at the NLO-QCD level from Ref.~\cite{Albrecht:2024oyn}. For the branching-ratio constraints, the SM factorization input is taken from the NNLO QCDF result~\cite{Huber:2016xod}.
\par
The experimental value for the $CP$ asymmetry
\begin{align}
    A_{\text{CP}}(B^+ \to \bar D^0 \pi^+) = (-3.2 \pm 3.5) \times10^{-3}
\end{align}
is taken from PDG \cite{ParticleDataGroup:2024cfk} based on the LHCb \cite{LHCb:2013jqb} and Belle \cite{Belle:2021nyg,Belle:2023yoe} experiments (see also previous measurement in the Belle experiment \cite{Belle:2006cuz}).

\subsection{Constraints on Wilson Coefficients}
In what follows, we present the $1\sigma$ and $2\sigma$ allowed regions for the NP Wilson coefficients $C_i^{\rm NP}(M_W)$ ($i=1,2$) in the complex plane. 
Since $C_1^{\rm NP}$ and $C_2^{\rm NP}$ are in general complex parameters, the full parameter space is four-dimensional. In this subsection, we consider simplified two-parameter scenarios as follows:
\begin{itemize}
    \item[(i)] The left panels show the constraints on $C_1^{\rm NP}(M_W)$ with the boundary condition $C_2^{\rm NP}(M_W)=0$.
    \item[(ii)] The right panels show the constraints on $C_2^{\rm NP}(M_W)$ with the boundary condition $C_1^{\rm NP}(M_W)=0$.
\end{itemize}
The observables are evaluated at $\mu = m_b$, and thus short-distance QCD corrections from $M_W$ to $m_b$ for the Wilson coefficients are considered in a relevant way. In this case, $C_1(m_b)$ and $C_2(m_b)$ are nonzero in general for both of the mentioned scenarios.
\par
For the unitarity-triangle and branching-ratio constraints, the observables are evaluated at $\mu=\bar m_b(\bar m_b)$, where $\alpha_s=0.224$. Regarding the lifetime-ratio constraints, we use the kinetic-scheme inputs, including $m_b^{\rm kin}$ and the HQE parameters $\mu_\pi^2$, $\mu_G^2$ and $\rho_D^3$, consistently with their extraction from inclusive semileptonic $B$-decay fits \cite{Bordone:2021oof}; the strong coupling is evaluated at $\mu=m_b^{\rm kin}/2$, giving $\alpha_s=0.281$. For the $B^0-\bar B^0$ mixing observables, we evaluate the relevant coefficients at the common scale $\mu_1=\mu_2=m_b^{pole}$, with $\alpha_s=0.215$. The scale evolution of $\alpha_s$ and quark masses is performed using the \texttt{RunDec} package~\cite{Chetyrkin:2000yt}. The detailed input values are listed in Tab.~\ref{Input parameters}.
\par
For constraints on the parameters, relevant observables are evaluated on a complex plane, with allowed regions of each quantity illustrated in figures, instead of performing a global $\chi^2$ fit. The resulting allowed regions are obtained from the overlap of the individual constraints. As for the treatment of uncertainties, we combine the experimental and the SM errors in the following ways: For the neutral meson mixing, based on Eq.~(\ref{uncertainty:mixing}), the combined objects of $\Delta \Gamma_q|_{\text{exp}} - \Delta \Gamma_q|_{\text{SM}}$ and $A_{\text{SL}}^q|_{\text{exp}} - A_{\text{SL}}^q|_{\text{SM}}$ are used to constrain the NP parameters, where Gaussian $10^{6}$ samples for both theoretical and experimental inputs are generated to evaluate the central values and uncertainties of those quantities. Regarding the lifetime ratio in Eq.~(\ref{eq:lifetime ratio}) and the branching ratios, similar ways of combining uncertainties are implemented with the SM uncertainties from the CKM matrix elements and form factors for the latter.
\par
To constrain possible NP effects, we use a set of independent observables from the channels $B_d \to D^{(*)-}\pi^+$ for $b\to c\bar u d$ and $B^+ \to \bar D^0K^+$ and $\bar{B}_s^0 \to D_s^+K^-$ for $b\to c\bar u s$. The resulting constraints are shown in the complex planes of the corresponding high-scale NP Wilson coefficients $C_i^{\rm NP}(M_W)$.
\par
\begin{figure}[htbp]
    \centering
    \begin{minipage}{0.3\textwidth}
        \centering
        \includegraphics[width=\linewidth]{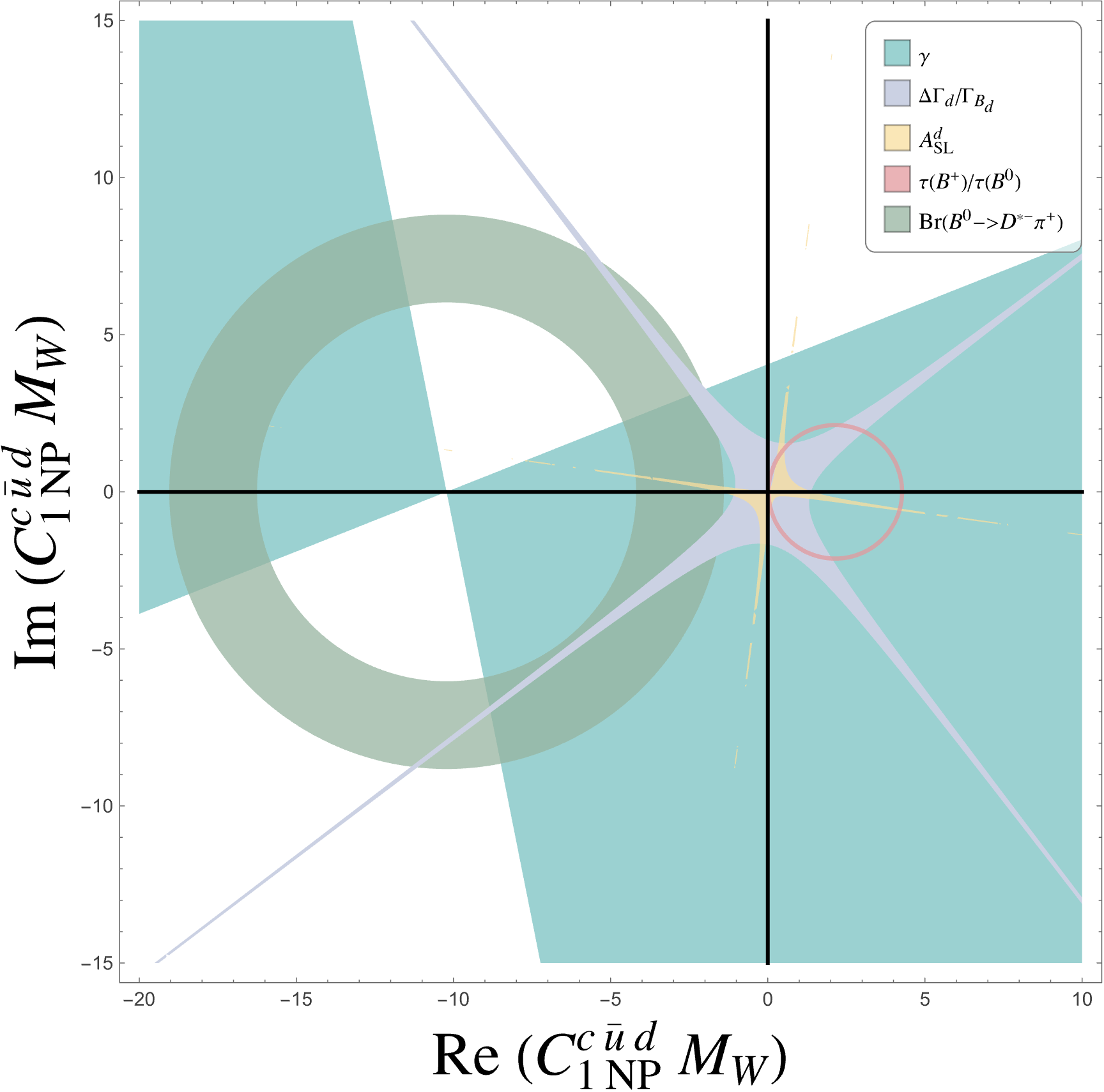}
    \end{minipage}
    \qquad \qquad \qquad 
    \begin{minipage}{0.3\textwidth}
        \centering
        \includegraphics[width=\linewidth]{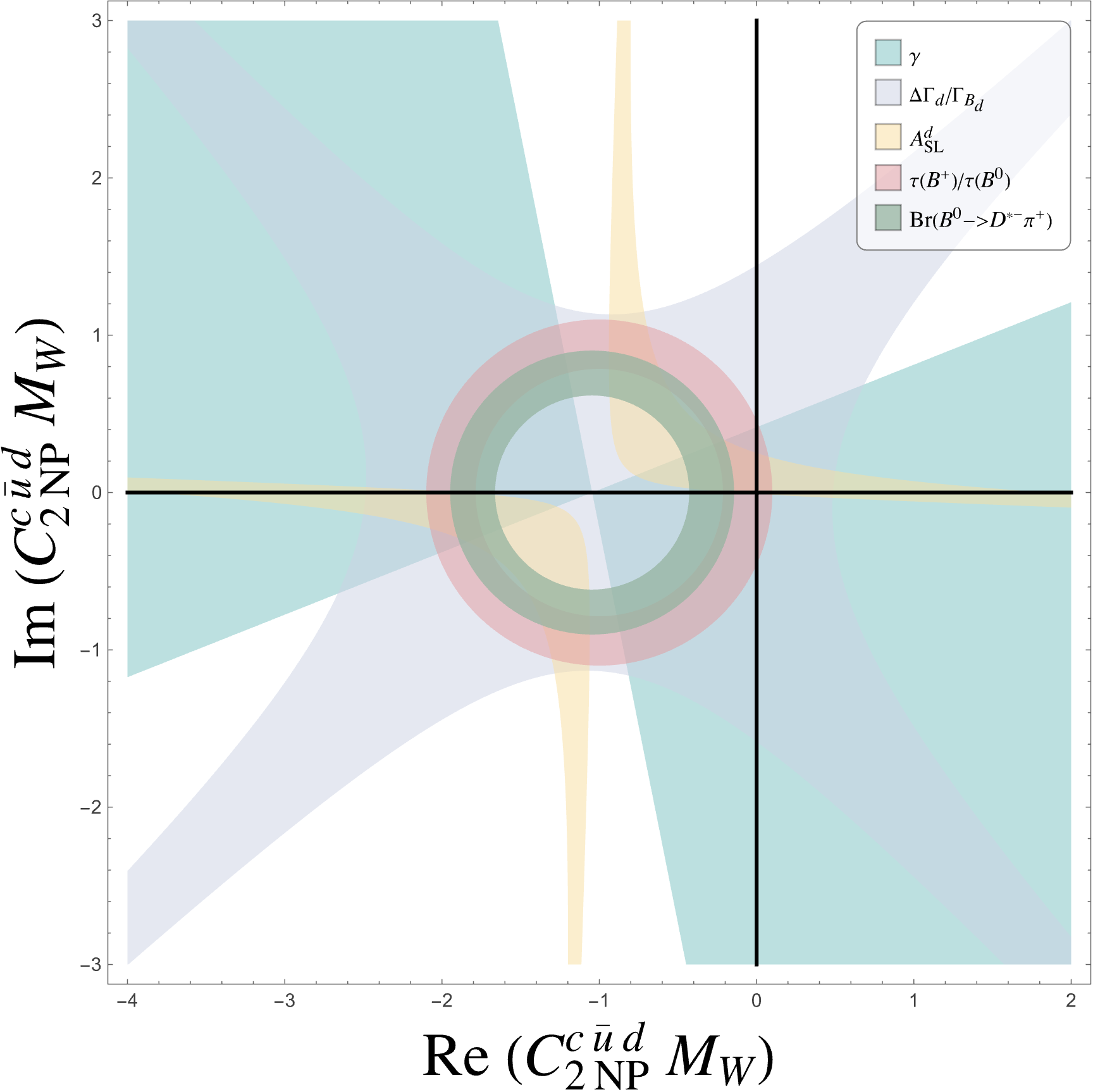}
    \end{minipage}
    \caption{\justifying{Combined constraints on the NP Wilson coefficient $C_{1,\text{NP}}^{c\bar{u}d}(M_W)$ and $C_{2,\text{NP}}^{c\bar{u}d}(M_W)$ from various flavor observables for $B_d\to D^{*-}\pi^+$ at $1\sigma$ level. The different colored regions represent the following constraints: 
    the cyan wedge for the CKM angle $\gamma$; the red and green bands for the lifetime ratio $\tau(B^+)/\tau(B^0)$ and the branching ratio $\text{Br}(B_d\to D^{*-}\pi^+)$, respectively; the light purple and yellow regions for the mixing observables $\Delta\Gamma_d/\Gamma_{B_d}$ and $A_{\text{SL}}^d$.}}
    \label{fig:constraints_BDstarPi}
\end{figure}
\begin{figure}[htbp]
    \centering
    \begin{minipage}{0.3\textwidth}
        \centering
        \includegraphics[width=\linewidth]{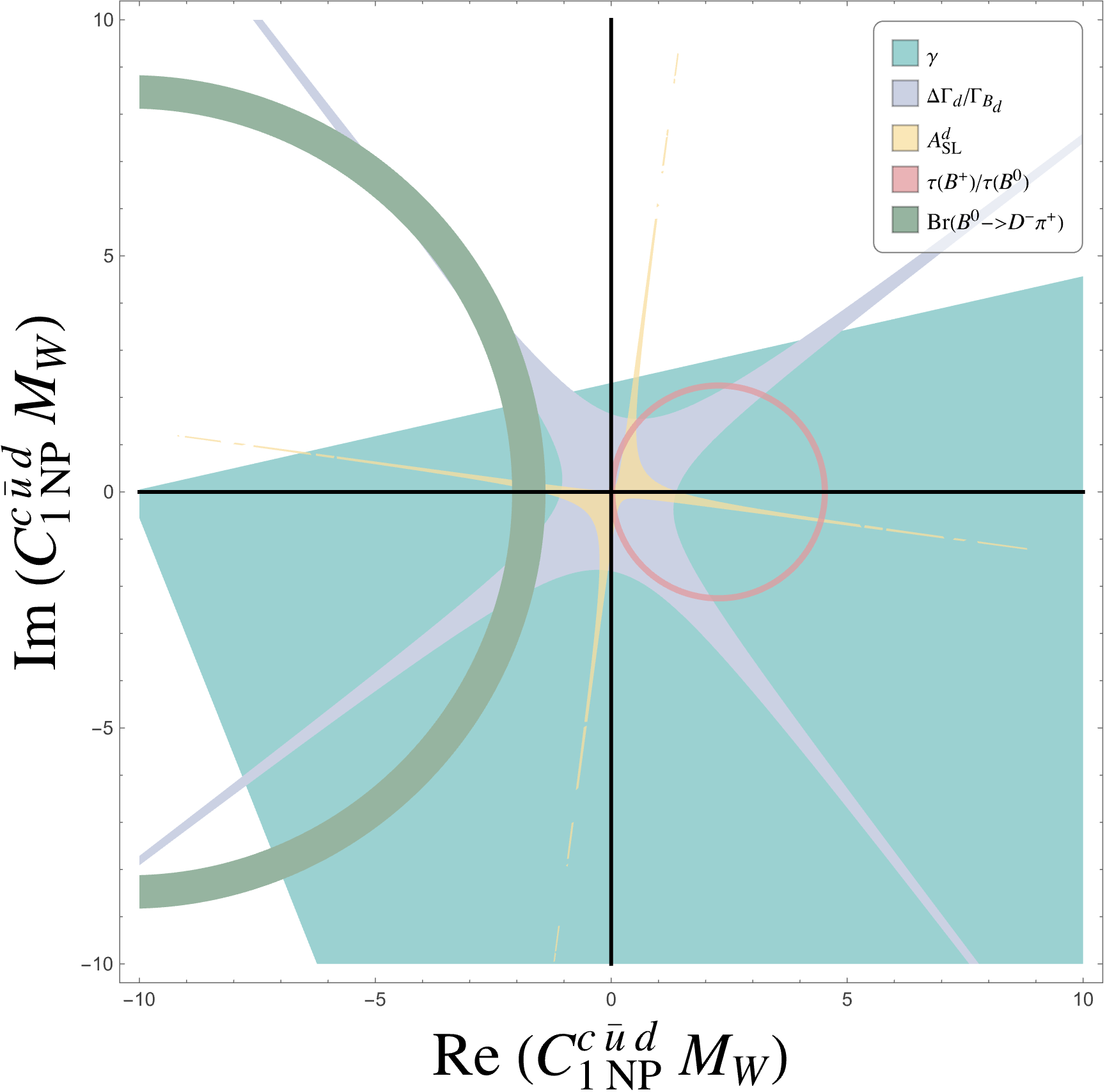}
    \end{minipage}
    \qquad \qquad \qquad 
    \begin{minipage}{0.3\textwidth}
        \centering
        \includegraphics[width=\linewidth]{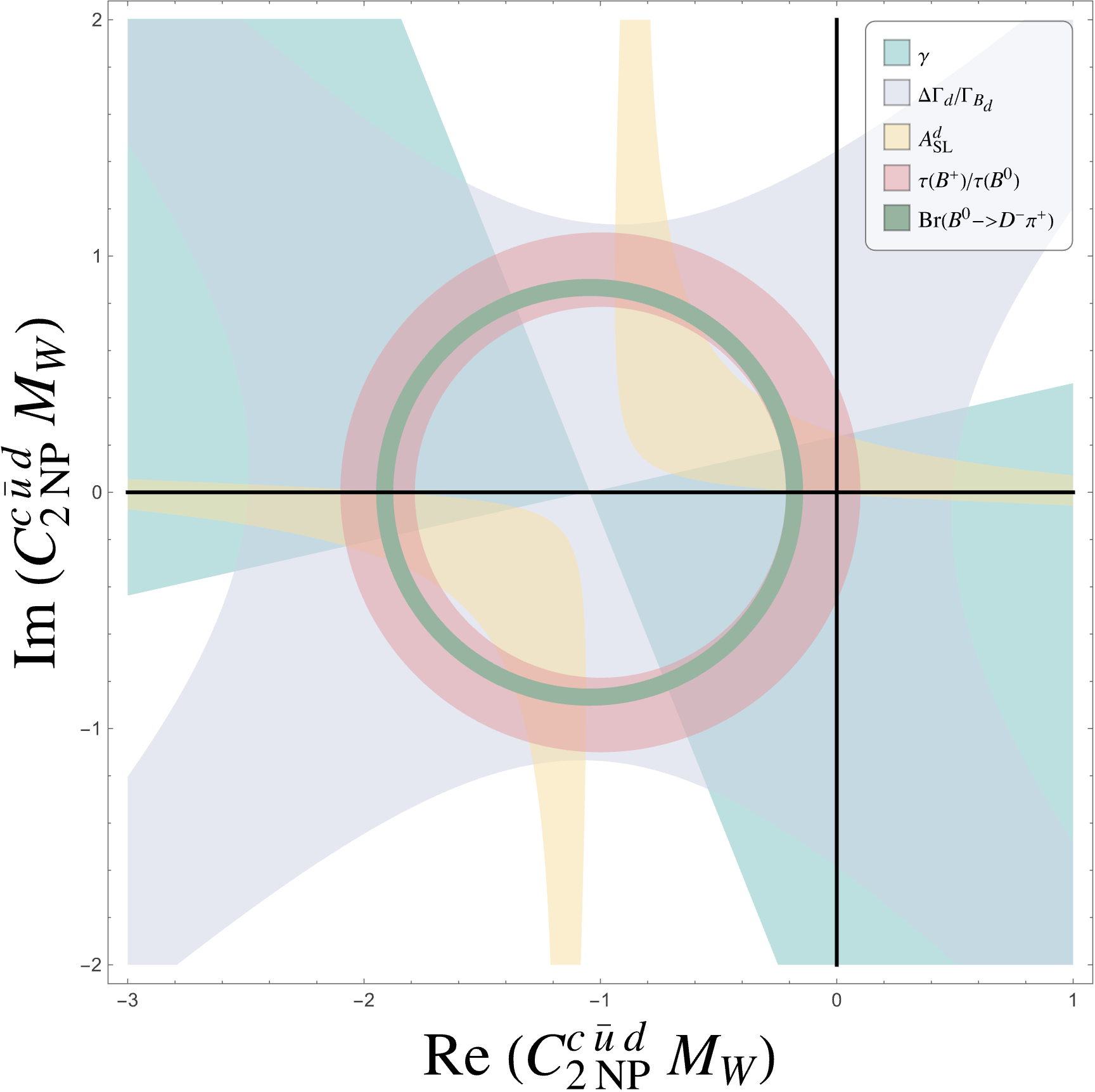}
    \end{minipage}
    \caption{\justifying{Combined constraints on the NP Wilson coefficient $C_{1,\text{NP}}^{c\bar{u}d}(M_W)$ and $C_{2,\text{NP}}^{c\bar{u}d}(M_W)$ from various flavor observables for $B_d\to D^{-}\pi^+$ at $1\sigma$ level.}}
    \label{fig:constraints_BDPi}
\end{figure}
\begin{figure}[htbp]
    \centering
    \begin{minipage}{0.3\textwidth}
        \centering
        \includegraphics[width=\linewidth]{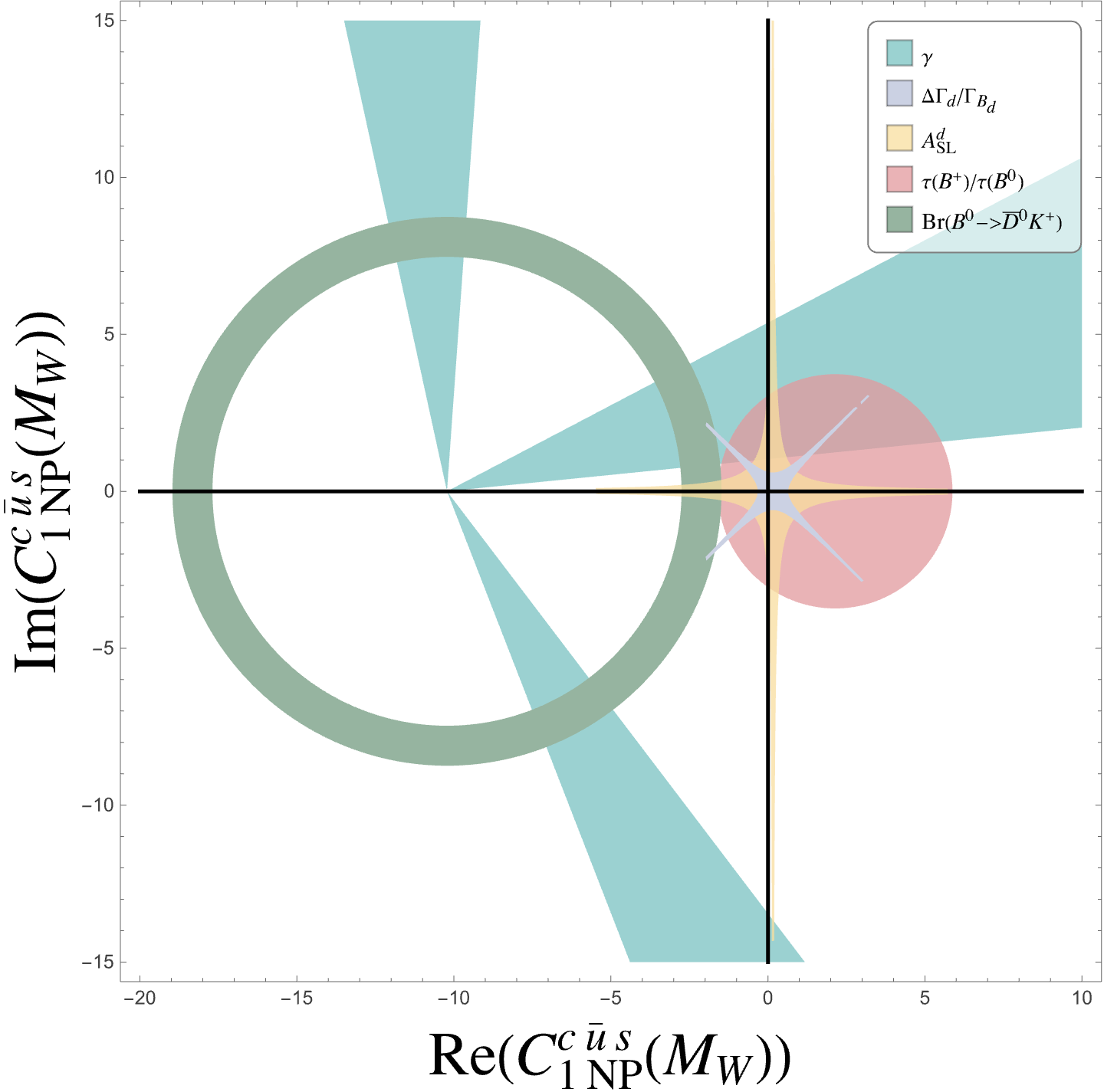}
    \end{minipage}
    \qquad \qquad \qquad  
    \begin{minipage}{0.3\textwidth}
        \centering
        \includegraphics[width=\linewidth]{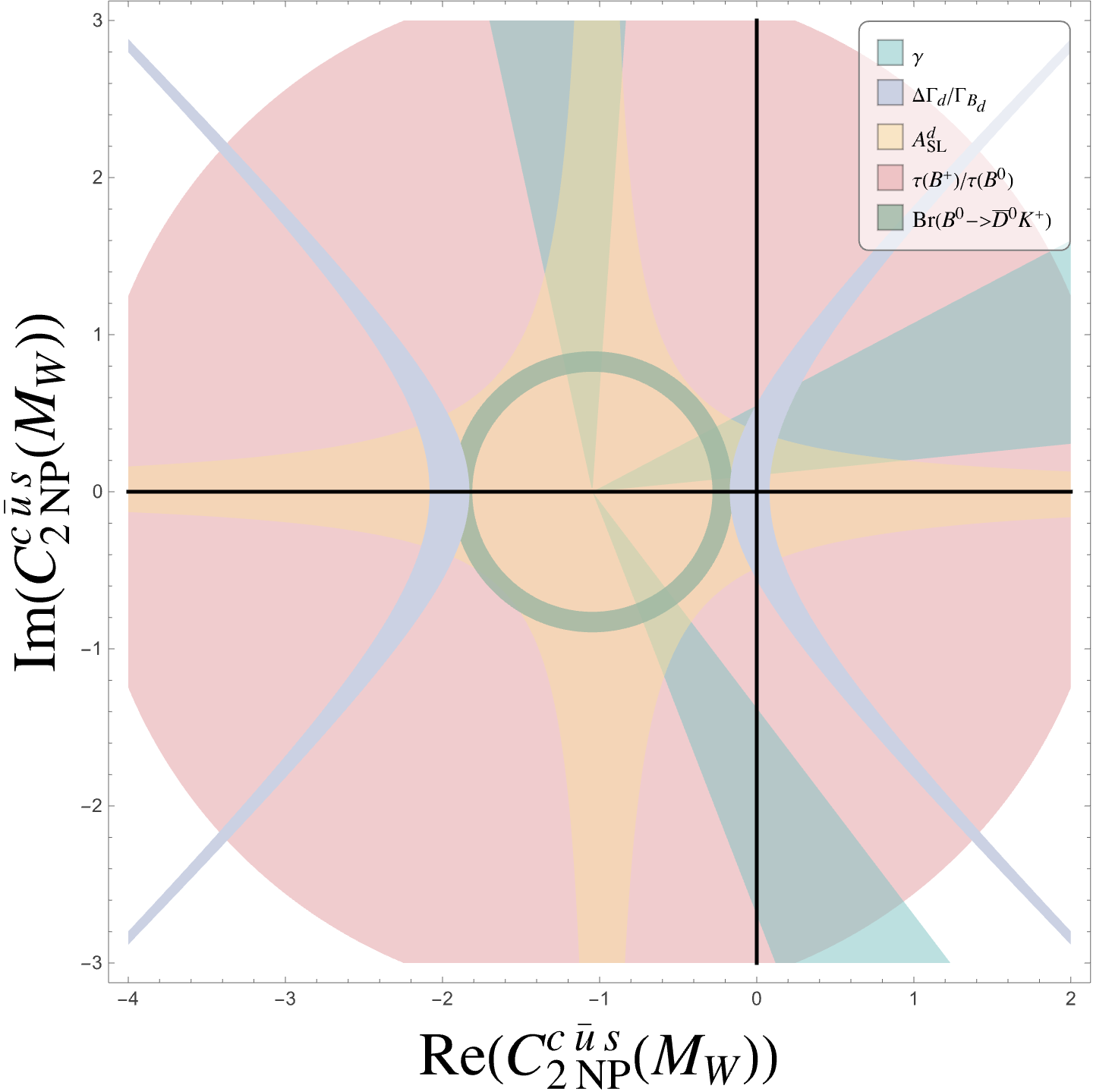}
    \end{minipage}
    \\
    \vspace{1.5cm}
    \centering
    \begin{minipage}{0.3\textwidth}
        \centering
        \includegraphics[width=\linewidth]{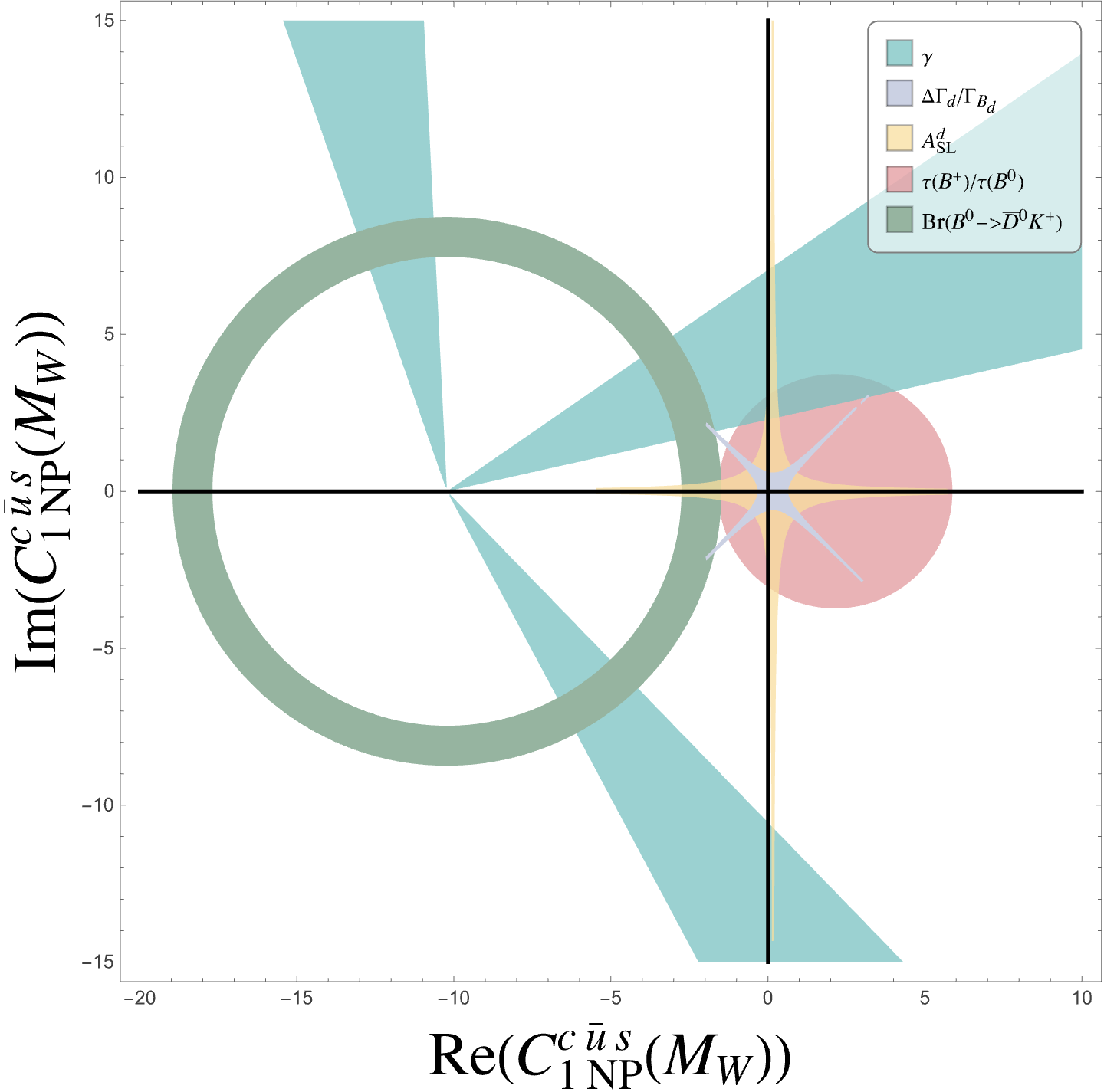}
    \end{minipage}
    \qquad \qquad \qquad 
    \begin{minipage}{0.3\textwidth}
        \centering
        \includegraphics[width=\linewidth]{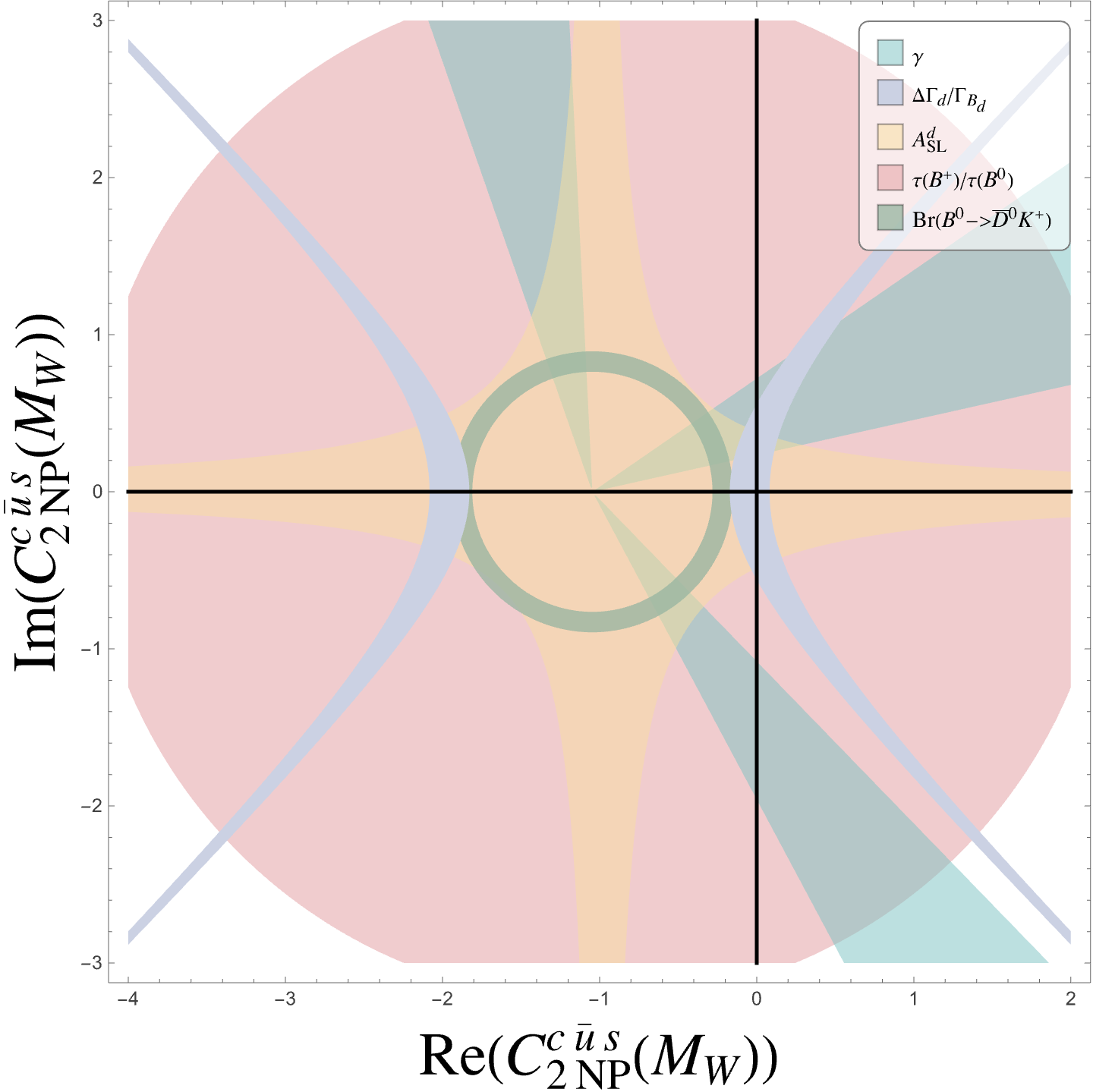}
    \end{minipage}
    \caption{\justifying{Combined constraints on the NP Wilson coefficients $C_{1,\text{NP}}^{c\bar{u}s}(M_W)$ and $C_{2,\text{NP}}^{c\bar{u}s}(M_W)$ from various flavor observables for $B^+\to \bar D^0K^+$ at $2\sigma$ level. The upper panels are obtained using $\gamma_{\text{indirect}} = 73.27^\circ$, while the lower panels serve as a robustness check using $\gamma_{\text{direct}} = 66.4^\circ$.}}
    \label{fig:constraints_BDK}
\end{figure}
\begin{figure}[H]
    \centering
    \begin{minipage}{0.37\textwidth}
        \centering
        \includegraphics[width=\linewidth]{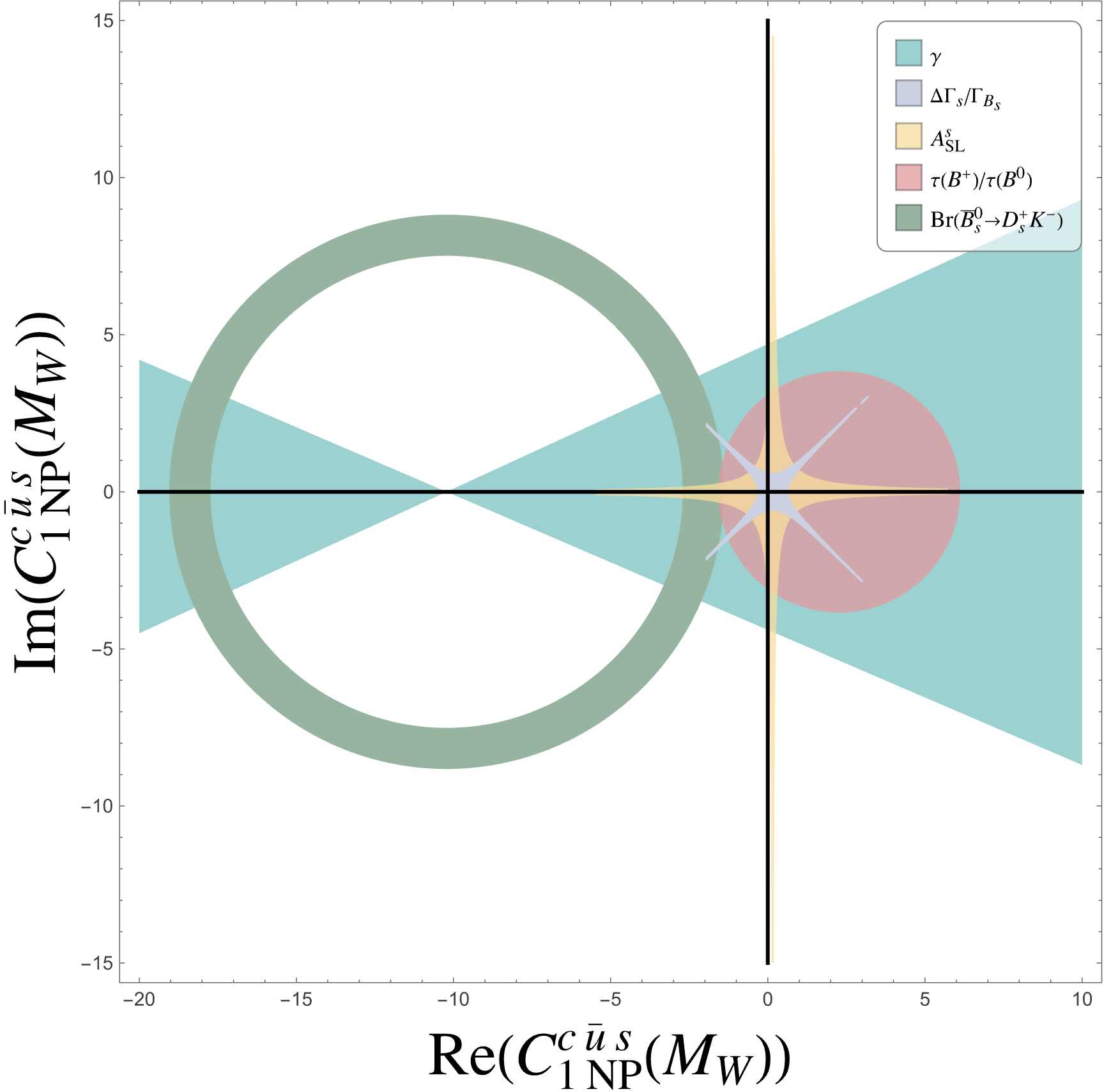} 
    \end{minipage}
    \qquad \qquad \qquad 
    \begin{minipage}{0.37\textwidth}
        \centering
        \includegraphics[width=\linewidth]{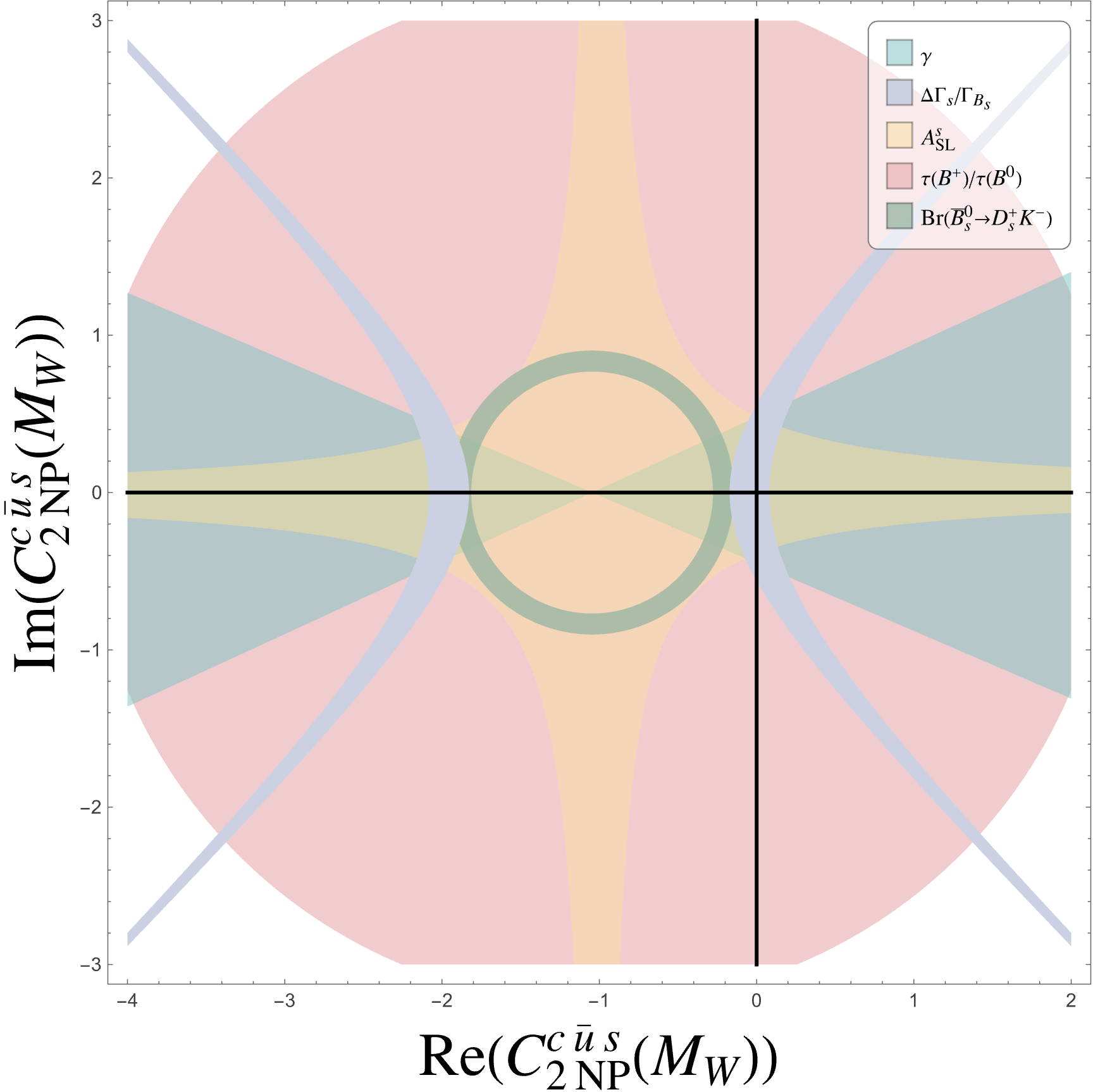}
    \end{minipage}
    \\
    \vspace{1.5cm}
    \begin{minipage}{0.37\textwidth}
        \centering
        \includegraphics[width=\linewidth]{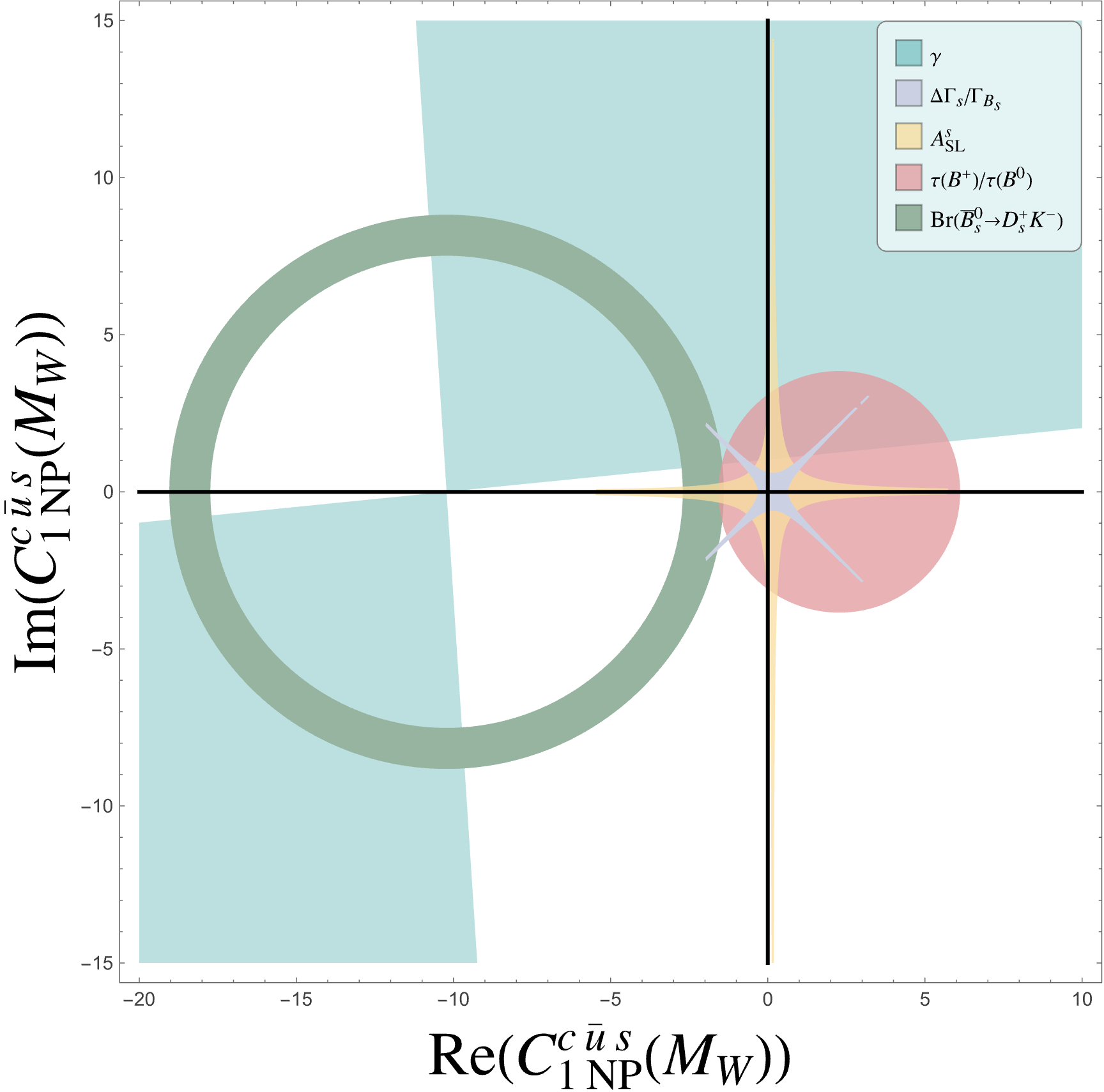} 
    \end{minipage}
    \qquad \qquad \qquad 
    \begin{minipage}{0.37\textwidth}
        \centering
        \includegraphics[width=\linewidth]{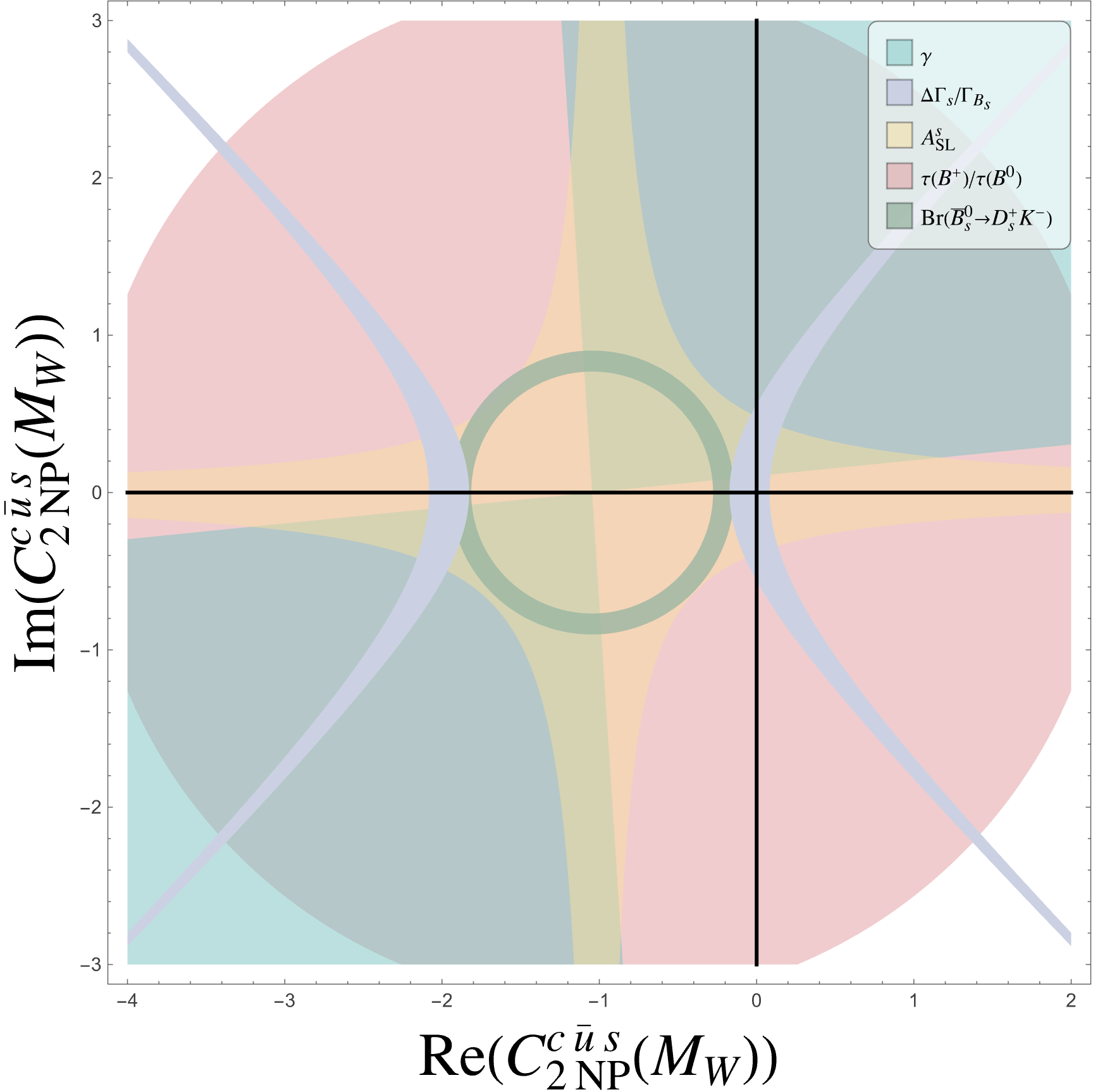}
    \end{minipage}
    \caption{\justifying{Combined constraints on the NP Wilson coefficients $C_{1,\text{NP}}^{c\bar{u}s}(M_W)$ and $C_{2,\text{NP}}^{c\bar{u}s}(M_W)$ for $\bar B_s^0 \to D^+_s K^-$ at the $2\sigma$ level. The upper panels correspond to the range $\gamma^{exp} \in [62^\circ, 86^\circ]$ \cite{LHCb:2017hkl}, while the lower panels correspond to $\gamma^{exp} \in [101^\circ, 145^\circ]$ \cite{LHCb:2024xyw}.}}
    \label{fig:constraints_BsDsK}
\end{figure}

For each channel, the combined allowed region is identified as the overlap of the individual constraints. For both the $b\to c\bar u s$ and $b\to c\bar u d$ transitions, the allowed regions show tension with the SM point~(the crossing point for black solid lines), in agreement with earlier observations, e.g. Ref.~\cite{Bordone:2020gao}. For the $b\to c\bar u d$ transition, a common overlap appears already at the $1\sigma$ level in the $C_1^{\rm NP}(M_W) = 0$ scenario, while the SM point lies outside this region; see the right panels of Figs.~\ref{fig:constraints_BDstarPi} and \ref{fig:constraints_BDPi}. For the $b\to c\bar u s$ transition, the individual $1\sigma$ constraints do not have a common overlap once the mixing observables are included, and we therefore show the viable region at the $2\sigma$ level in Figs.~\ref{fig:constraints_BDK} and \ref{fig:constraints_BsDsK}. Overall, the constraints tend to favor nonzero NP contributions, especially in the $C_1^{\rm NP}(M_W) = 0$ scenario.
\par
The two one-operator scenarios lead to visibly different constraint patterns. As shown representatively in Fig.~\ref{fig:constraints_BDK}, the constraints from $\gamma$ and the branching ratios are weaker in the $C_2^{\rm NP}(M_W)=0$ scenario than in the $C_1^{\rm NP}(M_W)=0$ scenario. This reflects the weaker sensitivity of these observables to NP contributions in the color-rearranged operator $\mathcal Q_1$ compared with the color-singlet operator $\mathcal Q_2$, due to the $1/N_c$ color suppression of the former.
\par
We also examine the impact of the different CKM angle $\gamma$ inputs. 
In addition to the results obtained with $\gamma_{\rm indirect}=73.27^\circ$, we show the constraints using $\gamma_{\rm direct}=66.4^\circ$ in Fig.~\ref{fig:constraints_BDK} for the $B^+ \to \bar D^0 K^+$ channel as a representative check. 
The change in $\gamma$ just rotates the corresponding allowed regions in the complex plane, while the overall overlap pattern remains qualitatively unchanged. 
As for $\gamma_{\text{CKMfitter}}$, it gives results visually indistinguishable from those obtained from $\gamma_{\rm direct}=66.4^\circ$; hence the plots only for $\gamma_{\rm indirect}$ and $\gamma_{\rm direct}$ are displayed. 
We have verified that the other decay channels show similar behavior when alternative $\gamma$ inputs are used.
\par
\subsection{$B_d - \bar B_d$ mixing and direct $CP$ violation}
\begin{figure}[htbp]
    \centering
    \begin{minipage}{0.85\textwidth}
        \centering
        \includegraphics[width=\linewidth]{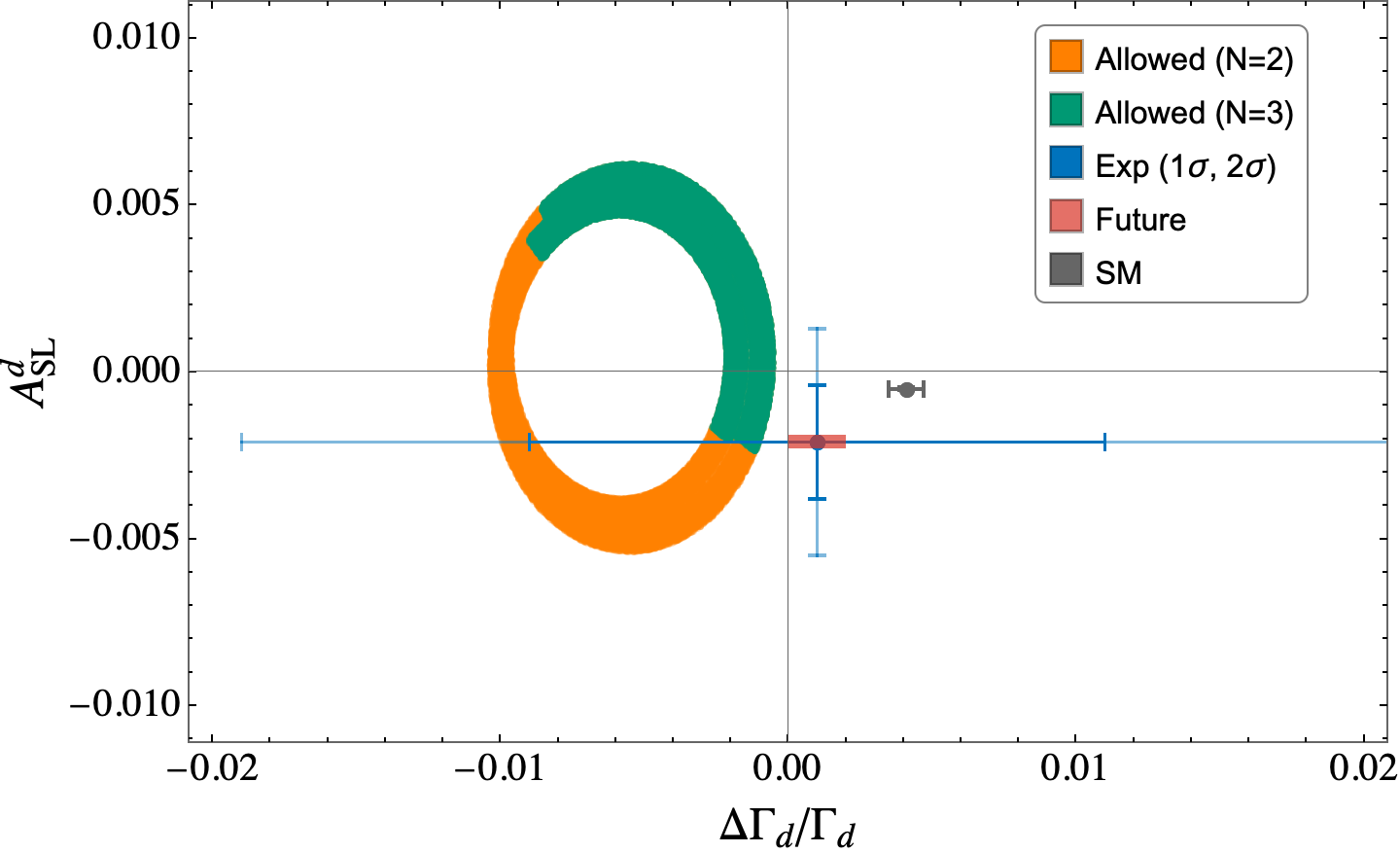}
    \end{minipage}
    \vspace{0.3cm} 

    \caption{\justifying{The predicted regions in the ($\Delta\Gamma_d/\Gamma_d$, $A_{\text{SL}}^d$) plane for the baseline scenario where $C_1^{\text{NP}}(M_W) = 0$ (i.e., NP only enters via $C_2^{\text{NP}}(M_W)$). The blue and gray error bars represent the current experimental constraints \cite{ParticleDataGroup:2024cfk, HeavyFlavorAveragingGroupHFLAV:2024ctg} and the SM predictions \cite{Albrecht:2024oyn}, respectively. The horizontal red bands correspond to the future experimental sensitivity \cite{LHCb:2018roe,Cerri:2018ypt}.}}
    \label{fig:Mixing_5Plots}
\end{figure}

\begin{figure}[H]
    \centering
    \setlength{\tabcolsep}{2pt}
    
    \includegraphics[height=0.21\textheight]{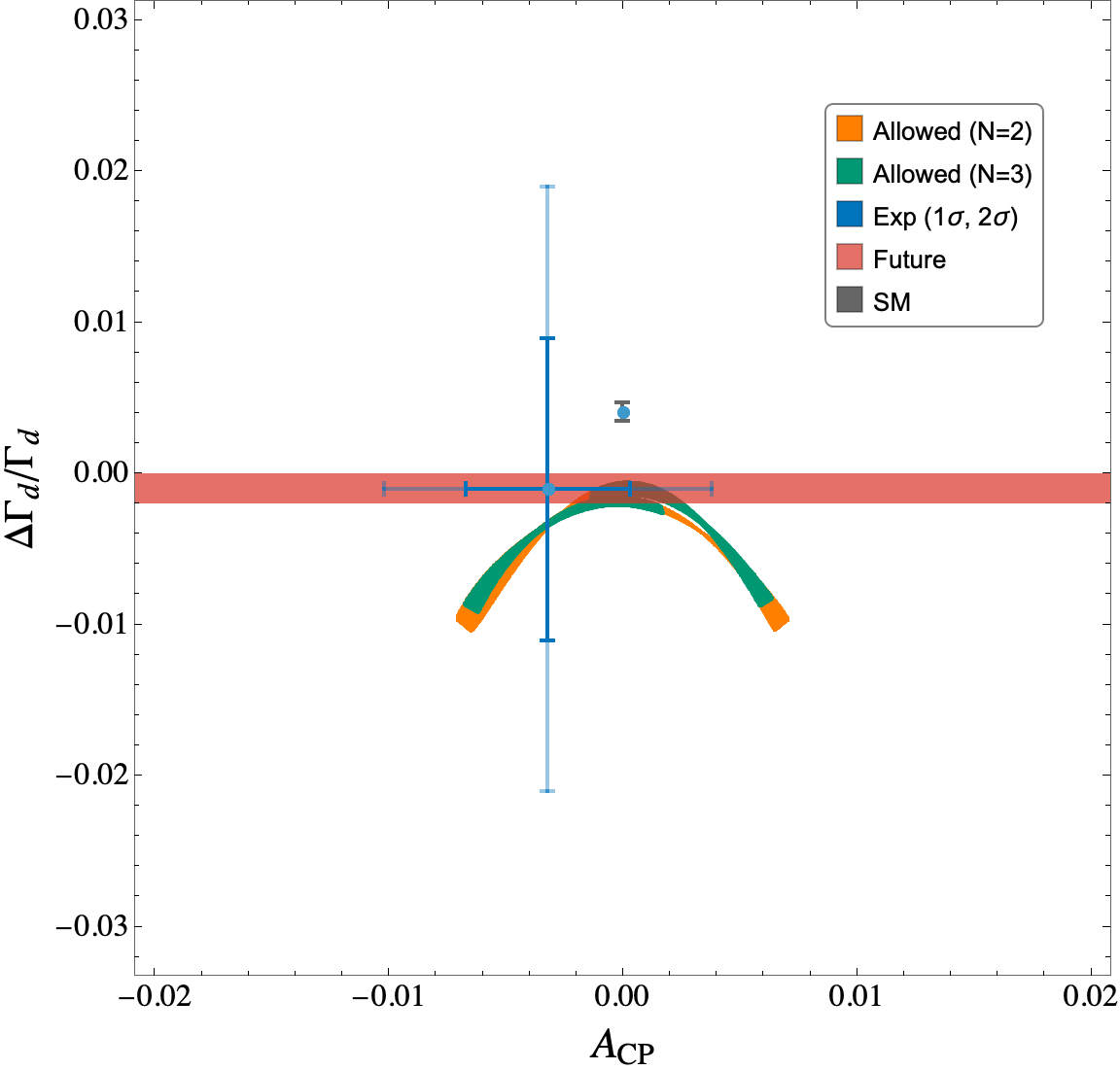}
    \includegraphics[height=0.21\textheight]{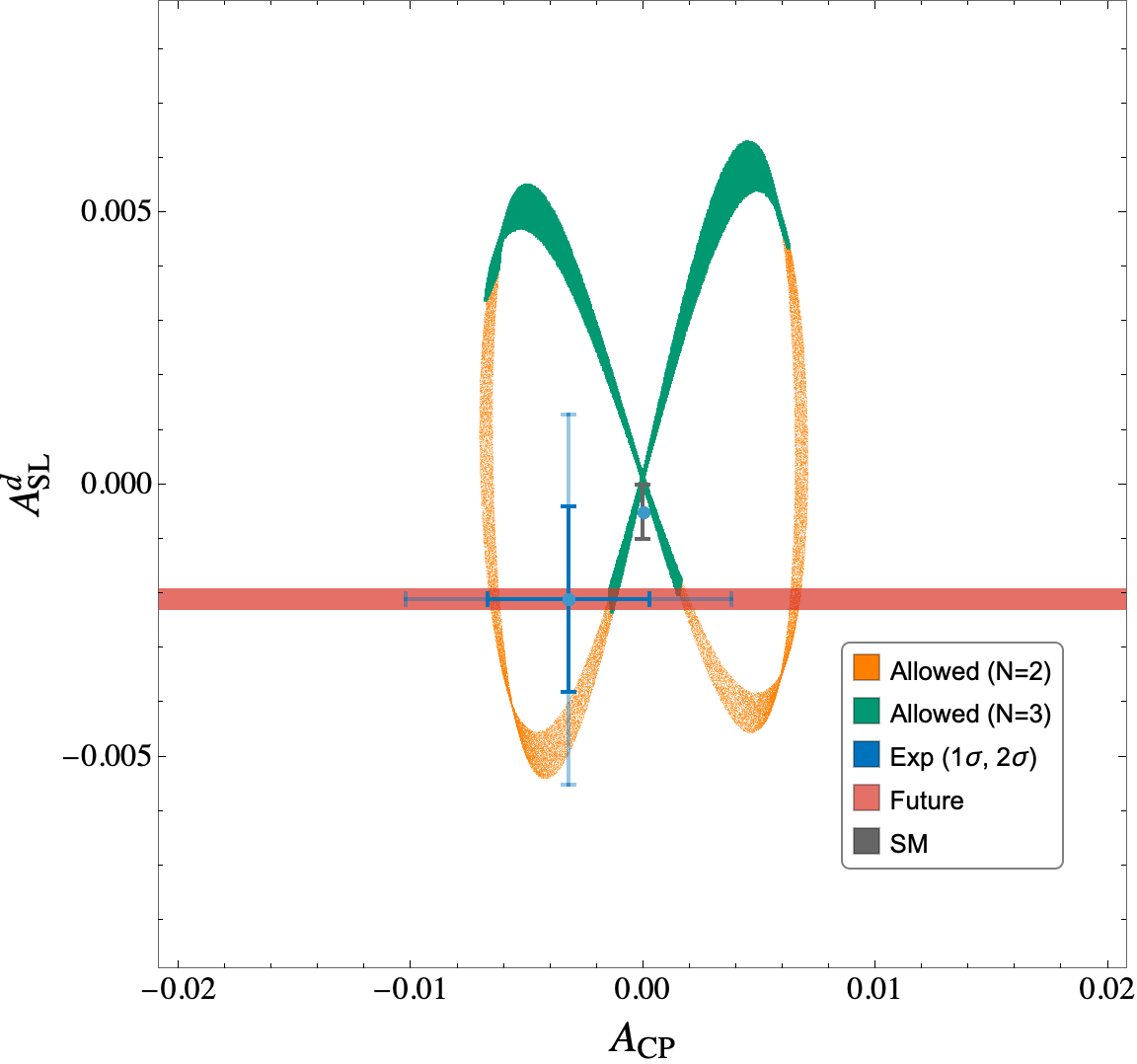}
    \\
    \vspace{-0.5cm}

    \includegraphics[height=0.21\textheight]{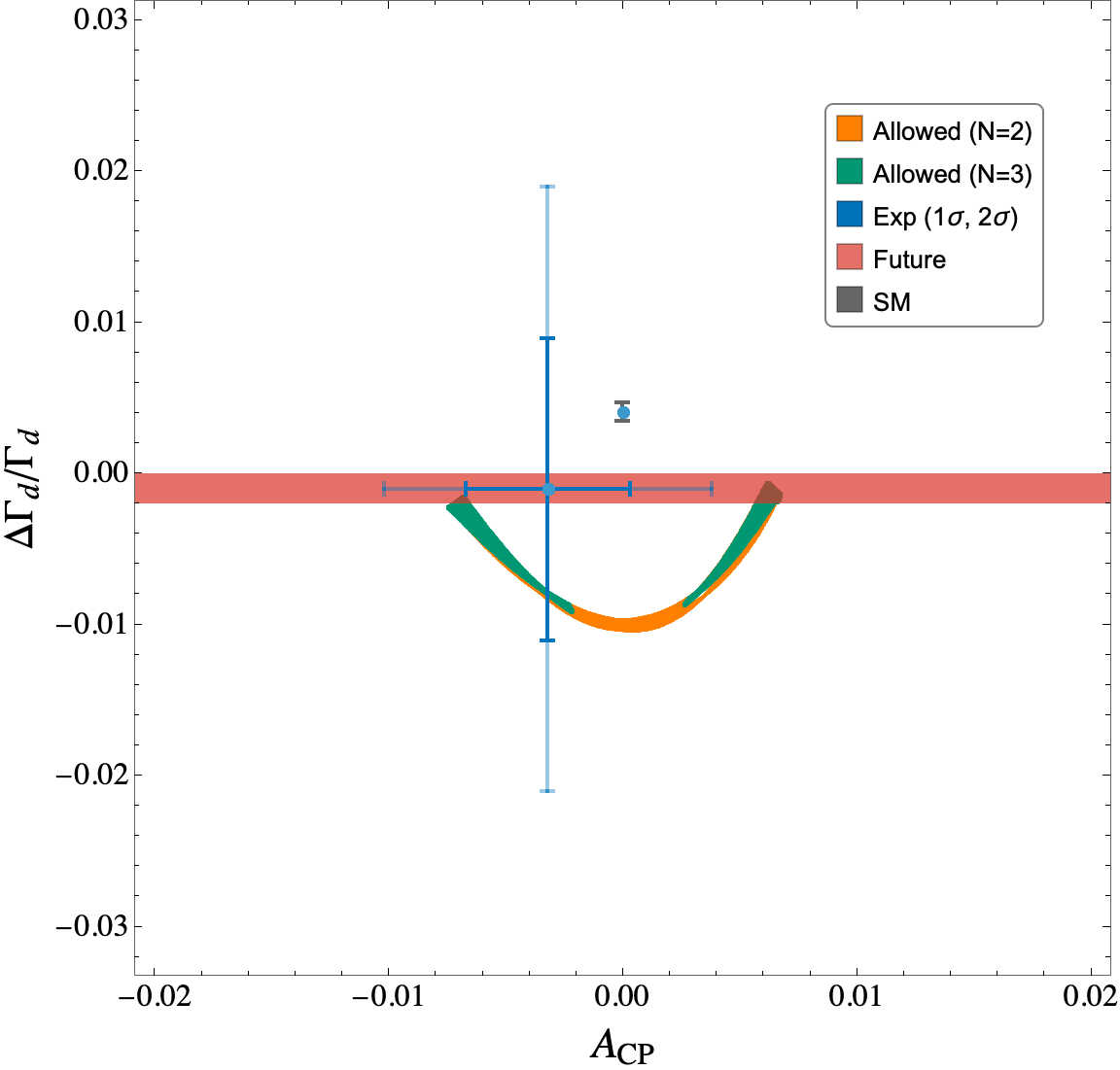}
    \includegraphics[height=0.21\textheight]{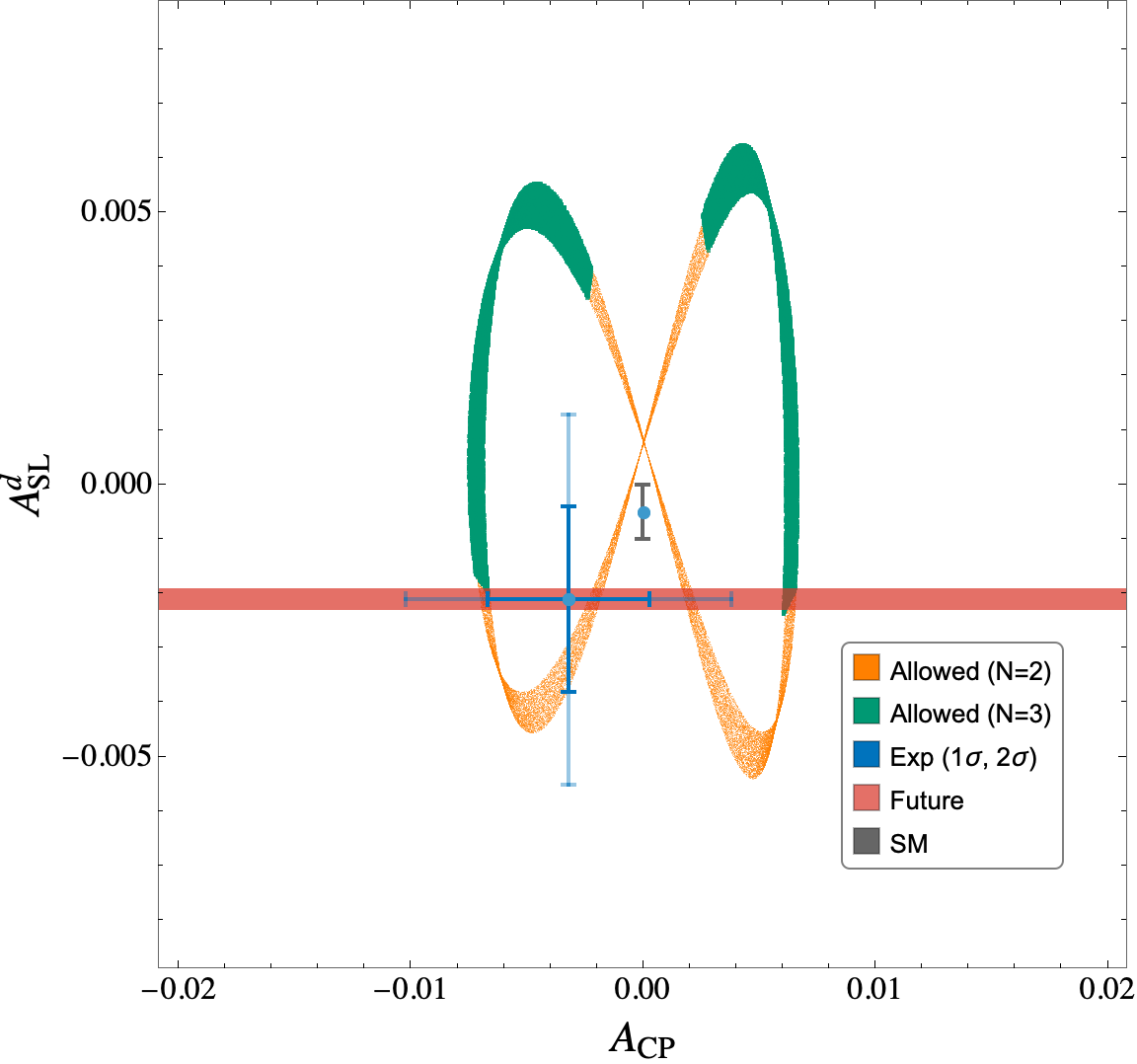}
    \\
    \vspace{-0.5cm}

    \includegraphics[height=0.21\textheight]{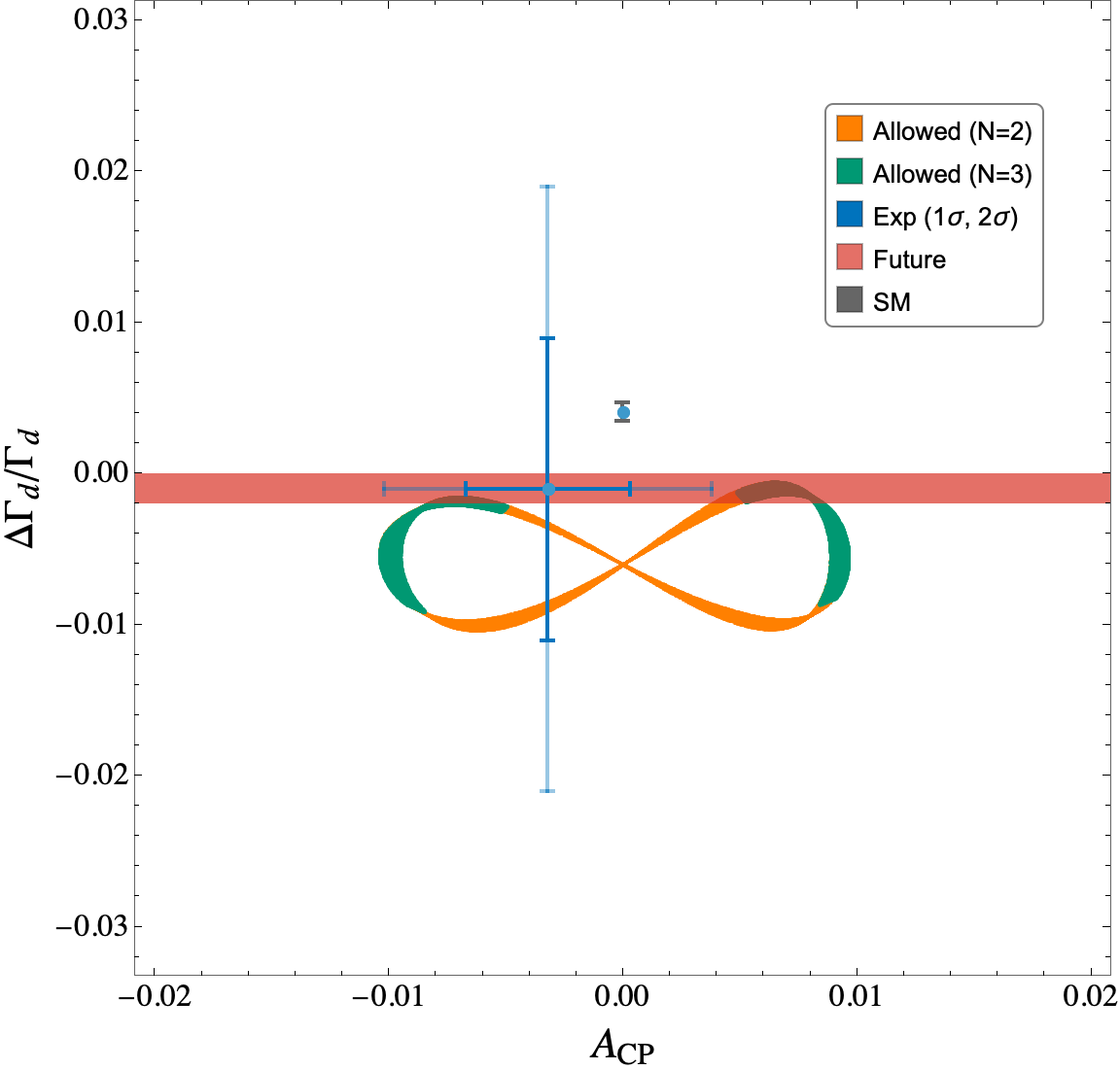}
    \includegraphics[height=0.21\textheight]{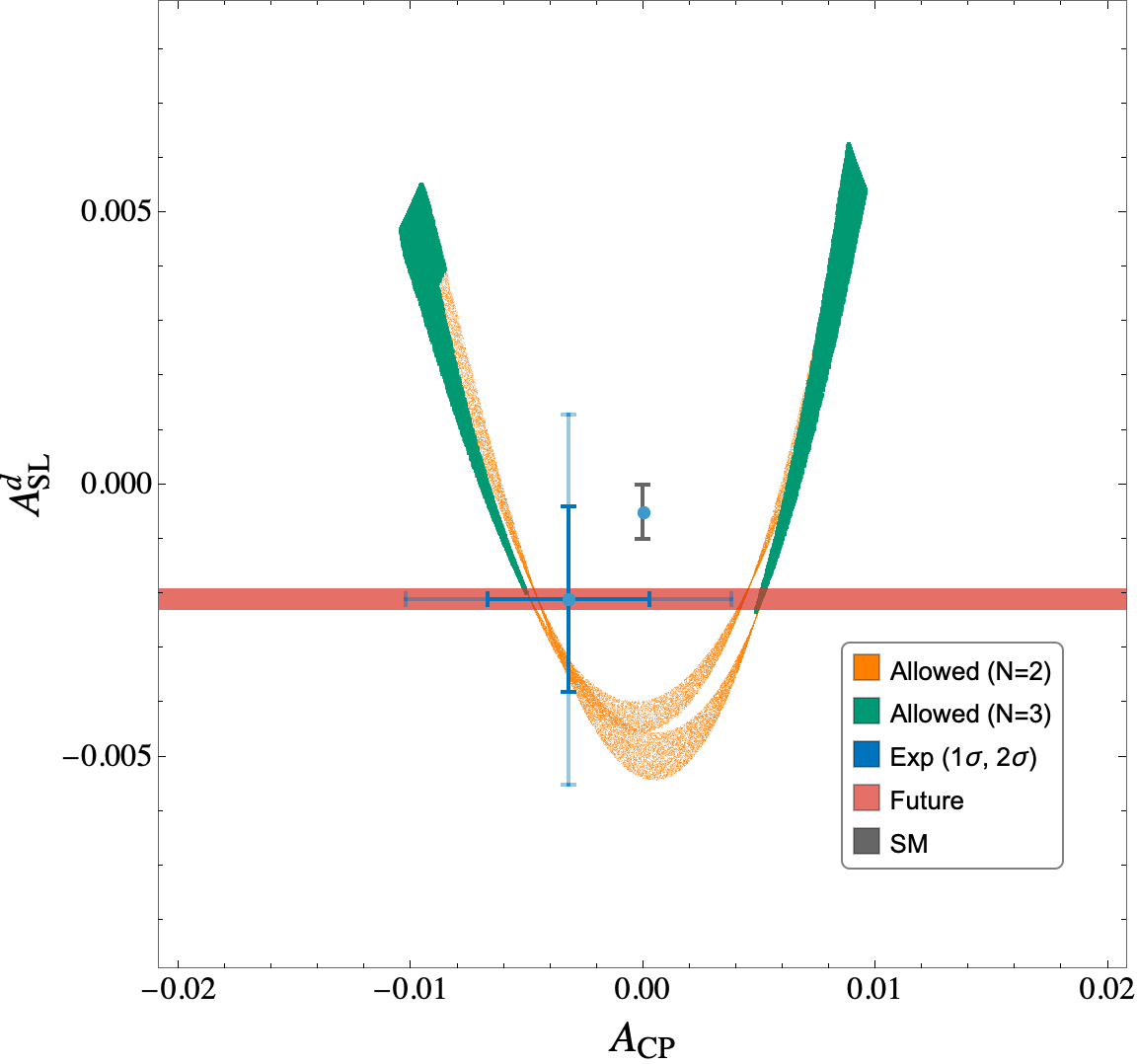} \\
    \vspace{-0.5cm}
    
    \includegraphics[height=0.21\textheight]{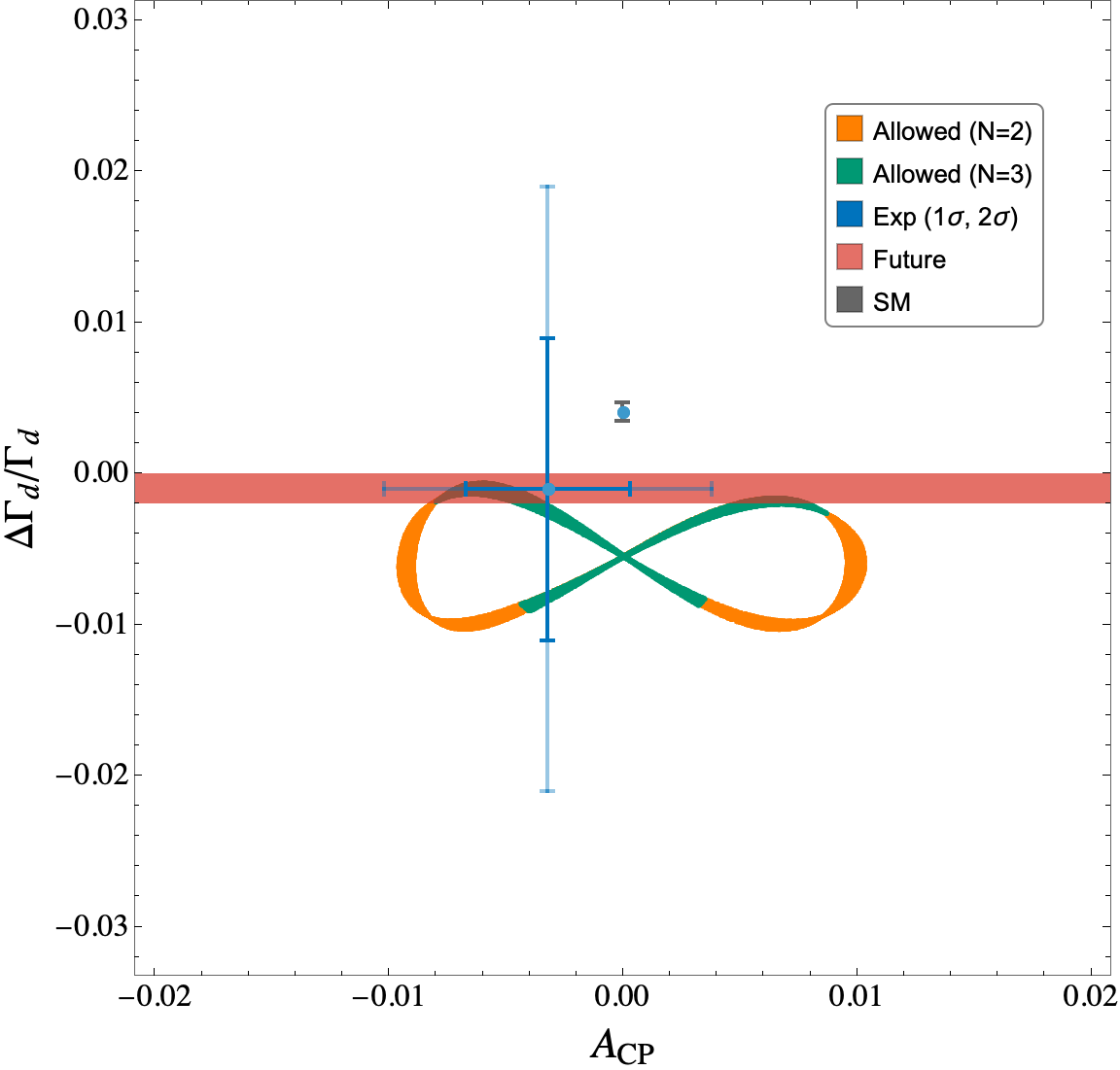}
    \includegraphics[height=0.21\textheight]{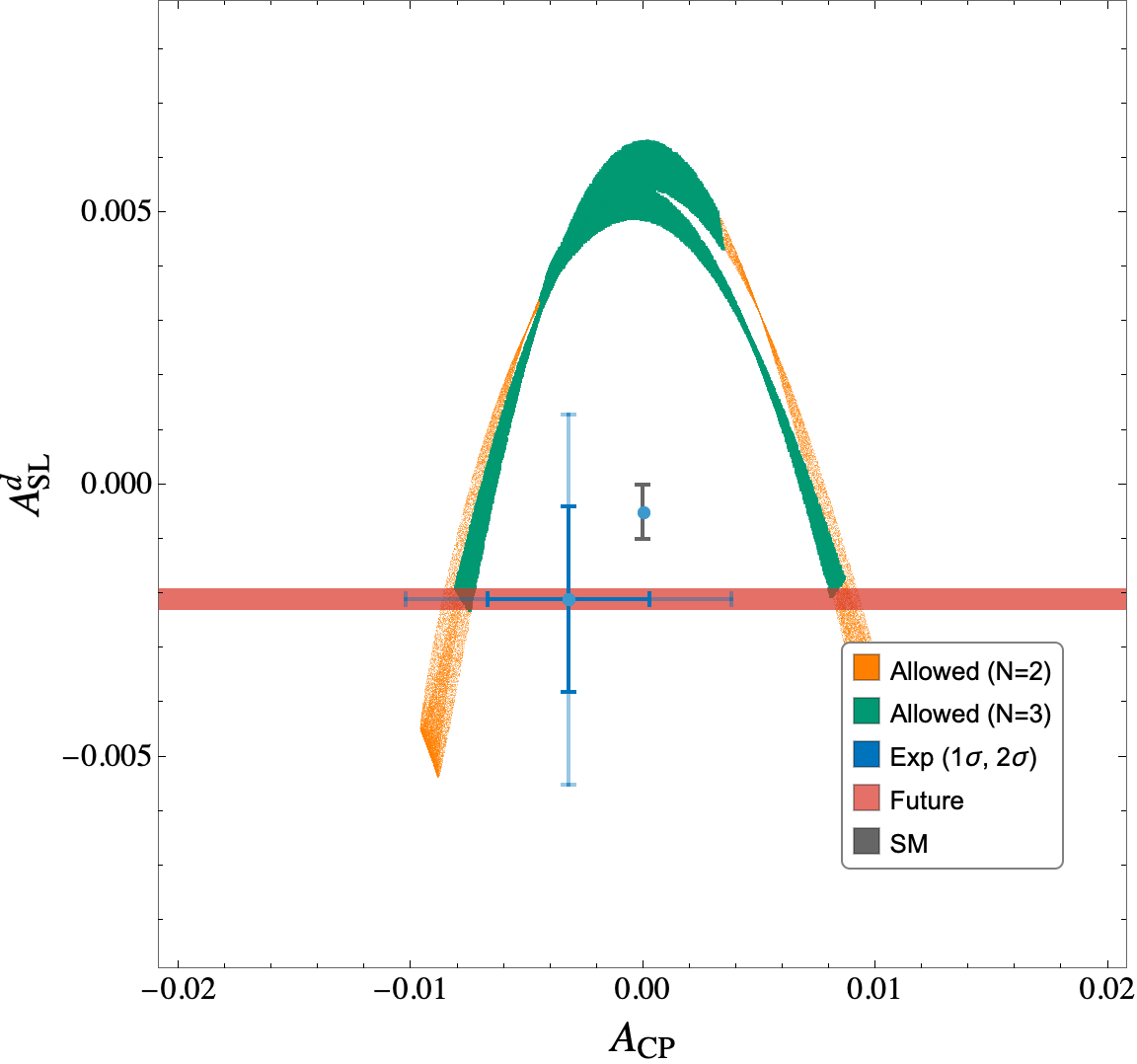} 
    
    \caption{\justifying{
    The predicted regions in the ($\Delta\Gamma_d/\Gamma_d$, $A_{\text{CP}}$) [left column] and ($A_{\text{SL}}^d$, $A_{\text{CP}}$) [right column] planes for four benchmark scenarios: $C_1^{\text{NP}}(M_W) = 0.01$, $0.01i$, $0.01+0.01i$, and $0.01-0.01i$ (from top to bottom). The blue and gray error bars represent the current experimental constraints \cite{ParticleDataGroup:2024cfk, HeavyFlavorAveragingGroupHFLAV:2024ctg} (at the $1\sigma$ and $2\sigma$ levels) and the SM predictions \cite{Albrecht:2024oyn} on the $y$-axis observables, respectively. The horizontal red bands correspond to the expectations for future experimental precision \cite{LHCb:2018roe,Cerri:2018ypt}.
    }}
    \label{fig:8}
\end{figure}

\begin{figure}[H]
    \centering
    \setlength{\tabcolsep}{2pt}
    
    \includegraphics[height=0.21\textheight]{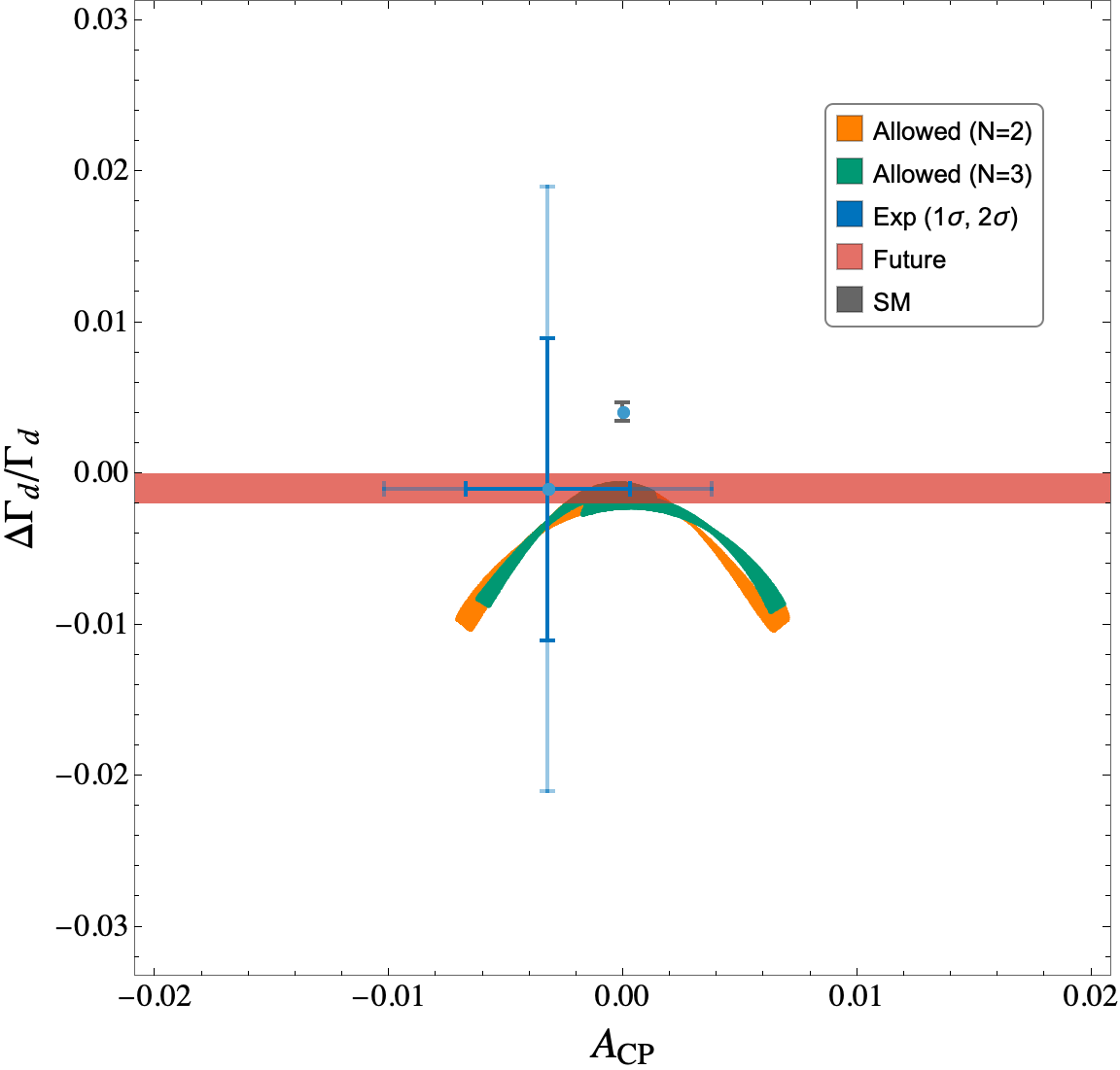}
    \includegraphics[height=0.21\textheight]{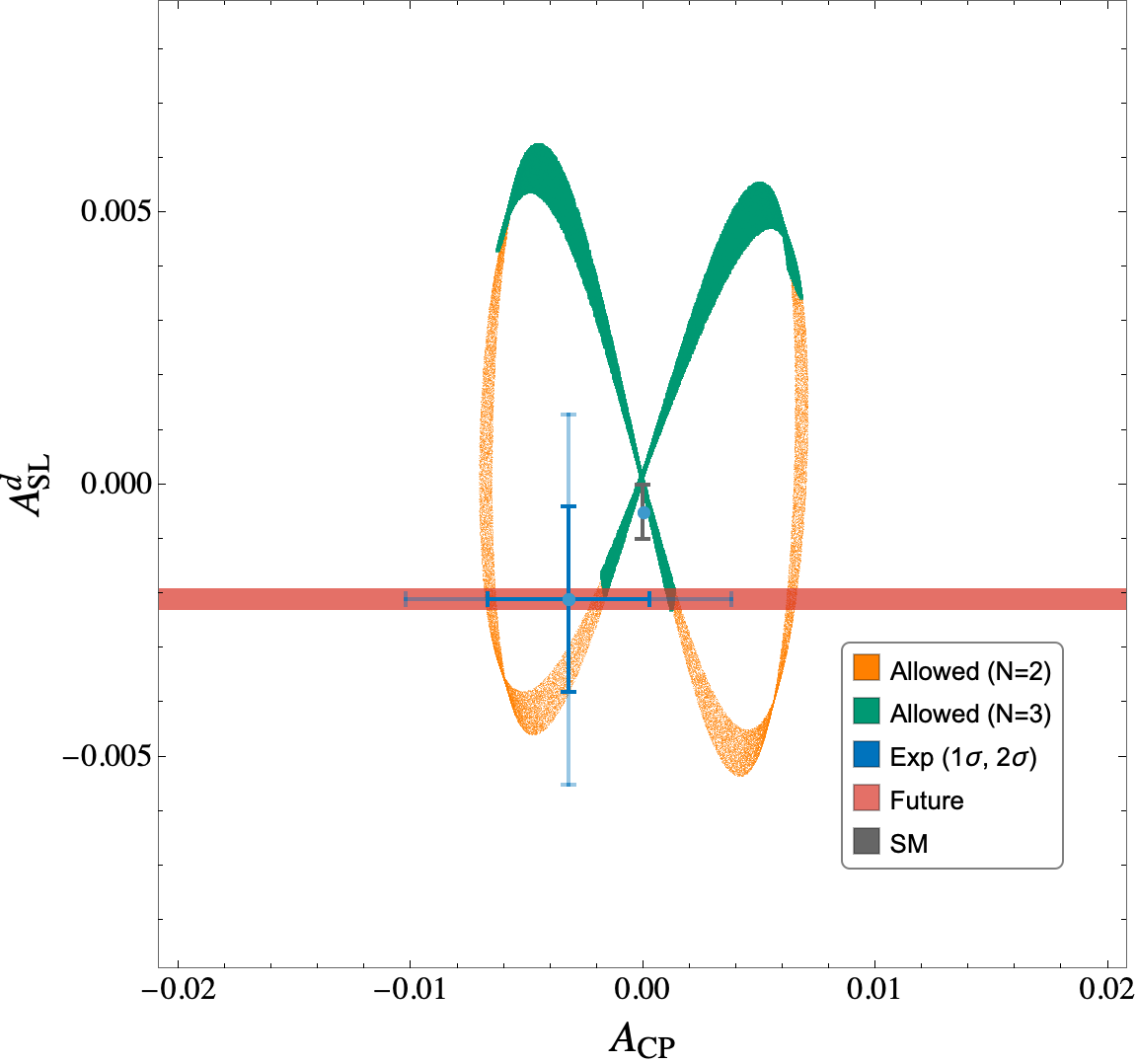}
    \\
    \vspace{-0.5cm}

    \includegraphics[height=0.21\textheight]{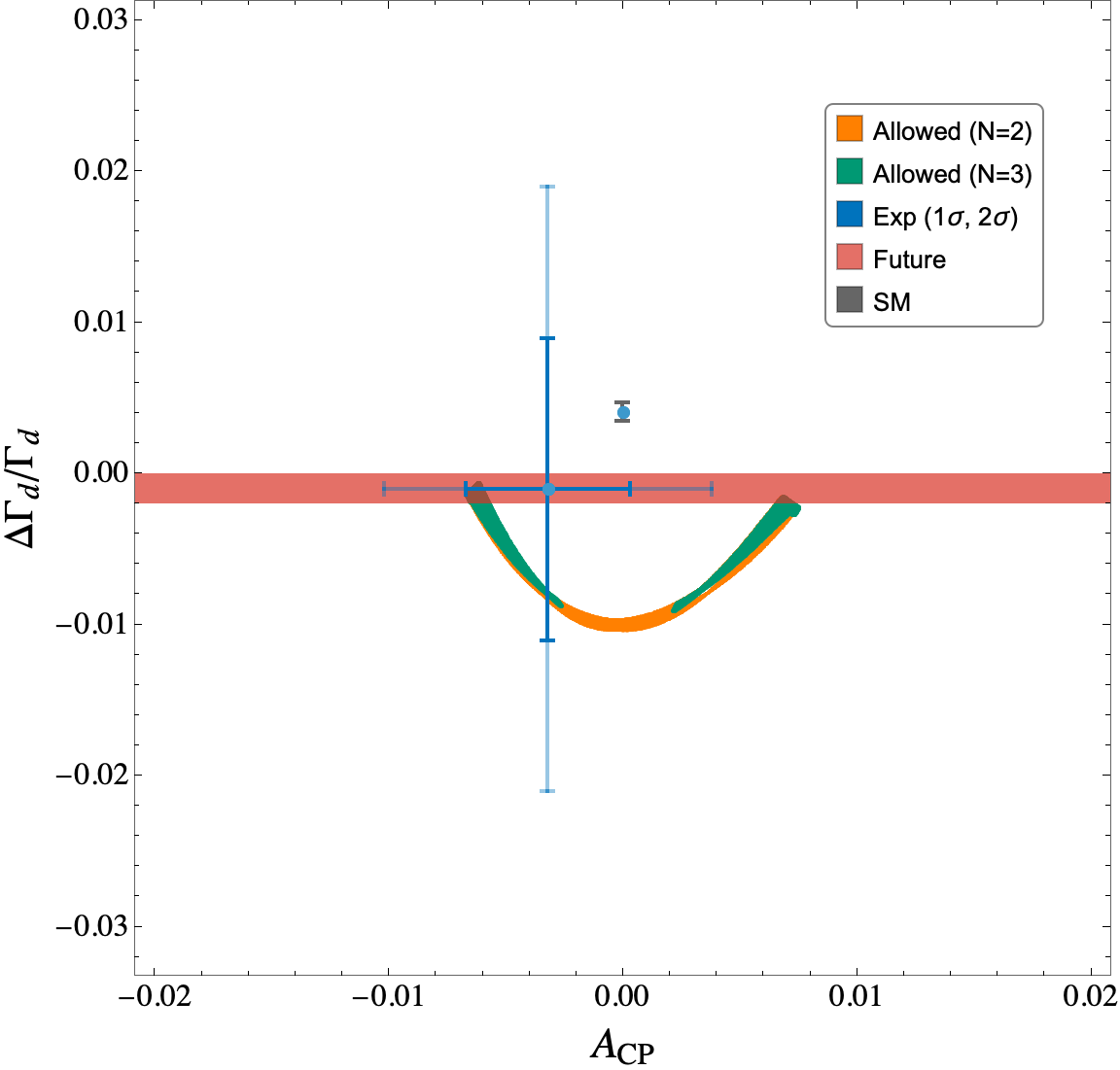}
    \includegraphics[height=0.21\textheight]{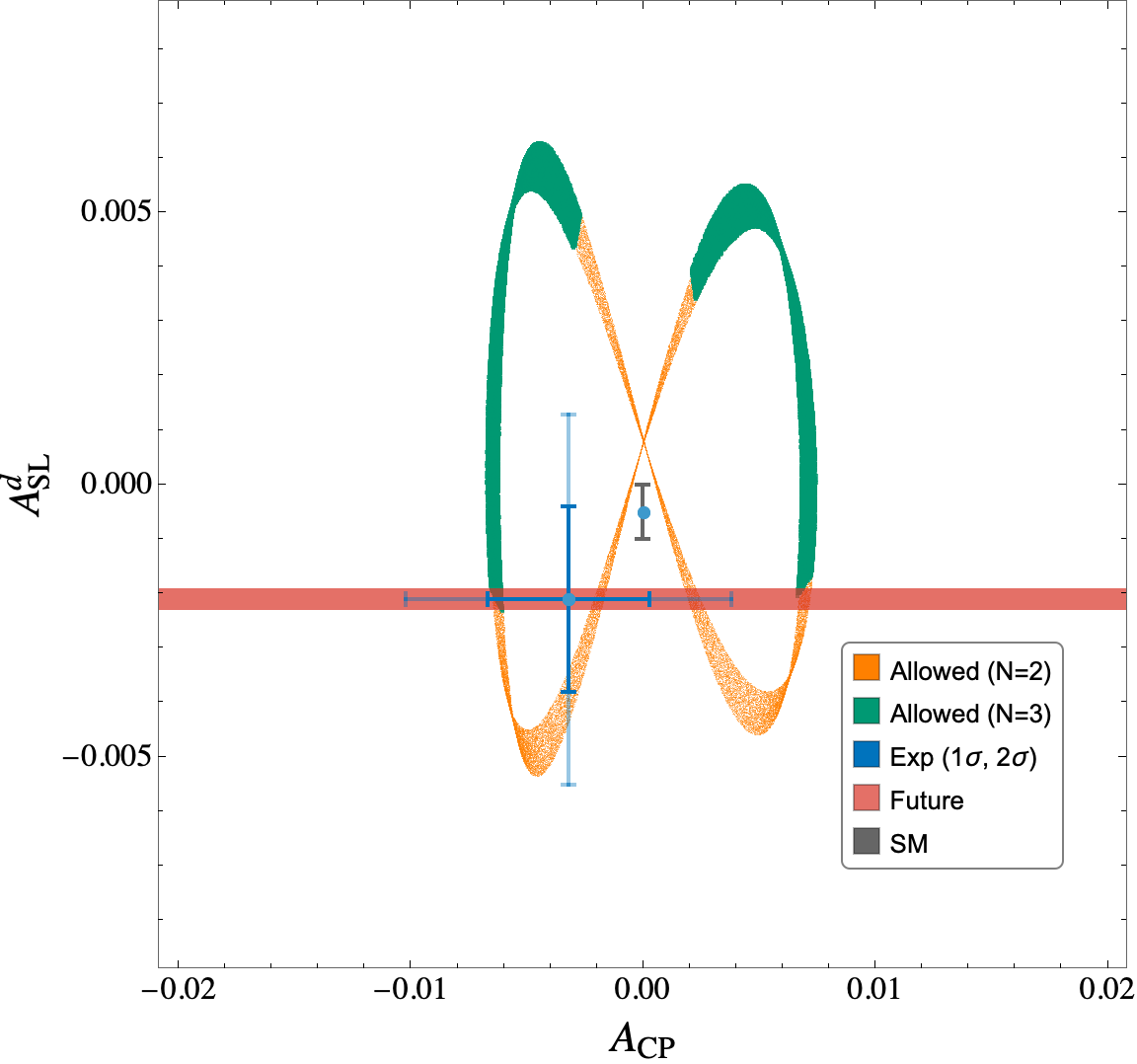}
    \\
    \vspace{-0.5cm}

    \includegraphics[height=0.21\textheight]{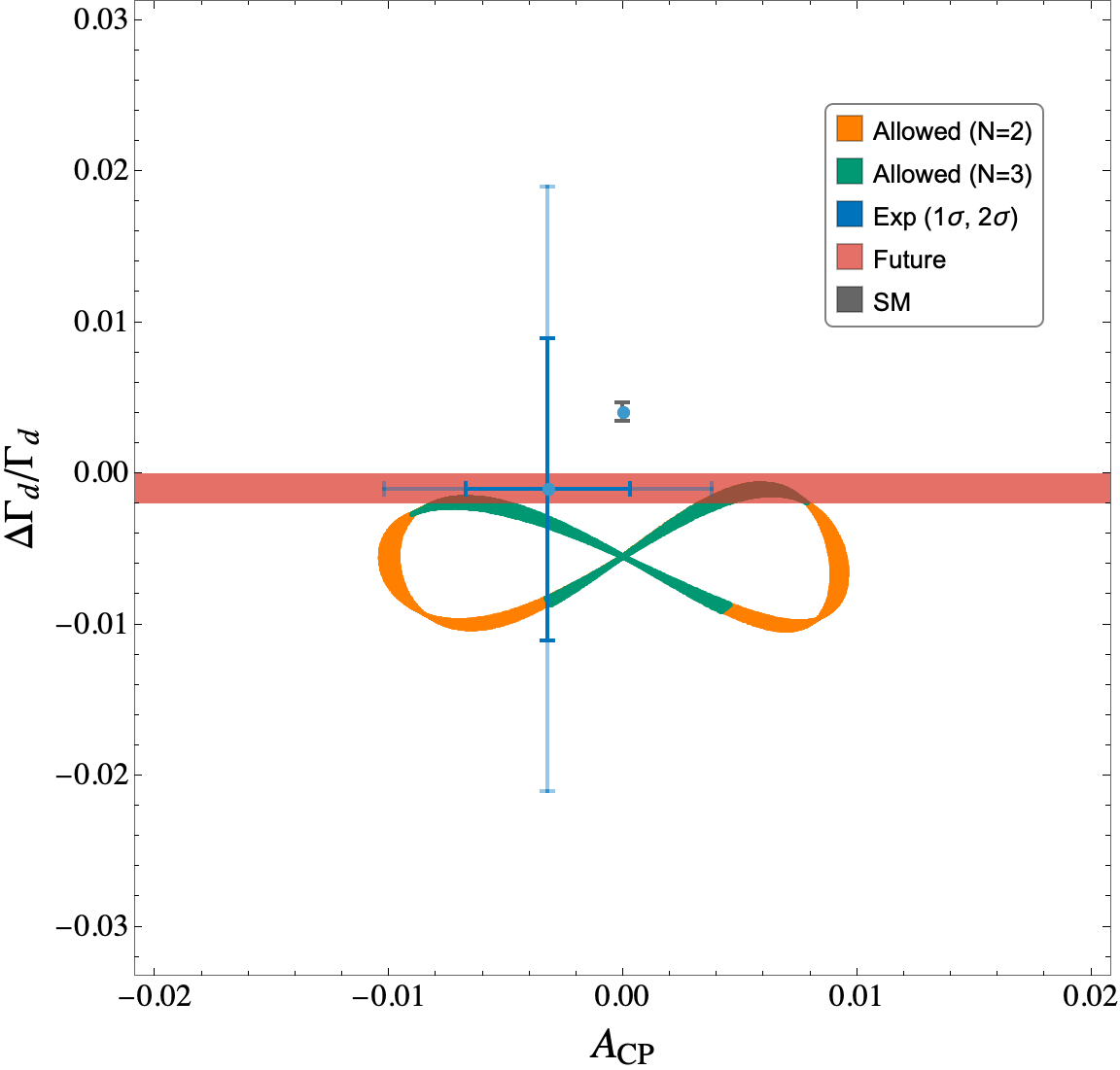}
    \includegraphics[height=0.21\textheight]{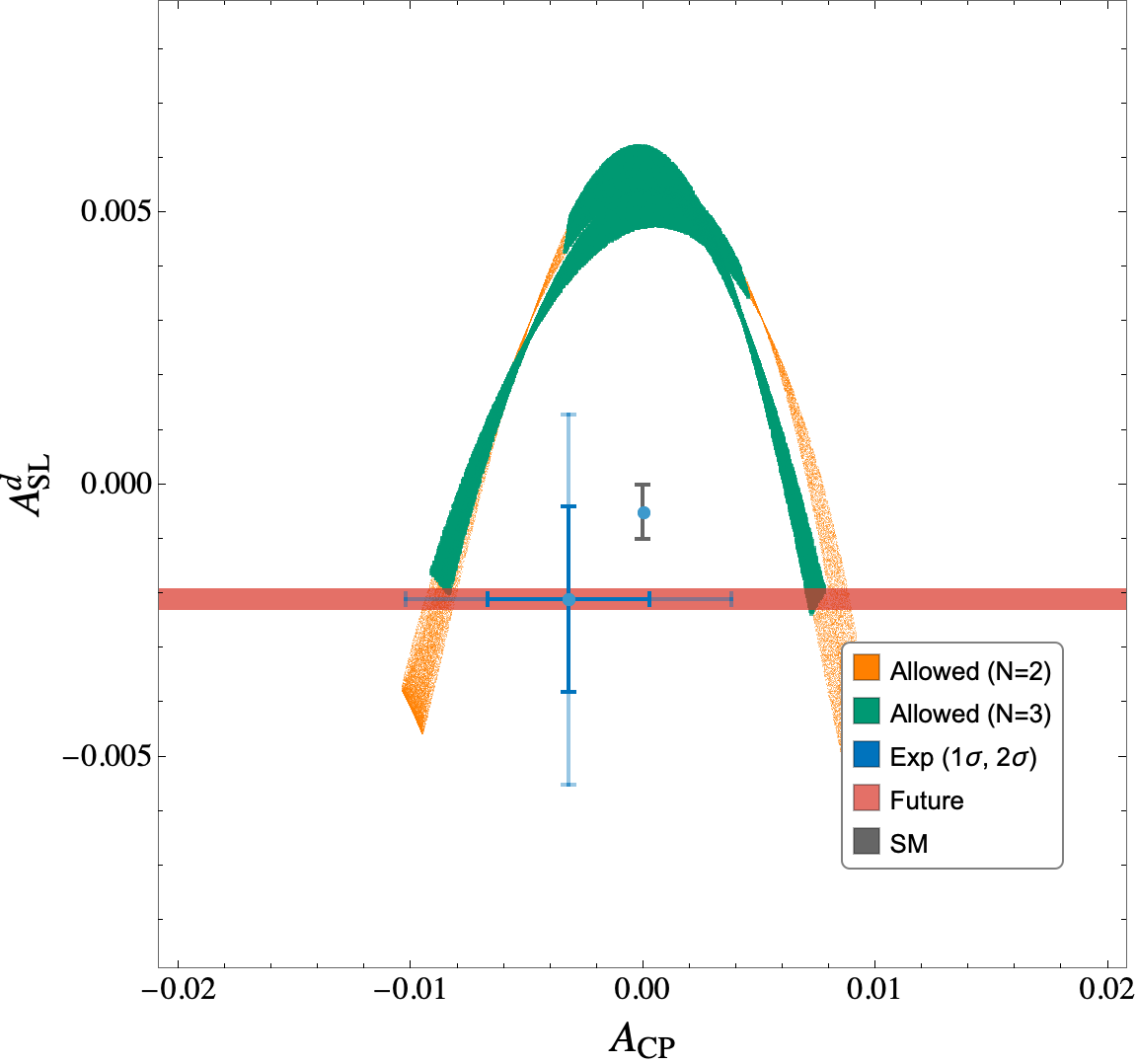} \\
    \vspace{-0.5cm}
    
    \includegraphics[height=0.21\textheight]{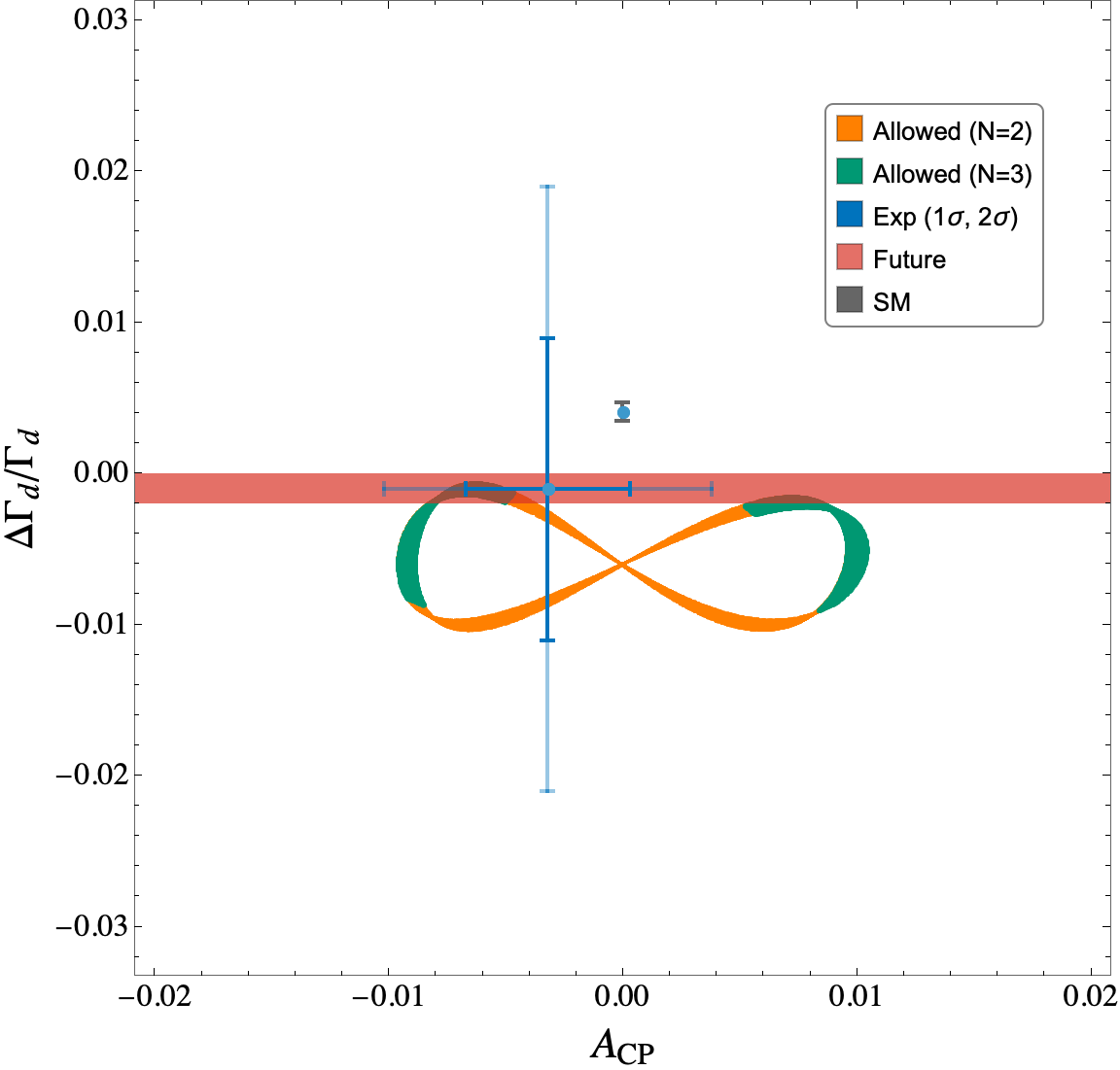}
    \includegraphics[height=0.21\textheight]{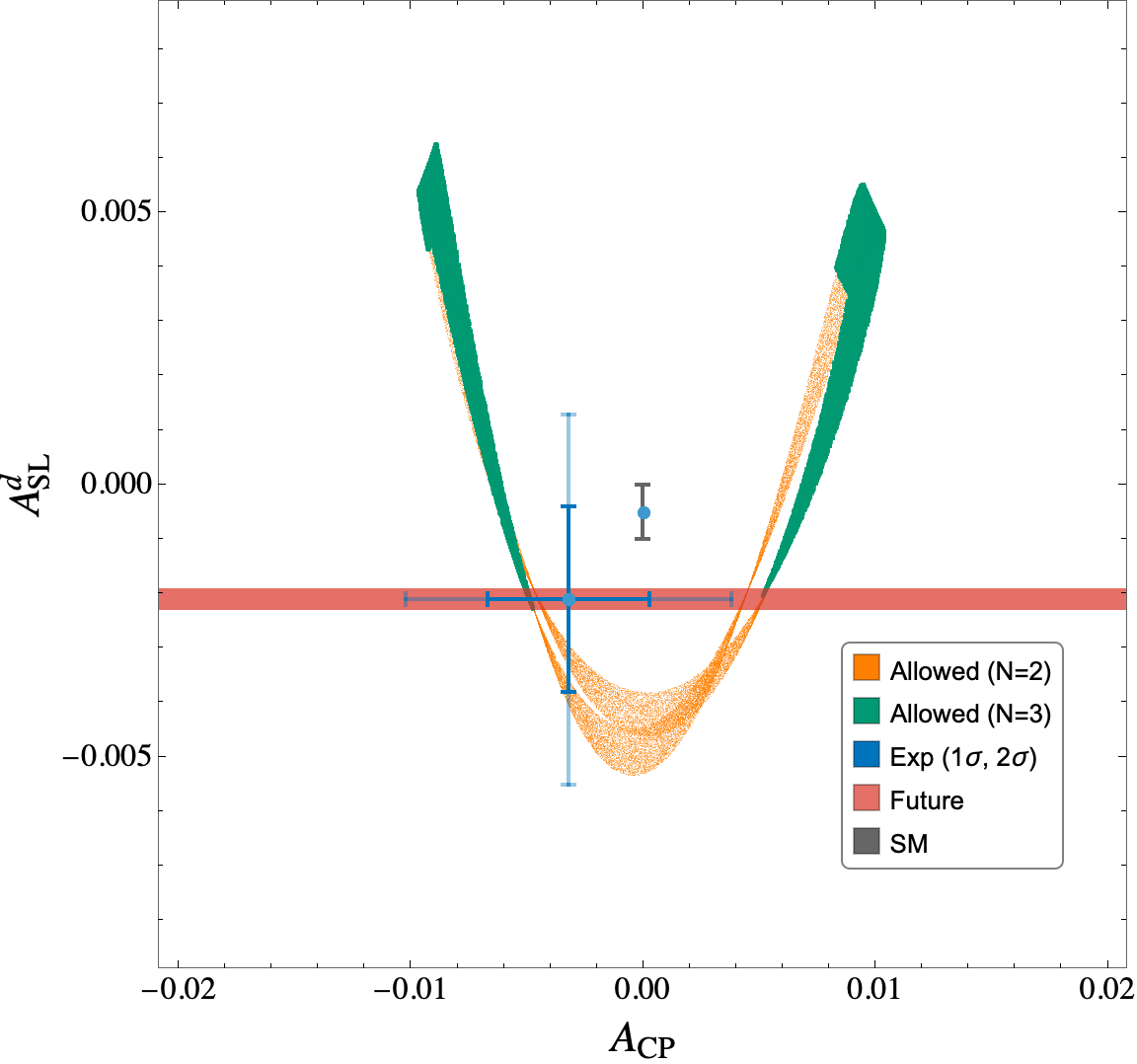} 
    
    \caption{\justifying{
    The predicted regions in the ($\Delta\Gamma_d/\Gamma_d$, $A_{\text{CP}}$) [left column] and ($A_{\text{SL}}^d$, $A_{\text{CP}}$) [right column] planes for four benchmark scenarios: $C_1^{\text{NP}}(M_W) = -0.01$, $-0.01i$, $-0.01+0.01i$, and $-0.01-0.01i$ (from top to bottom). The blue and gray error bars represent the current experimental constraints \cite{ParticleDataGroup:2024cfk, HeavyFlavorAveragingGroupHFLAV:2024ctg} (at the $1\sigma$ and $2\sigma$ levels) and the SM predictions \cite{Albrecht:2024oyn} on the $y$-axis observables, respectively. The horizontal red bands correspond to the expectations for future experimental precision \cite{LHCb:2018roe,Cerri:2018ypt}.
    }}
    \label{fig:App_CP}
\end{figure}

In Fig.~\ref{fig:Mixing_5Plots}, the predicted correlation between the width difference normalized by the total width $\Delta\Gamma_d/\Gamma_d$ and the semileptonic asymmetry $A_{\mathrm{SL}}^d$ is shown as allowed regions for the baseline scenario $C_1^{\text{NP}}(M_W) = 0$ with $C_2^{\text{NP}}(M_W) \neq 0$. The green region is derived from a parameter space that satisfies three independent constraints: (i) the direct measurement of the CKM angle $\gamma$, (ii) the branching ratio of the $B_d \to D^{-} \pi^+$ decay, and (iii) the lifetime ratio $\tau(B^+)/\tau(B_d^0)$, simultaneously at $68\%$ CL. For comparison, the orange region shows the prediction when the constraint from the CKM angle $\gamma$ is excluded from the set of constraints. It should be noted that the orange region also lies behind the green ones in Figs.~\ref{fig:Mixing_5Plots}, \ref{fig:8} and \ref{fig:App_CP}. The resulting allowed regions for $C_2^{\text{NP}}(M_W)$ are then mapped onto the $(\Delta\Gamma_d/\Gamma_d, A_{\mathrm{SL}}^d)$ plane to generate the displayed predictions. Also shown are values of the current experimental measurements, the SM predictions, and the projected future experimental sensitivity. As can be seen in Fig.~\ref{fig:Mixing_5Plots} (and also in Figs.~\ref{fig:8} and \ref{fig:App_CP} discussed later), predicted regions of the observables are significantly reduced if the constraint from $\gamma$ is considered additionally.

\newpage
\indent Furthermore, results including the direct $CP$ asymmetry, $A_{\text{CP}}$, in $B^- \to D^0 \pi^-$ decays are discussed in what follows. It is well known that this observable can be regarded as a clean signal of NP since it vanishes in the SM. Crucially, a nonzero $A_{\text{CP}}$ fundamentally requires the interference of at least two amplitudes possessing different weak phases and different strong phases. To obtain the prediction for $A_{\text{CP}}$, the PQCD expressions in Sec.~\ref{Sec:IVA} for evaluating the $T+C$ amplitude are used, where the necessary strong phases are naturally generated through the decay dynamics. We find that a scenario with $C_2^{\text{NP}}(M_W) \neq 0$ and $C_1^{\text{NP}}(M_W) = 0$ fails to provide the necessary relative weak phase difference, yielding $A_{\text{CP}} = 0$, which we both confirmed numerically and analytically. Therefore, introducing a nonvanishing $C_1^{\text{NP}}(M_W)$ is mandatory to generate the direct $CP$ asymmetry. We focus on eight representative benchmarks: $C_1^{\text{NP}}(M_W) = \pm 0.01$, $\pm 0.01i$, $\pm 0.01 \pm 0.01i$, and $\mp 0.01 \pm 0.01i$, which give testable size  of $A_{\text{CP}}$ in the future. The region of $C_2^{\text{NP}}(M_W)$ is scanned over the possible range as was done before. We also verified that under this small variation of $C_1^{\text{NP}}(M_W)$, Fig.~\ref{fig:Mixing_5Plots} is not significantly changed. In Figs.~\ref{fig:8} and \ref{fig:App_CP}, numerical predictions for $A_{\text{CP}}$ are displayed with correlated results for the $B_d^0-\bar{B}_d^0$ mixing observables for these four selected scenarios. As illustrated in Figs.~\ref{fig:8} and \ref{fig:App_CP}, a small NP contribution ($|C_1^{\text{NP}}(M_W)| \sim 0.01$) produces an $A_{\text{CP}}$ that is mostly within the $2\sigma$ range of the current PDG, while some of the regions are outside the range. Future improvements in the experimental precision of $A_{\text{CP}}$, together with those for $\Delta \Gamma_d/\Gamma_d$ and $A_{\mathrm{SL}}^d$, will enable us to test this NP scenario.

\section{conclusion}\label{Sec:VI}
In this work, we have performed a comprehensive analysis of the $b \to c\overline{u}q$ ($q=d,s$) transitions, aiming to address the persistent tension between the QCD factorization predictions and experimental data for color-allowed nonleptonic $B$ meson decays. Extending the effective Hamiltonian to include complex-valued NP contributions to the Wilson coefficients $C_1$ and $C_2$, we derived constraints by analyzing the combined allowed regions from branching ratios, the CKM angle $\gamma/\phi_3$, the lifetime ratio $\tau(B^+)/\tau(B^0)$, and neutral meson mixing observables ($\Delta\Gamma_q$ and $A_{\mathrm{SL}}^q$).
\par
For the branching ratios of the channels where amplitudes are proportional to $T+C$, the contribution of the color-suppressed diagram is model independently extracted from data, including the experimental values of branching ratios, factorization input for the ratio of $T$ diagrams, and meson masses, as in the discussion of $T+E$ processes \cite{Fleischer:2021cct}. This method enables us to discuss constraints on the NP Wilson coefficients for the combined analysis of branching ratios and $\gamma$ in a robust way. Furthermore, $CP$ violation in $B^-\to D^0\pi^-$ decays from the $CP$ phase of NP contributions is analyzed in the PQCD approach. With a relevant strong-phase difference determined in this method, it is shown that the direct $CP$ violation can be $\mathcal{O}(10^{-2})$, and partially already excluded by the measurements.
\par
Within the channels that we have studied, the combined analysis reveals a tension with the SM predictions across both $b \to c\overline{u}d$ and $b \to c\overline{u}s$ sectors. This tension manifests differently in the two channels: for the $b \to c \bar ud$ transition ($B_d \to D^{(*)-}\pi^+$), the SM point (i.e., $C_{i, \text{NP}}(M_W)=0$) lies outside the common $1\sigma$ allowed region. For the $b \to c \bar us$ transition ($B^+ \to \bar D^0K^+$ and $\bar{B}^0_s \to D^+_sK^-$), the tension is more pronounced, with the individual $1\sigma$ constraints failing to find a common region that simultaneously satisfies the experimental constraints. In both cases, this suggests that current data favors nonzero NP contributions to the tree-level current-current operators. Furthermore, the compatibility of the constraints across different decay channels points toward a potential universal NP effect governing the $b \to c\overline{u}q$ transitions, independent of $q = d,s$.
\par
Our study reveals distinct phenomenological patterns associated with the color-singlet ($C_2$) and color-rearranged ($C_1$) scenarios. While branching ratios and the CKM angle $\gamma$ provide stringent bounds on $C_{2}^{\text{NP}}$, they are comparatively less restrictive for $C_{1}^{\text{NP}}$ due to $1/N_c$ suppression for the latter. Crucially, we demonstrated that the constraints from the lifetime ratio and $B-\overline{B}$ mixing observables play a decisive role in narrowing the allowed region for $C_{1}^{\text{NP}}$. In order to obtain nonzero $A_{\text{CP}}$, different amplitudes that have relative weak and strong phases need to interfere. It can be found that scenarios such as $C_2^{\text{NP}}(M_W) \neq 0$ and $C_1^{\text{NP}}(M_W) = 0$ cannot generate nonzero $A_{\text{CP}}$. To illustrate $A_{\text{CP}}$ testable in the future, $C_2^{\text{NP}}(M_W) = \mathcal{O}(1)$ and $C_1^{\text{NP}}(M_W) = \mathcal{O}(0.01)$ are considered with some benchmark points for the latter. Based on this constrained parameter space, we provided detailed predictions for the pairwise correlations among the width difference $\Delta\Gamma_d/\Gamma_d$, the semileptonic asymmetry $A_{\text{SL}}^d$, and the direct $CP$ asymmetry $A_{\text{CP}}$.

\par
The precision of this theoretical framework is currently limited by hadronic uncertainties and leading-order (LO) approximations for the NP contributions. Future refinements should incorporate higher-order perturbative corrections and a more rigorous treatment of SU(3) flavor symmetry breaking. On the experimental side, the anticipated precision from the LHCb upgrade II~\cite{LHCb:2018roe,Cerri:2018ypt}, particularly for $\Delta\Gamma_d/\Gamma_d$ and $A_{\text{SL}}^d$, will be pivotal in further narrowing the NP parameter space and clarifying the tensions observed in $b \to c \bar u q$ decays.
\begin{acknowledgments}
    The work of H.~U. is supported by the National Science Foundation of China under Grant
No.~12405111 and the Seeds Funding of Jilin University.
\end{acknowledgments}
\section*{DATA AVAILABILITY}
There are no publicly available research data or software supporting this manuscript. Requests for further information or data should be sent to the authors.
\newpage
\appendix
\section{Input Parameters and Experimental Data}
The ranges of $\gamma$ denoted as $\gamma = \gamma^{exp}$ in Eq. (\ref{eq:new def. of rb and gamma}), measured from different decay channels are listed in Table \ref{Gamma}.
\begin{table}[h]
    \centering
    \setlength{\tabcolsep}{12pt}
    \begin{tabular}{|c|c|c|}
        \hline
         & $8.5^\circ<\gamma<16.5^\circ$ & \\
        $B^+\to \bar D^0K^+$ & $84.5^\circ<\gamma<95.5^\circ$ & \cite{Belle:2023yoe}\\
         & $163.3^\circ<\gamma<171.5^\circ$ & \\\hline
        \multirow{2}{*}{$\bar B_s^0\to D^+_sK^-$} & $62^\circ<\gamma<86^\circ \quad modulo \quad 180^\circ$ & \cite{LHCb:2017hkl}\\
        \cline{2-3}
         & $101^\circ<\gamma<145^\circ \quad modulo \quad 180^\circ$ & \cite{LHCb:2024xyw}\\\hline
         & $-6.96^\circ<\gamma<96.44^\circ \quad modulo \quad 180^\circ$ & \cite{BaBar:2005jis}\\\cline{2-3}
        $B_d\to D^{*-}\pi^+$ & $-19.16^\circ<\gamma<108.64^\circ \quad modulo \quad 180^\circ$ & \cite{Belle:2006lts}\\\cline{2-3}
         & $-5.46^\circ<\gamma<94.94^\circ \quad modulo \quad 180^\circ$ & \cite{BaBar:2006slj}\\\hline
        $B^0\to D^{-}\pi^+$ & $5^\circ<\gamma<86^\circ \quad modulo \quad 180^\circ$ & \cite{LHCb:2018zap}\\\hline
    \end{tabular}
    \caption{\justifying{The ranges of $\gamma$. For $\bar B_s^0\to D^+_sK^-$, two ranges are used separately. The third range is used for $B_d\to D^{*-}\pi^+$.}}
    \label{Gamma}
\end{table}

\noindent The branching ratios of the different decay channels used in our analysis are listed in Table \ref{Branching Ratio}. The left column includes the main research objects, while the right columns are used only for calculating the ratios of amplitudes.
\begin{table}[h]
    \centering
    \setlength{\tabcolsep}{8pt}
    \begin{tabular}{|cc|cc|}
        \hline
        $Br\left(B^+\to \bar D^0K^+\right)$ & $(3.64 \pm 0.15)\times10^{-4}$ & $Br\left(B^0\to D^-K^+\right)$& $(2.05 \pm 0.08)\times10^{-4}$\\
        $Br\left(\bar B_s^0\to D^+_sK^-\right)$ &  $(2.25 \pm 0.12)\times10^{-4}$&  $Br\left(\bar B^0\to D^+K^-\right)$& $(2.05 \pm 0.08)\times10^{-4}$\\
        $Br\left(B^0\to D^{*-}\pi^+\right)$& $(2.66 \pm 0.07)\times10^{-3}$ & $Br(B^0_s \to D^{*-}_s \pi^+)$& $\left(1.9^{+0.5}_{-0.4}\right)\times10^{-3}$\\
        $Br\left(B^0\to D^{-}\pi^+\right)$ & $(2.51 \pm 0.08)\times10^{-3}$ &  $Br(B^0_s \to D^-_s \pi^+)$& $(2.98 \pm 0.14)\times10^{-3}$\\\hline
    \end{tabular}
    \caption{Experimental branching ratios of different decay channels from PDG~\cite{ParticleDataGroup:2024cfk}.}
    \label{Branching Ratio}
\end{table}

\noindent Other data used in our analysis are listed in Table \ref{Input parameters}. The NNLO values of $a_1$ are taken from Ref.~\cite{Cai:2021mlt}. Regarding the CKM input in the $B^0-\bar B^0$ mixing, the Wolfenstein parameters in Table \ref{Input parameters} are used while $|V_{cb}|$, $|V_{ud}|$ and $|V_{us}|$ for the lifetime ratio and branching ratio analyses are from the value listed in Table \ref{Input parameters}.
\begin{table}
    \centering
     \setlength{\tabcolsep}{8pt}
    \begin{tabular}{|ccc|ccc|}
    \hline
         $F_0^{B \to D}(m_{K^-}^2)$&  $0.671 \pm 0.011$& \cite{Cai:2021mlt} & $A_0^{B \to D^*}(m_{K^-}^2)$ & $0.664 \pm 0.018$ & \cite{Cai:2021mlt}\\
        $F_0^{B_s \to D_s}(m_{\pi^-}^2)$ & $0.666 \pm 0.012$ & \cite{Cai:2021mlt} & $A_0^{B_s \to D_s^*}(m_{\pi^-}^2)$ & $0.630 \pm 0.069$ & \cite{Cai:2021mlt}\\
        $|a_1(D^+_sK^-)|$ & $1.075^{+0.007}_{-0.011}$ & \cite{Cai:2021mlt} & $|a_1(D^+K^-)|$ & $1.075^{+0.007}_{-0.011}$ & \cite{Cai:2021mlt}\\
        $|a_1(D^{*+}_s\pi^-)|$ & $1.075^{+0.006}_{-0.011}$ & \cite{Cai:2021mlt} & $|a_1(D^{*+}\pi^-)|$ & $1.075^{+0.006}_{-0.011}$ & \cite{Cai:2021mlt}\\
        $|a_1(D^{+}_s\pi^-)|$ & $1.073^{+0.005}_{-0.010}$ & \cite{Cai:2021mlt} & $|a_1(D^{+}\pi^-)|$ & $1.073^{+0.005}_{-0.010}$ & \cite{Cai:2021mlt}\\
        $f_{\pi^{\pm}}$ & $130.2 \pm 0.8$ MeV & \cite{FlavourLatticeAveragingGroupFLAG:2024oxs} & $f_{K^{\pm}}$ & $155.7 \pm 0.3$ MeV & \cite{FlavourLatticeAveragingGroupFLAG:2024oxs}\\
        $\tau(B^+)$ & $1.638 \times 10^{-12}$ s & \cite{ParticleDataGroup:2024cfk} & $\tau(B^0)$ & $1.517 \times 10^{-12}$ s & \cite{ParticleDataGroup:2024cfk}\\
        $\tau(B_s^0)$ & $1.516 \times 10^{-12}$ s & \cite{ParticleDataGroup:2024cfk} & $\lambda$ & $0.22501 \pm 0.00068$ & \cite{ParticleDataGroup:2024cfk} \\
        A & $0.826^{+0.016}_{-0.015}$ & \cite{ParticleDataGroup:2024cfk} & $\bar \rho$ & $0.1591 \pm 0.0094$ & \cite{ParticleDataGroup:2024cfk} \\
        $\bar \eta$ & $0.3523^{+0.0073}_{-0.0071}$ & \cite{ParticleDataGroup:2024cfk} & $|V_{cb}|$ & $0.0411 \pm 0.0012$ & \cite{ParticleDataGroup:2024cfk} \\
        $|V_{ud}|$ & $0.97367 \pm 0.00032$ & \cite{ParticleDataGroup:2024cfk} & $|V_{us}|$ & $0.22431 \pm 0.00085$ & \cite{ParticleDataGroup:2024cfk} \\
        \hline
        $\bar B_1(\bar m_b(\bar m_b))$ & $1.013^{+0.066}_{-0.059}$ & \cite{Black:2024bus} & $\bar B_2(\bar m_b(\bar m_b))$ & $1.004^{+0.085}_{-0.081}$ & \cite{Black:2024bus}\\
        $\bar \epsilon_1(\bar m_b(\bar m_b))$ & $-0.098^{+0.029}_{-0.032}$ & \cite{Black:2024bus} & $\bar \epsilon_2(\bar m_b(\bar m_b))$ & $-0.037^{+0.019}_{-0.020}$ & \cite{Black:2024bus}\\
        $\rho^i_q$ & 1 &  & $\sigma^i_q$ & 0 & \\
        $m_b^{kin}$ & $(4.573 \pm 0.012)$ GeV & \cite{Bordone:2021oof} & $\bar m_c$(2 GeV) & $1.092 \pm 0.008$ GeV & \cite{Bordone:2021oof}\\
        $\Gamma^{SM}(B^0)$ & $0.636^{+0.028}_{-0.037}$ ps$^{-1}$ & \cite{Egner:2024lay} & $\Gamma^{SM}(B^+)$ & $0.587^{+0.025}_{-0.035}$ ps$^{-1}$ & \cite{Egner:2024lay}\\
        $\mu_{\pi}^2$ & $0.477 \pm 0.056$ GeV$^2$ & \cite{Bordone:2021oof} & $\mu_G^2$ & $0.306 \pm 0.050$ GeV$^2$ & \cite{Bordone:2021oof}\\
        $f_{B^+}$ & $189.4 \pm 1.4$ MeV & \cite{FlavourLatticeAveragingGroupFLAG:2024oxs} & $\alpha_s (M_Z)$ & $0.1180 \pm 0.0009$ & \cite{ParticleDataGroup:2024cfk}\\
        \hline
        $m_t^{pole}$ & 172.4 GeV & \cite{ParticleDataGroup:2024cfk} & $M_W$ & 80.3692 GeV & \cite{ParticleDataGroup:2024cfk}\\
        $m_b^{pole}$ & 4.78 GeV & \cite{ParticleDataGroup:2024cfk} & $\bar{m}_b(\bar{m}_b)$ & 4.183 GeV & \cite{ParticleDataGroup:2024cfk}\\
        $M_{B^0}$ & 5279.72 MeV & \cite{ParticleDataGroup:2024cfk} & $M_{B_s^0}$ & 5366.93 MeV & \cite{ParticleDataGroup:2024cfk}\\
        $f_{B^0}$ & $190.5 \pm 1.3$ MeV& \cite{FlavourLatticeAveragingGroupFLAG:2024oxs} & $f_{B_s^0}$ & $230.7 \pm 1.2$ MeV & \cite{FlavourLatticeAveragingGroupFLAG:2024oxs}\\
        $B_{R_2}^s$ & $0.89 \pm 0.38$ & \cite{Davies:2019gnp} & $B_{R_3}^s$ & $1.07 \pm 0.42$ & \cite{Davies:2019gnp}\\
        $B_1^d$ & $0.835 \pm 0.028$ & \cite{DiLuzio:2019jyq} & $B_2^d$ & $0.791 \pm 0.034$ & \cite{DiLuzio:2019jyq}\\
        $B_3^d$ & $0.775 \pm 0.054$ & \cite{DiLuzio:2019jyq} & $B_4^d$ & $1.063 \pm 0.041$ & \cite{DiLuzio:2019jyq}\\
        $B_5^d$ & $0.994 \pm 0.037$ & \cite{DiLuzio:2019jyq} & $\eta_B$(4.75 GeV) & $0.85 \pm 0.02$ & \cite{Ciuchini:2003ww}\\
        \hline
    \end{tabular}
    \caption{\justifying{Parameters used in the analysis. The values $\rho^i_q = 1$ and $\sigma^i_q = 0$ correspond to the vacuum-insertion approximation (VIA).}}
    \label{Input parameters}
\end{table}

\section{Formalism for \texorpdfstring{$B_q - \bar B_q$}{Bq-Bqbar} Mixing}
\subsection{Effective Operator Basis}
For the analysis of $B$ meson mixing, we employ the complete operator basis introduced in Ref.~\cite{Becirevic:2001xt}:
\begin{equation}
    \begin{aligned}
            \mathcal{O}_1^q &= (\bar{b}_i q_i)_{V-A} (\bar{b}_j q_j)_{V-A},  &  \mathcal{O}_2^q &= (\bar{b}_i q_i)_{S-P} (\bar{b}_j q_j)_{S-P},\\
            \mathcal{O}_3^q &= (\bar{b}_i q_j)_{S-P} (\bar{b}_j q_i)_{S-P},  &  \mathcal{O}_4^q &= (\bar{b}_i q_i)_{S-P} (\bar{b}_j q_j)_{S+P},\\
            \mathcal{O}_5^q &= (\bar{b}_i q_j)_{S-P} (\bar{b}_j q_i)_{S+P}.
    \end{aligned}
\end{equation}
The hadronic matrix elements are parametrized in terms of bag parameters as follows~\cite{Ciuchini:2003ww}:
\begin{equation}
    \begin{aligned}
        \langle \overline{B}_q | O_1^q | B_q \rangle &= \frac{8}{3} f_{B_q}^2 M_{B_q}^2 B_1^q, &
        \langle \overline{B}_q | O_2^q | B_q \rangle &= -\frac{5}{3} \frac{f_{B_q}^2 M_{B_q}^4}{(m_b+m_q)^2} B_2^q, \\
        \langle \overline{B}_q | \mathcal{O}_3^q | B_q \rangle &= \frac{1}{3} \left( \frac{m_{B_q}}{m_b+m_q} \right)^2 m_{B_q}^2 f_{B_q}^2 B_3^q, &
        \langle \overline{B}_q | \mathcal{O}_4^q | B_q \rangle &= 2 \left( \frac{m_{B_q}}{m_b+m_q} \right)^2 m_{B_q}^2 f_{B_q}^2 B_4^q, \\
        \langle \overline{B}_q | \mathcal{O}_5^q | B_q \rangle &= \frac{2}{3} \left( \frac{m_{B_q}}{m_b+m_q} \right)^2 m_{B_q}^2 f_{B_q}^2 B_5^q.\\
    \end{aligned}
\end{equation}
In addition to the operators $\mathcal{O}_i^q$, we also consider the four QCD operators:
\begin{equation}
    \begin{aligned}
        R_1^q &= \frac{m_q}{m_b} (\bar{b}_i q_i)_{S-P} (\bar{b}_j q_j)_{S+P}, \\
        R_2^q &= \frac{1}{m_b^2} (\bar{b}_i \overleftarrow{D}_\rho \gamma^\mu (1-\gamma_5) D^\rho q_i) (\bar{b}_j \gamma_\mu (1-\gamma_5) q_j), \\
        R_3^q &= \frac{1}{m_b^2} (\bar{b}_i \overleftarrow{D}_\rho (1-\gamma_5) D^\rho q_i) (\bar{b}_j (1-\gamma_5) q_j), \\
        R_4^q &= \frac{1}{m_b} (\bar{b}_i (1-\gamma_5) i \overleftarrow{D}_\mu q_i) (\bar{b}_j \gamma^\mu (1-\gamma_5) q_j).
    \end{aligned}
\end{equation}
The matrix elements of these operators are parametrized as~\cite{Ciuchini:2003ww}:
\begin{equation}
    \begin{aligned}
        \langle \overline{B}_q | R_1^q | B_q \rangle &= \frac{7}{3} \frac{m_q}{m_b} f_{B_q}^2 M_{B_q}^2 B_{R_1}^q, &
        \langle \overline{B}_q | R_2^q | B_q \rangle &= -\frac{2}{3} f_{B_q}^2 M_{B_q}^2 \left( \frac{M_{B_q}^2}{m_b^2} - 1 \right) B_{R_2}^q, \\
        \langle \overline{B}_q | R_3^q | B_q \rangle &= \frac{7}{6} f_{B_q}^2 M_{B_q}^2 \left( \frac{M_{B_q}^2}{m_b^2} - 1 \right) B_{R_3}^q, &
        \langle \overline{B}_q | R_4^q | B_q \rangle &= -f_{B_q}^2 M_{B_q}^2 \left( \frac{M_{B_q}^2}{m_b^2} - 1 \right) B_{R_4}^q.\\
    \end{aligned}
\end{equation}
The operators $R_1^q$ and $R_4^q$ are related to the basis operators $\mathcal{O}_i^q$ via the relations:
\begin{align}
    R_1^q = \frac{m_q}{m_b} \mathcal{O}_4^q, \qquad \qquad
    2R_4^q = \mathcal{O}_3^q + \mathcal{O}_1^q/2 + \mathcal{O}_2^q - 2\frac{m_q}{m_b} \mathcal{O}_5^q + R_2^q.
\end{align}
\subsection{Analytical Expressions for the Mixing Coefficients}
Adopting the formalism from Ref.~\cite{Ciuchini:2003ww}, the general expressions for the coefficient functions $D_k$ are decomposed into current-current ($F$) and penguinlike ($P$) contributions:
\begin{equation}
    \begin{aligned}
        D_k^{uu}(\mu_2) &= \sum_{i,j=1,2} C_i^*(\mu_1) C_j^*(\mu_1) F_{k,ij}^{uu}(\mu_1, \mu_2) + \frac{\alpha_s}{4\pi} (C_2^*(\mu_1))^2 P_{k,22}^{uu}(\mu_1, \mu_2) \\
        &\quad + 2\frac{\alpha_s}{4\pi} C_2^* C_{8G}^* P_{k,28}^{u} + 2 \sum_{i=1,2} \sum_{r=3,6} C_i^* C_r ^*P_{k,ir}^{u}, \\[3ex]
        D_k^{cu}(\mu_2) &= \sum_{i,j=1,2} C_i^*(\mu_1) C_j^*(\mu_1) F_{k,ij}^{cu}(\mu_1, \mu_2) + \frac{\alpha_s}{4\pi} (C_2^*(\mu_1))^2  P_{k,22}^{cu}(\mu_1, \mu_2) \\
        &\quad + \frac{\alpha_s}{4\pi} C_2^* C_{8G}^* (P_{k,28}^{c} + P_{k,28}^{u}) + \sum_{i=1,2} \sum_{r=3,6} C_i^* C_r^*  (P_{k,ir}^{c} + P_{k,ir}^{u}), \\[3ex]
        D_k^{cc}(\mu_2) &= \sum_{i,j=1,2} C_i^*(\mu_1) C_j^*(\mu_1) F_{k,ij}^{cc}(\mu_1, \mu_2) + \frac{\alpha_s}{4\pi} (C_2^*(\mu_1))^2  P_{k,22}^{cc}(\mu_1, \mu_2) \\
        &\quad + 2\frac{\alpha_s}{4\pi} C_2^* C_{8G}^* P_{k,28}^{c} + 2 \sum_{i=1,2} \sum_{r=3,6} C_i^* C_r^*  P_{k,ir}^{c}.
    \end{aligned}
\end{equation}
It should be noted that complex conjugate is taken for the Wilson coefficients  for our convention in Eq. (\ref{eq:Hamiltonian}). In this work, we restrict the calculation of contributions induced by NP to LO accuracy, and discard the penguin contributions represented by the $P$ terms and focus on the current-current functions $F_{k,ij}^{qq'}$. The perturbative expansion of these functions is formally written as:
\begin{equation}
F_{k,ij}^{qq'} = A_{k,ij}^{qq'} + \frac{\alpha_s}{4\pi} B_{k,ij}^{qq'} \, ,
\end{equation}
where the indices $(qq')$ run over $\{(uu), (cu), (cc)\}$. To maintain consistency with the LO approximation, we retain only the zeroth-order coefficients $A_{k,ij}^{qq'}$ and neglect the $\mathcal{O}(\alpha_s)$ corrections $B_{k,ij}^{qq'}$. 

The analytical forms for the relevant $A_{k,ij}^{qq'}$ coefficients are listed below. They are expressed in terms of the mass ratio $z = m_c^2/m_b^2$. We note that the coefficients for the $uu$ sector can be derived directly from the $cc$ sector by taking the massless charm limit ($m_c \to 0$)\cite{Ciuchini:2003ww}.
\begin{equation}
    \begin{aligned}
    A_{1,11}^{cu} &= \frac{3}{2} (2 - 3 z + z^3) \, ,&\qquad \qquad A_{2,11}^{cu} &= 3 (1 - z)^2 (1 + 2 z)  \, ,\\ 
    A_{1,12}^{cu} &= \frac{1}{2} (2 - 3 z + z^3)  \, ,& A_{2,12}^{cu} &= (1 - z)^2 (1 + 2 z) \, , \\
     A_{1,22}^{cu} &= \frac{1}{2} (1 - z)^3  \, ,&  A_{2,22}^{cu} &= -(1 - z)^2 (1 + 2 z) \, ,\\
    A_{1,11}^{cc} &= 3\sqrt{1-4z}\,(1-z) \, ,&  A_{2,11}^{cc} &= 3\sqrt{1-4z}\,(1+2z) \, , \\
     A_{1,12}^{cc} &= \sqrt{1-4z}\,(1-z) \, ,& A_{2,12}^{cc} &= \sqrt{1-4z}\,(1+2z) \, ,\\
     A_{1,22}^{cc} &= \frac{1}{2}(1-4z)^{3/2} \, ,&  A_{2,22}^{cc} &= -\sqrt{1-4z}\,(1+2z) \, .
    \end{aligned}
\end{equation}
Finally, we incorporate the $1/m_b$ power corrections through the parameters $\delta_{1/m}^{qq', q}$. These corrections depend on the matrix elements of the dimension-7 operators and coefficients $K_{1,2}$, defined as follows \cite{Ciuchini:2003ww}:
\begin{align}
\delta_{1/m}^{ccq} &= 
\begin{aligned}[t]
    &\sqrt{1-4z}((1+2z)[K_2(\langle R_2^q \rangle + 2\langle R_4^q \rangle) - 2K_1(\langle R_1^q \rangle + \langle R_2^q \rangle)] \\
    &-\frac{12z^2}{1-4z}[K_1(\langle R_2^q \rangle + 2\langle R_3^q \rangle) + 2K_2\langle R_3^q \rangle])\space,\nonumber
\end{aligned} \\
\delta_{1/m}^{cuq} &= 
\begin{aligned}[t]
    &(1-z)^2((1+2z)[K_2(\langle R_2^q \rangle + 2\langle R_4^q \rangle) - 2K_1(\langle R_1^q \rangle + \langle R_2^q \rangle)] \\
    &-\frac{6z^2}{(1-z)^2}(K_1(\langle R_2^q \rangle + 2\langle R_3^q \rangle) + 2K_2\langle R_3^q \rangle))\space,
\end{aligned} \\\nonumber
\delta_{1/m}^{uuq} &= [K_2(\langle R_2^q \rangle + 2\langle R_4^q \rangle) - 2K_1(\langle R_1^q \rangle + \langle R_2^q \rangle)]\space,
\end{align}
where $\langle R_i^q \rangle=\langle \bar{B}_q | R_i^q | B_q \rangle$ and the combinations $K_{1,2}$ are determined by the $\Delta B=1$ Wilson coefficients:
\begin{align}
    K_1=3C_1^{*2}+2Re(C_1^*C_2^*) \space,\qquad K_2=C_2^{*2}\space.
\end{align}
\section{Formulas For $B \to D \pi$ in PQCD}
\label{Appendix:PQCD}
In this appendix, we collect the explicit PQCD expressions for the building blocks
entering the class-\uppercase\expandafter{\romannumeral 3} amplitude $B^- \to D^0\pi^-$ in Eq.~(\ref{eq:PQCD}).
Unless specified otherwise, the formulas, definitions, and conventions in this appendix are compiled from Ref.~\cite{Keum:2003js}, to which we refer the reader for further details.
Only the quantities used in the direct $CP$ analysis are listed below, including the
factorizable amplitudes, nonfactorizable amplitudes, hard kernels, evolution factors,
and meson distribution amplitudes.

We use the light-cone momentum assignment
\begin{align}
    P_1 &= \frac{m_B}{\sqrt{2}}(1,1,\mathbf 0_T), \\
    P_2 &= \frac{m_B}{\sqrt{2}}(1,r^2,\mathbf 0_T), \\
    P_3 &= \frac{m_B}{\sqrt{2}}(0,1-r^2,\mathbf 0_T),
\end{align}
where $P_1$, $P_2$ and $P_3$ denote the momenta of the $B$, $D$ and $\pi$
mesons, respectively. The shorthand notations are
\begin{equation}
\label{def: r,r0,[dx]}
    r=\frac{m_D}{m_B},\qquad
    r_0=\frac{m_0}{m_B},\qquad
    [dx]=dx_1dx_2dx_3,
\end{equation}
where $m_0 = 1.4$ GeV is the chiral enhancement scale for the pion twist-3 distribution amplitudes. The variables $x_i$ denote the
longitudinal momentum fractions of the light partons in the corresponding
mesons, and $b_i$ are the impact parameters conjugate to the transverse
momenta. We also use
\begin{equation}
    N=N_c=3,\qquad
    C_F=\frac{N_c^2-1}{2N_c}=\frac{4}{3}.
\end{equation}
The effective combinations of Wilson coefficients are
\begin{equation}
    a_1(t)=C_2(t)+\frac{C_1(t)}{N_c},\qquad
    a_2(t)=C_1(t)+\frac{C_2(t)}{N_c}.
\end{equation}
In the presence of NP, the coefficients $C_{1,2}(t)$ appearing below are
understood as the low-scale Wilson coefficients obtained after RGE evolution
from the boundary values at $M_W$.

The factorizable external and internal emission amplitudes are
\begin{align}
\xi_{\rm ext}
=&\,
16\pi C_F \sqrt r\, m_B^2
\int_0^1 dx_1 dx_2
\int_0^{1/\Lambda} b_1 db_1\, b_2 db_2\,
\phi_B(x_1,b_1)\phi_D(x_2)
\nonumber\\
&\times
\left[
E_e(t_e^{(1)})
h(x_1,x_2,b_1,b_2)
S_t(x_2)
+
r\,E_e(t_e^{(2)})
h(x_2,x_1,b_2,b_1)
S_t(x_1)
\right],
\label{eq:app_xi_ext}
\\[8pt]
\xi_{\rm int}
=&\,
16\pi C_F \sqrt r\, m_B^2
\int_0^1 dx_1 dx_3
\int_0^{1/\Lambda} b_1 db_1\, b_3 db_3\,
\phi_B(x_1,b_1)
\nonumber\\
&\times
\Bigg\{
\left[
(1+x_3)\phi_\pi(x_3)
+
r_0(1-2x_3)
\left(\phi_\pi^p(x_3)+\phi_\pi^t(x_3)\right)
\right]
\nonumber\\
&\hspace{1.0cm}\times
E_i(t_i^{(1)})
h\!\left(x_1,x_3(1-r^2),b_1,b_3\right)
S_t(x_3)
\nonumber\\
&\hspace{1.0cm}
+
2r_0\phi_\pi^p(x_3)
E_i(t_i^{(2)})
h\!\left(x_3,x_1(1-r^2),b_3,b_1\right)
S_t(x_1)
\Bigg\}.
\label{eq:app_xi_int}
\end{align}
The corresponding evolution factors are
\begin{align}
    E_e(t) &= \alpha_s(t)a_1(t)\exp[-S_B(t)-S_D(t)],\\
    E_i(t) &= \alpha_s(t)a_2(t)\exp[-S_B(t)-S_\pi(t)].
\end{align}

The nonfactorizable amplitudes are
\begin{align}
\mathcal M_{\rm ext}
=&\,
32\pi\sqrt{2N}\, C_F \sqrt r\,m_B^2
\int_0^1 [dx]
\int_0^{1/\Lambda} b_1db_1\,b_3db_3\,
\phi_B(x_1,b_1)\phi_D(x_2)\phi_\pi(x_3)
\nonumber\\
&\times
\left[
x_3 E_b(t_b^{(1)})h_b^{(1)}(x_i,b_i)
-
(1-x_3+x_2)E_b(t_b^{(2)})h_b^{(2)}(x_i,b_i)
\right],
\label{eq:app_M_ext}
\\[8pt]
\mathcal M_{\rm int}
=&\,
32\pi\sqrt{2N}\, C_F \sqrt r\,m_B^2
\int_0^1 [dx]
\int_0^{1/\Lambda} b_1db_1\,b_2db_2\,
\phi_B(x_1,b_1)\phi_D(x_2)
\nonumber\\
&\times
\left[
(-x_2-x_3)\phi_\pi(x_3)E_d(t_d^{(1)})h_d^{(1)}(x_i,b_i)
+
(1-x_2)\phi_\pi(x_3)E_d(t_d^{(2)})h_d^{(2)}(x_i,b_i)
\right].
\label{eq:app_M_int}
\end{align}
The evolution factors for the nonfactorizable terms are
\begin{align}
    E_b(t) &= \alpha_s(t)\frac{C_1(t)}{N_c}
    \exp\!\left[-S(t)\big|_{b_2=b_1}\right],\\
    E_d(t) &= \alpha_s(t)\frac{C_2(t)}{N_c}
    \exp\!\left[-S(t)\big|_{b_3=b_1}\right],
\end{align}
with
\begin{equation}
    S(t)=S_B(t)+S_D(t)+S_\pi(t).
\end{equation}

The hard functions entering the factorizable amplitudes are
\begin{align}
h(x_1,x_2,b_1,b_2)
=&\,
K_0\!\left(\sqrt{x_1x_2}\,m_Bb_1\right)
\nonumber\\
&\times
\Big[
\theta(b_1-b_2)
K_0\!\left(\sqrt{x_2}\,m_Bb_1\right)
I_0\!\left(\sqrt{x_2}\,m_Bb_2\right)
\nonumber\\
&\hspace{0.7cm}
+
\theta(b_2-b_1)
K_0\!\left(\sqrt{x_2}\,m_Bb_2\right)
I_0\!\left(\sqrt{x_2}\,m_Bb_1\right)
\Big].
\end{align}
For the nonfactorizable amplitudes, the hard functions are
\begin{align}
h_b^{(j)}(x_i,b_i)
=&\,
\Big[
\theta(b_1-b_3)
K_0(Bm_Bb_1)I_0(Bm_Bb_3)
+
\theta(b_3-b_1)
K_0(Bm_Bb_3)I_0(Bm_Bb_1)
\Big]
\nonumber\\
&\times
\begin{cases}
K_0(B_jm_Bb_3), & B_j^2\geq 0,\\[3pt]
\dfrac{i\pi}{2}H_0^{(1)}
\!\left(\sqrt{|B_j^2|}\,m_Bb_3\right), & B_j^2\leq 0,
\end{cases}
\label{eq:app_hb}
\\[8pt]
h_d^{(j)}(x_i,b_i)
=&\,
\Big[
\theta(b_1-b_2)
K_0(Dm_Bb_1)I_0(Dm_Bb_2)
+
\theta(b_2-b_1)
K_0(Dm_Bb_2)I_0(Dm_Bb_1)
\Big]
\nonumber\\
&\times
\begin{cases}
K_0(D_jm_Bb_2), & D_j^2\geq 0,\\[3pt]
\dfrac{i\pi}{2}H_0^{(1)}
\!\left(\sqrt{|D_j^2|}\,m_Bb_2\right), & D_j^2\leq 0.
\end{cases}
\label{eq:app_hd}
\end{align}
Here $j=1,2$, $K_0$ and $I_0$ are modified Bessel functions, $H_0^{(1)}$
is the Hankel function of the first kind, and $\theta$ is the step function.
The variables entering these functions are
\begin{align}
    B^2 &= x_1x_2,\\
    B_1^2 &= x_1x_2-x_2x_3(1-r^2),\\
    B_2^2 &= x_1x_2-x_2(1-x_3)(1-r^2),\\
    D^2 &= x_1x_3(1-r^2),\\
    D_1^2 &= (x_1-x_2)x_3(1-r^2),\\
    D_2^2 &= (x_1+x_2)r^2-(1-x_1-x_2)x_3(1-r^2).
\end{align}

The hard scales are chosen as
\begin{align}
t_e^{(1)}
&=
\max\!\left(\sqrt{x_2}\,m_B,\frac{1}{b_1},\frac{1}{b_2}\right),
&
t_e^{(2)}
&=
\max\!\left(\sqrt{x_1}\,m_B,\frac{1}{b_1},\frac{1}{b_2}\right),
\\[5pt]
t_i^{(1)}
&=
\max\!\left(\sqrt{x_3(1-r^2)}\,m_B,\frac{1}{b_1},\frac{1}{b_3}\right),
&
t_i^{(2)}
&=
\max\!\left(\sqrt{x_1(1-r^2)}\,m_B,\frac{1}{b_1},\frac{1}{b_3}\right),
\\[5pt]
t_b^{(j)}
&=
\max\!\left(Bm_B,\sqrt{|B_j^2|}\,m_B,\frac{1}{b_1},\frac{1}{b_3}\right),
&
t_d^{(j)}
&=
\max\!\left(Dm_B,\sqrt{|D_j^2|}\,m_B,\frac{1}{b_1},\frac{1}{b_2}\right).
\end{align}

The meson distribution amplitudes used in the calculation are
\begin{align}
\phi_B(x,b)
&=
N_B x^2(1-x)^2
\exp\!\left[
-\frac{1}{2}\left(\frac{xm_B}{\omega_B}\right)^2
-\frac{\omega_B^2b^2}{2}
\right],
\\
\phi_D(x)
&=
\frac{3}{\sqrt{2N_c}}f_D x(1-x)
\left[1+C_D(1-2x)\right],
\\
\phi_\pi(x)
&=
\frac{3f_\pi}{\sqrt{2N_c}}x(1-x)
\left[
1+0.44\,C_2^{3/2}(2x-1)
+0.25\,C_4^{3/2}(2x-1)
\right],
\\
\phi_\pi^p(x)
&=
\frac{f_\pi}{2\sqrt{2N_c}}
\left[
1+0.43\,C_2^{1/2}(2x-1)
+0.09\,C_4^{1/2}(2x-1)
\right],
\\
\phi_\pi^t(x)
&=
\frac{f_\pi}{2\sqrt{2N_c}}(1-2x)
\left[
1+0.55(10x^2-10x+1)
\right].
\end{align}
The normalization constant $N_B$ is fixed by
\begin{equation}
    \int_0^1 dx\,\phi_B(x,0)=\frac{f_B}{2\sqrt{2N_c}}.
\end{equation}
The shape parameters are taken as
\begin{equation}
    \omega_B=0.4~{\rm GeV},\qquad
    C_D=0.8\pm0.2,
\end{equation}
and the infrared cutoff is chosen as
\begin{equation}
    \Lambda=\Lambda_{\rm QCD}^{(4)}=0.25~{\rm GeV}.
\end{equation}
For the Gegenbauer polynomials, we define $z=2x-1$ and use
\begin{align}
    C_2^{1/2}(z) &= \frac{1}{2}(3z^2-1),
&
    C_4^{1/2}(z) &= \frac{1}{8}(35z^4-30z^2+3),
\\
    C_2^{3/2}(z) &= \frac{3}{2}(5z^2-1),
&
    C_4^{3/2}(z) &= \frac{15}{8}(21z^4-14z^2+1).
\end{align}
The threshold resummation factor is parametrized as
\begin{equation}
    S_t(x)
    =
    \frac{2^{1+2c}\Gamma(3/2+c)}
    {\sqrt{\pi}\Gamma(1+c)}
    [x(1-x)]^c,
    \qquad c=0.35.
\end{equation}
The Sudakov exponents are~\cite{Keum:2000wi,Lu:2002iv}
\begin{align}
S_B(t)
&=
s(x_1P_1^+,b_1)
+
2\int_{1/b_1}^{t}
\frac{d\mu'}{\mu'}\gamma_q(\mu'),
\\
S_D(t)
&=
s(x_2P_2^+,b_2)
+
2\int_{1/b_2}^{t}
\frac{d\mu'}{\mu'}\gamma_q(\mu'),
\\
S_\pi(t)
&=
s(x_3P_3^-,b_3)
+
s((1-x_3)P_3^-,b_3)
+
2\int_{1/b_3}^{t}
\frac{d\mu'}{\mu'}\gamma_q(\mu').
\end{align}
The exponent $s(Q,b)$ is given by~\cite{Lu:2002iv}
\begin{align}
s(Q,b)
=&
\int_{1/b}^{Q}\frac{d\mu'}{\mu'}
\Bigg[
\left\{
\frac{2}{3}(2\gamma_E-1-\ln2)
+
C_F\ln\frac{Q}{\mu'}
\right\}
\frac{\alpha_s(\mu')}{\pi}
\nonumber\\
&\hspace{1.0cm}
+
\left\{
\frac{67}{9}
-
\frac{\pi^2}{3}
-
\frac{10}{27}n_f
+
\frac{2}{3}\beta_0\ln\frac{\gamma_E}{2}
\right\}
\left(\frac{\alpha_s(\mu')}{\pi}\right)^2
\ln\frac{Q}{\mu'}
\Bigg],
\end{align}
where
\begin{equation}
    \gamma_q(\mu')=-\frac{\alpha_s(\mu')}{\pi},
    \qquad
    \beta_0=11-\frac{2}{3}n_f,
    \qquad
    n_f=4.
\end{equation}
Here $\gamma_E$ denotes Euler's constant.

\clearpage 
\bibliography{apssamp}

@PREAMBLE{
 "\providecommand{\noopsort}[1]{}" 
 # "\providecommand{\singleletter}[1]{#1}%" 
}

@article{Gronau:1990ra,
    author = "Gronau, Michael and London, David",
    title = "{How to determine all the angles of the unitarity triangle from B(d)0 ---{\ensuremath{>}} D K(s) and B(s)0 ---{\ensuremath{>}} D0}",
    reportNumber = "DESY-90-125, TECHNION-PH-90-26, UDEM-LPN-TH-34",
    doi = "10.1016/0370-2693(91)91756-L",
    journal = "Phys. Lett. B",
    volume = "253",
    pages = "483--488",
    year = "1991"
}

@article{Gronau:1991dp,
    author = "Gronau, Michael and Wyler, Daniel",
    title = "{On determining a weak phase from CP asymmetries in charged B decays}",
    reportNumber = "TECHNION-PH-91-14, ZU-TH-6-91",
    doi = "10.1016/0370-2693(91)90034-N",
    journal = "Phys. Lett. B",
    volume = "265",
    pages = "172--176",
    year = "1991"
}

@article{Atwood:1996ci,
    author = "Atwood, David and Dunietz, Isard and Soni, Amarjit",
    title = "{Enhanced CP violation with B ---{\ensuremath{>}} K D0 (anti-D0) modes and extraction of the CKM angle gamma}",
    eprint = "hep-ph/9612433",
    archivePrefix = "arXiv",
    reportNumber = "FERMILAB-PUB-96-448-T, JLAB-THY-96-19",
    doi = "10.1103/PhysRevLett.78.3257",
    journal = "Phys. Rev. Lett.",
    volume = "78",
    pages = "3257--3260",
    year = "1997"
}

@article{Cai:2021mlt,
    author = "Cai, Fang-Min and Deng, Wei-Jun and Li, Xin-Qiang and Yang, Ya-Dong",
    title = "{Probing new physics in class-I B-meson decays into heavy-light final states}",
    eprint = "2103.04138",
    archivePrefix = "arXiv",
    primaryClass = "hep-ph",
    doi = "10.1007/JHEP10(2021)235",
    journal = "J. High Energy Phys.",
    volume = "10",
    pages = "235",
    year = "2021"
}

@article{Cheng:2018rkz,
    author = "Cheng, Hai-Yang",
    title = "{Phenomenological Study of Heavy Hadron Lifetimes}",
    eprint = "1807.00916",
    archivePrefix = "arXiv",
    primaryClass = "hep-ph",
    doi = "10.1007/JHEP11(2018)014",
    journal = "J. High Energy Phys.",
    volume = "11",
    pages = "014",
    year = "2018"
}

@article{FlavourLatticeAveragingGroupFLAG:2024oxs,
    author = "Aoki, Y. and others",
    collaboration = "Flavour Lattice Averaging Group (FLAG)",
    title = "{FLAG review 2024}",
    eprint = "2411.04268",
    archivePrefix = "arXiv",
    primaryClass = "hep-lat",
    reportNumber = "CERN-TH-2024-192, FERMILAB-PUB-24-0785-T",
    doi = "10.1103/nfzp-p5dn",
    journal = "Phys. Rev. D",
    volume = "113",
    number = "1",
    pages = "014508",
    year = "2026"
}

@article{Black:2024bus,
    author = {Black, Matthew and Lang, Martin and Lenz, Alexander and W{\"u}thrich, Zachary},
    title = "{HQET sum rules for matrix elements of dimension-six four-quark operators for meson lifetimes within and beyond the Standard Model}",
    eprint = "2412.13270",
    archivePrefix = "arXiv",
    primaryClass = "hep-ph",
    reportNumber = "P3H-24-098, SI-HEP-2024-29",
    doi = "10.1007/JHEP04(2025)081",
    journal = "J. High Energy Phys.",
    volume = "04",
    pages = "081",
    year = "2025"
}

@article{Belle:2023yoe,
    author = "Adachi, I. and others",
    collaboration = "Belle, Belle-II",
    title = "{Measurement of branching-fraction ratios and CP asymmetries in $B^{\pm} \to D_{CP\pm}K^{\pm}$ decays at Belle and Belle II}",
    eprint = "2308.05048",
    archivePrefix = "arXiv",
    primaryClass = "hep-ex",
    reportNumber = "Belle II Preprint 2023-011;KEK Preprint 2023-9",
    doi = "10.1007/JHEP05(2024)212",
    journal = "J. High Energy Phys.",
    volume = "05",
    pages = "212",
    year = "2024"
}

@article{LHCb:2017hkl,
    author = "Aaij, Roel and others",
    collaboration = "LHCb",
    title = "{Measurement of $CP$ asymmetry in $B_s^0 \to D_s^{\mp} K^{\pm}$ decays}",
    eprint = "1712.07428",
    archivePrefix = "arXiv",
    primaryClass = "hep-ex",
    reportNumber = "LHCB-PAPER-2017-047, CERN-EP-2017-315",
    doi = "10.1007/JHEP03(2018)059",
    journal = "J. High Energy Phys.",
    volume = "03",
    pages = "059",
    year = "2018"
}

@article{LHCb:2024xyw,
    author = "Aaij, Roel and others",
    collaboration = "LHCb",
    title = "{Measurement of CP asymmetry in $ {\textrm{B}}_{\textrm{s}}^0\to {\textrm{D}}_{\textrm{s}}^{\mp }{\textrm{K}}^{\pm } $ decays}",
    eprint = "2412.14074",
    archivePrefix = "arXiv",
    primaryClass = "hep-ex",
    reportNumber = "LHCb-PAPER-2024-020, CERN-EP-2024-219",
    doi = "10.1007/JHEP03(2025)139",
    journal = "J. High Energy Phys.",
    volume = "03",
    pages = "139",
    year = "2025"
}

@article{BaBar:2005jis,
    author = "Aubert, Bernard and others",
    collaboration = "BABAR",
    title = "{Measurement of time-dependent CP-violating asymmetries and constraints on $\sin(2\beta+\gamma)$ with partial reconstruction of $B \to D^{*\mp} \pi^\pm$ decays}",
    eprint = "hep-ex/0504035",
    archivePrefix = "arXiv",
    reportNumber = "SLAC-PUB-11136, BABAR-PUB-05-013",
    doi = "10.1103/PhysRevD.71.112003",
    journal = "Phys. Rev. D",
    volume = "71",
    pages = "112003",
    year = "2005"
}

@article{Belle:2006lts,
    author = "Ronga, F. J. and others",
    collaboration = "Belle",
    title = "{Measurements of CP violation in B0 ---{\ensuremath{>}} D*- pi+ and B0 ---{\ensuremath{>}} D- pi+ decays}",
    eprint = "hep-ex/0604013",
    archivePrefix = "arXiv",
    doi = "10.1103/PhysRevD.73.092003",
    journal = "Phys. Rev. D",
    volume = "73",
    pages = "092003",
    year = "2006"
}

@article{BaBar:2006slj,
    author = "Aubert, Bernard and others",
    collaboration = "BABAR",
    title = "{Measurement of time-dependent CP asymmetries in $B^0 \to D^{(*)}$ +- $\pi^\mp$ and $B^0 \to D^\pm \rho^\mp$ decays}",
    eprint = "hep-ex/0602049",
    archivePrefix = "arXiv",
    reportNumber = "BABAR-PUB-06-001, SLAC-PUB-11719",
    doi = "10.1103/PhysRevD.73.111101",
    journal = "Phys. Rev. D",
    volume = "73",
    pages = "111101",
    year = "2006"
}

@article{LHCb:2018zap,
    author = "Aaij, Roel and others",
    collaboration = "LHCb",
    title = "{Measurement of $CP$ violation in $B^{0}\rightarrow D^{\mp}\pi^{\pm}$ decays}",
    eprint = "1805.03448",
    archivePrefix = "arXiv",
    primaryClass = "hep-ex",
    reportNumber = "LHCb-PAPER-2018-009, CERN-EP-2018-084, LHCB-PAPER-2018-009",
    doi = "10.1007/JHEP06(2018)084",
    journal = "J. High Energy Phys.",
    volume = "06",
    pages = "084",
    year = "2018"
}

@article{Egner:2024lay,
    author = {Egner, Manuel and Fael, Matteo and Lenz, Alexander and Piscopo, Maria Laura and Rusov, Aleksey V. and Sch{\"o}nwald, Kay and Steinhauser, Matthias},
    title = "{Total decay rates of B mesons at NNLO-QCD}",
    eprint = "2412.14035",
    archivePrefix = "arXiv",
    primaryClass = "hep-ph",
    reportNumber = "TUM-HEP-1545/24, P3H-24-101, SI-HEP-2024-31, TTP24-046, Nikhef
  2024-019, ZU-TH 67/24",
    doi = "10.1007/JHEP04(2025)106",
    journal = "J. High Energy Phys.",
    volume = "04",
    pages = "106",
    year = "2025"
}

@article{DiLuzio:2019jyq,
    author = "Di Luzio, Luca and Kirk, Matthew and Lenz, Alexander and Rauh, Thomas",
    title = "{$\Delta M_s$ theory precision confronts flavour anomalies}",
    eprint = "1909.11087",
    archivePrefix = "arXiv",
    primaryClass = "hep-ph",
    reportNumber = "IPPP/19/70",
    doi = "10.1007/JHEP12(2019)009",
    journal = "J. High Energy Phys.",
    volume = "12",
    pages = "009",
    year = "2019"
}

@article{Davies:2019gnp,
    author = "Davies, Christine T. H. and Harrison, Judd and Lepage, G. Peter and Monahan, Christopher J. and Shigemitsu, Junko and Wingate, Matthew",
    collaboration = "HPQCD",
    title = "{Lattice QCD matrix elements for the ${B_s^0-\bar{B}_s^0}$ width difference beyond leading order}",
    eprint = "1910.00970",
    archivePrefix = "arXiv",
    primaryClass = "hep-lat",
    reportNumber = "INT-PUB-19-044, IPPP/19/77, JLAB-THY-19-3052",
    doi = "10.1103/PhysRevLett.124.082001",
    journal = "Phys. Rev. Lett.",
    volume = "124",
    number = "8",
    pages = "082001",
    year = "2020"
}

@article{ParticleDataGroup:2024cfk,
    author = "Navas, S. and others",
    collaboration = "Particle Data Group",
    title = "{Review of particle physics}",
    doi = "10.1103/PhysRevD.110.030001",
    journal = "Phys. Rev. D",
    volume = "110",
    number = "3",
    pages = "030001",
    year = "2024"
}

@article{Albrecht:2024oyn,
    author = "Albrecht, Johannes and Bernlochner, Florian and Lenz, Alexander and Rusov, Aleksey",
    title = "{Lifetimes of b-hadrons and mixing of neutral B-mesons: theoretical and experimental status}",
    eprint = "2402.04224",
    archivePrefix = "arXiv",
    primaryClass = "hep-ph",
    reportNumber = "SI-HEP-2024-04",
    doi = "10.1140/epjs/s11734-024-01124-3",
    journal = "Eur. Phys. J. ST",
    volume = "233",
    number = "2",
    pages = "359--390",
    year = "2024"
}

@article{Bordone:2021oof,
    author = "Bordone, Marzia and Capdevila, Bernat and Gambino, Paolo",
    title = "{Three loop calculations and inclusive Vcb}",
    eprint = "2107.00604",
    archivePrefix = "arXiv",
    primaryClass = "hep-ph",
    doi = "10.1016/j.physletb.2021.136679",
    journal = "Phys. Lett. B",
    volume = "822",
    pages = "136679",
    year = "2021"
}

@article{Beneke:2000ry,
    author = "Beneke, M. and Buchalla, G. and Neubert, M. and Sachrajda, Christopher T.",
    title = "{QCD factorization for exclusive, nonleptonic B meson decays: General arguments and the case of heavy light final states}",
    eprint = "hep-ph/0006124",
    archivePrefix = "arXiv",
    reportNumber = "CERN-TH-2000-159, CLNS-00-1675, PITHA-00-06, SHEP-00-06",
    doi = "10.1016/S0550-3213(00)00559-9",
    journal = "Nucl. Phys. B",
    volume = "591",
    pages = "313--418",
    year = "2000"
}

@article{Huber:2016xod,
    author = {Huber, Tobias and Kr{\"a}nkl, Susanne and Li, Xin-Qiang},
    title = "{Two-body non-leptonic heavy-to-heavy decays at NNLO in QCD factorization}",
    eprint = "1606.02888",
    archivePrefix = "arXiv",
    primaryClass = "hep-ph",
    reportNumber = "SI-HEP-2016-12, QFET-2016-06",
    doi = "10.1007/JHEP09(2016)112",
    journal = "J. High Energy Phys.",
    volume = "09",
    pages = "112",
    year = "2016"
}

@article{Bordone:2020gao,
    author = "Bordone, Marzia and Gubernari, Nico and Huber, Tobias and Jung, Martin and van Dyk, Danny",
    title = "{A puzzle in $\bar{B}_{(s)}^0 \to D_{(s)}^{(*)+} \lbrace \pi^-, K^-\rbrace$ decays and extraction of the $f_s/f_d$ fragmentation fraction}",
    eprint = "2007.10338",
    archivePrefix = "arXiv",
    primaryClass = "hep-ph",
    reportNumber = "TUM-HEP 1271/20, P3H-20-034, SI-HEP-2020-17",
    doi = "10.1140/epjc/s10052-020-08512-8",
    journal = "Eur. Phys. J. C",
    volume = "80",
    number = "10",
    pages = "951",
    year = "2020"
}

@article{Piscopo:2023opf,
    author = "Piscopo, Maria Laura and Rusov, Aleksey V.",
    title = "{Non-factorisable effects in the decays $ {\overline{B}}_s^0\to {D}_s^{+}{\pi}^{-} $ and $ {\overline{B}}^0\to {D}^{+}{K}^{-} $ from LCSR}",
    eprint = "2307.07594",
    archivePrefix = "arXiv",
    primaryClass = "hep-ph",
    reportNumber = "SI-HEP-2023-15",
    doi = "10.1007/JHEP10(2023)180",
    journal = "J. High Energy Phys.",
    volume = "10",
    pages = "180",
    year = "2023"
}

@article{Endo:2021ifc,
    author = "Endo, Motoi and Iguro, Syuhei and Mishima, Satoshi",
    title = "{Revisiting rescattering contributions to $ \overline{B} _{(s)}$ {\textrightarrow} $ {D}_{(s)}^{\left(\ast \right)}M $ decays}",
    eprint = "2109.10811",
    archivePrefix = "arXiv",
    primaryClass = "hep-ph",
    reportNumber = "IPMU21-0057, KEK-TH-2348",
    doi = "10.1007/JHEP01(2022)147",
    journal = "J. High Energy Phys.",
    volume = "01",
    pages = "147",
    year = "2022"
}

@article{Chua:2001br,
    author = "Chua, Chun-Khiang and Hou, Wei-Shu and Yang, Kwei-Chou",
    title = "{Final state rescattering and color suppressed anti-B0 ---{\ensuremath{>}} D0(*)0 h0 decays}",
    eprint = "hep-ph/0112148",
    archivePrefix = "arXiv",
    doi = "10.1103/PhysRevD.65.096007",
    journal = "Phys. Rev. D",
    volume = "65",
    pages = "096007",
    year = "2002"
}

@article{Chua:2005dt,
    author = "Chua, Chun-Khiang and Hou, Wei-Shu",
    title = "{Implications of anti-B ---{\ensuremath{>}} D0 h0 decays on anti-B ---{\ensuremath{>}} D anti-K, anti-D anti-K decays}",
    eprint = "hep-ph/0504084",
    archivePrefix = "arXiv",
    doi = "10.1103/PhysRevD.72.036002",
    journal = "Phys. Rev. D",
    volume = "72",
    pages = "036002",
    year = "2005"
}

@article{Chua:2007qw,
    author = "Chua, Chun-Khiang and Hou, Wei-Shu",
    title = "{Rescattering effects in anti-B(u,d,s) ---{\ensuremath{>}} DP, anti-DP decays}",
    eprint = "0712.1882",
    archivePrefix = "arXiv",
    primaryClass = "hep-ph",
    doi = "10.1103/PhysRevD.77.116001",
    journal = "Phys. Rev. D",
    volume = "77",
    pages = "116001",
    year = "2008"
}

@article{Chua:2018ikx,
    author = "Chua, Chun-Khiang",
    title = "{Revisiting final state interaction in charmless $B_q\to PP$ decays}",
    eprint = "1802.00155",
    archivePrefix = "arXiv",
    primaryClass = "hep-ph",
    doi = "10.1103/PhysRevD.97.093004",
    journal = "Phys. Rev. D",
    volume = "97",
    number = "9",
    pages = "093004",
    year = "2018"
}

@article{Iguro:2020ndk,
    author = "Iguro, Syuhei and Kitahara, Teppei",
    title = "{Implications for new physics from a novel puzzle in $\bar{B}_{(s)}^0 \to D^{(\ast)+}_{(s)} \lbrace \pi^-, K^- \rbrace$ decays}",
    eprint = "2008.01086",
    archivePrefix = "arXiv",
    primaryClass = "hep-ph",
    doi = "10.1103/PhysRevD.102.071701",
    journal = "Phys. Rev. D",
    volume = "102",
    number = "7",
    pages = "071701",
    year = "2020"
}

@article{Fleischer:2021cct,
    author = "Fleischer, Robert and Malami, Eleftheria",
    title = "{Using $B^0_s\to D_s^\mp K^\pm$ Decays as a Portal to New Physics}",
    eprint = "2109.04950",
    archivePrefix = "arXiv",
    primaryClass = "hep-ph",
    reportNumber = "Nikhef-2021-017",
    doi = "10.1103/PhysRevD.106.056004",
    journal = "Phys. Rev. D",
    volume = "106",
    number = "5",
    pages = "056004",
    year = "2022"
}

@article{Lenz:2022pgw,
    author = {Lenz, Alexander and M{\"u}ller, Jakob and Piscopo, Maria Laura and Rusov, Aleksey V.},
    title = "{Taming new physics in b {\textrightarrow} c{\={u}}d(s) with {\ensuremath{\tau}}(B$^{+}$)/{\ensuremath{\tau}}(B$_{d}$) and $ {a}_{sl}^d $}",
    eprint = "2211.02724",
    archivePrefix = "arXiv",
    primaryClass = "hep-ph",
    reportNumber = "SI-HEP-2022-32",
    doi = "10.1007/JHEP09(2023)028",
    journal = "J. High Energy Phys.",
    volume = "09",
    pages = "028",
    year = "2023"
}

@article{Panuluh:2024cxc,
    author = "Panuluh, Albertus Hariwangsa and Tanaka, Satoshi and Umeeda, Hiroyuki",
    title = "{B(s){\textrightarrow}D(s)(*)M decays in the presence of final-state interactions}",
    eprint = "2408.15466",
    archivePrefix = "arXiv",
    primaryClass = "hep-ph",
    reportNumber = "HUPD-2406",
    doi = "10.1103/PhysRevD.111.095020",
    journal = "Phys. Rev. D",
    volume = "111",
    number = "9",
    pages = "095020",
    year = "2025"
}

@article{Meiser:2024zea,
    author = "Meiser, Stefan and van Dyk, Danny and Virto, Javier",
    title = "{Towards a global analysis of the $b \to c\bar{u}q$ puzzle}",
    eprint = "2411.09458",
    archivePrefix = "arXiv",
    primaryClass = "hep-ph",
    reportNumber = "EOS-2024-04, IPPP/24/71",
    doi = "10.1007/JHEP06(2025)019",
    journal = "J. High Energy Phys.",
    volume = "06",
    pages = "019",
    year = "2025"
}

@article{Lenz:2022rbq,
    author = "Lenz, Alexander and Piscopo, Maria Laura and Rusov, Aleksey V.",
    title = "{Disintegration of beauty: a precision study}",
    eprint = "2208.02643",
    archivePrefix = "arXiv",
    primaryClass = "hep-ph",
    reportNumber = "SI-HEP-2022-22",
    doi = "10.1007/JHEP01(2023)004",
    journal = "J. High Energy Phys.",
    volume = "01",
    pages = "004",
    year = "2023"
}

@article{Fleischer:2003yb,
    author = "Fleischer, Robert",
    title = "{New strategies to obtain insights into CP violation through B(s) ---{\ensuremath{>}} D(s)+- K-+, D(s)*+- K-+, ... and B(d) ---{\ensuremath{>}} D+- pi-+, D*+- pi-+, ... decays}",
    eprint = "hep-ph/0304027",
    archivePrefix = "arXiv",
    reportNumber = "CERN-TH-2003-084",
    doi = "10.1016/j.nuclphysb.2003.08.010",
    journal = "Nucl. Phys. B",
    volume = "671",
    pages = "459--482",
    year = "2003"
}

@unpublished{LHCb:2018roe,
    author        = {Aaij, Roel and others},
    collaboration = {LHCb},
    title         = {Physics case for an {LHCb} {Upgrade II} --- {Opportunities} in flavour physics, and beyond, in the {HL-LHC} era},
    year          = {2018},
    month         = aug,
    eprint        = {1808.08865},
    archivePrefix = {arXiv},
    primaryClass  = {hep-ex},
    note          = {}
}

@article{Cerri:2018ypt,
    author = "Cerri, A. and others",
    editor = "Dainese, Andrea and Mangano, Michelangelo and Meyer, Andreas B. and Nisati, Aleandro and Salam, Gavin and Vesterinen, Mika Anton",
    title = "{Report from Working Group 4}: {Opportunities in Flavour Physics at the HL-LHC and HE-LHC}",
    eprint = "1812.07638",
    archivePrefix = "arXiv",
    primaryClass = "hep-ph",
    reportNumber = "CERN-LPCC-2018-06",
    doi = "10.23731/CYRM-2019-007.867",
    journal = "CERN Yellow Rep. Monogr.",
    volume = "7",
    pages = "867--1158",
    year = "2019"
}

@article{Becirevic:2001xt,
    author = "Becirevic, D. and Gimenez, V. and Martinelli, G. and Papinutto, M. and Reyes, J.",
    title = "{B parameters of the complete set of matrix elements of delta B = 2 operators from the lattice}",
    eprint = "hep-lat/0110091",
    archivePrefix = "arXiv",
    reportNumber = "IFUP-TH-2001-32, FTUV-IFIC-01-0927, ROMA-1323-01",
    doi = "10.1088/1126-6708/2002/04/025",
    journal = "J. High Energy Phys.",
    volume = "04",
    pages = "025",
    year = "2002"
}

@article{Ciuchini:2003ww,
    author = "Ciuchini, M. and Franco, E. and Lubicz, V. and Mescia, F. and Tarantino, C.",
    title = "{Lifetime differences and CP violation parameters of neutral B mesons at the next-to-leading order in QCD}",
    eprint = "hep-ph/0308029",
    archivePrefix = "arXiv",
    reportNumber = "RM3-TH-03-9, ROME1-1355-03, SHEP-0318",
    doi = "10.1088/1126-6708/2003/08/031",
    journal = "J. High Energy Phys.",
    volume = "08",
    pages = "031",
    year = "2003"
}

@unpublished{Lang:2025ios,
    author        = {Lang, Martin and Lenz, Alexander and Mohamed, Ali and Piscopo, Maria Laura and Rusov, Aleksey V.},
    title         = {{$B$}-meson decay width up to $1/m_b^3$ corrections within and beyond the {Standard Model}},
    year          = {2025},
    month         = dec,
    eprint        = {2512.14635},
    archivePrefix = {arXiv},
    primaryClass  = {hep-ph},
    note          = {}
}

@article{Nierste:2025muk,
    author = "Nierste, Ulrich and Reeck, Pascal and Shtabovenko, Vladyslav and Steinhauser, Matthias",
    title = "{Complete next-to-next-to-leading order QCD corrections to the decay matrix in B-meson mixing at leading power}",
    eprint = "2512.07949",
    archivePrefix = "arXiv",
    primaryClass = "hep-ph",
    reportNumber = "P3H-25-107, SI-HEP-2025-27, TTP25-054",
    doi = "10.1007/JHEP03(2026)094",
    journal = "J. High Energy Phys.",
    volume = "03",
    pages = "094",
    year = "2026"
}

@article{Inami:1980fz,
    author = "Inami, T. and Lim, C. S.",
    title = "{Effects of Superheavy Quarks and Leptons in Low-Energy Weak Processes k(L) ---{\ensuremath{>}} mu anti-mu, K+ ---{\ensuremath{>}} pi+ Neutrino anti-neutrino and K0 {\ensuremath{<}}---{\ensuremath{>}} anti-K0}",
    reportNumber = "UT-KOMABA-80-8",
    doi = "10.1143/PTP.65.297",
    journal = "Prog. Theor. Phys.",
    volume = "65",
    pages = "297",
    year = "1981",
    note = "[Erratum: Prog.Theor.Phys. 65, 1772 (1981)]"
}

@article{HeavyFlavorAveragingGroupHFLAV:2024ctg,
    author = "Banerjee, Sw. and others",
    collaboration = "Heavy Flavor Averaging Group (HFLAV)",
    title = "{Averages of b-hadron, c-hadron, and {\ensuremath{\tau}}-lepton properties as of 2023}",
    eprint = "2411.18639",
    archivePrefix = "arXiv",
    primaryClass = "hep-ex",
    doi = "10.1103/x87q-tld5",
    journal = "Phys. Rev. D",
    volume = "113",
    number = "1",
    pages = "012008",
    year = "2026"
}

@article{Chetyrkin:2000yt,
    author = "Chetyrkin, K. G. and Kuhn, Johann H. and Steinhauser, M.",
    title = "{RunDec: A Mathematica package for running and decoupling of the strong coupling and quark masses}",
    eprint = "hep-ph/0004189",
    archivePrefix = "arXiv",
    reportNumber = "DESY-00-034, TTP-00-05",
    doi = "10.1016/S0010-4655(00)00155-7",
    journal = "Comput. Phys. Commun.",
    volume = "133",
    pages = "43--65",
    year = "2000"
}

@article{Keum:2003js,
    author = "Keum, Yong-Yeon and Kurimoto, T. and Li, Hsiang Nan and Lu, Cai-Dan and Sanda, A. I.",
    title = "{Nonfactorizable contributions to $B \to D^{(*)}M$ decays}",
    eprint = "hep-ph/0305335",
    archivePrefix = "arXiv",
    doi = "10.1103/PhysRevD.69.094018",
    journal = "Phys. Rev. D",
    volume = "69",
    pages = "094018",
    year = "2004"
}

@article{Lu:2002iv,
    author = "Lu, Cai-Dian and Ukai, Kazumasa",
    title = "{Branching ratios of B ---{\ensuremath{>}} D(s) K decays in perturbative QCD approach}",
    eprint = "hep-ph/0210206",
    archivePrefix = "arXiv",
    reportNumber = "BIHEP-TH-2002-47, DPNU-02-31",
    doi = "10.1140/epjc/s2003-01150-4",
    journal = "Eur. Phys. J. C",
    volume = "28",
    pages = "305--312",
    year = "2003"
}

@article{Keum:2000wi,
    author = "Keum, Y. Y. and Li, Hsiang-Nan and Sanda, A. I.",
    title = "{Penguin enhancement and $B \to K \pi$ decays in perturbative QCD}",
    eprint = "hep-ph/0004173",
    archivePrefix = "arXiv",
    reportNumber = "NCKU-HEP-00-01A, APCTP-00-05, DPNU-00-14",
    doi = "10.1103/PhysRevD.63.054008",
    journal = "Phys. Rev. D",
    volume = "63",
    pages = "054008",
    year = "2001"
}

@misc{HFLAV:Online,
    author = "{HFLAV Collaboration}",
    title = "{Summer 2025 updates}",
    howpublished = "\url{https://hflav.web.cern.ch/}",
    year = "2025",
    note = "[Accessed: 2026-01-21]"
}

@article{Gershon:2021pnc,
    author = "Gershon, Tim and Lenz, Alexander and Rusov, Aleksey V. and Skidmore, Nicola",
    title = "{Testing the Standard Model with CP asymmetries in flavor-specific nonleptonic decays}",
    eprint = "2111.04478",
    archivePrefix = "arXiv",
    primaryClass = "hep-ph",
    reportNumber = "SI-HEP-2021-030",
    doi = "10.1103/PhysRevD.105.115023",
    journal = "Phys. Rev. D",
    volume = "105",
    number = "11",
    pages = "115023",
    year = "2022"
}

@article{Fleischer:2021cwb,
    author = "Fleischer, Robert and Malami, Eleftheria",
    title = "{Revealing new physics in $B^{0}_{s}\to D_s^{\mp } K^{\pm }$ decays}",
    eprint = "2110.04240",
    archivePrefix = "arXiv",
    primaryClass = "hep-ph",
    reportNumber = "Nikhef-2021-023",
    doi = "10.1140/epjc/s10052-023-11588-7",
    journal = "Eur. Phys. J. C",
    volume = "83",
    number = "5",
    pages = "420",
    year = "2023"
}

@unpublished{LHCb:2026gyt,
    author = "Aaij, Roel and others",
    collaboration = "LHCb",
    title = "{First measurement of the decay-time-integrated $C\!P$ asymmetry in $B_s^0 \to D_s^- \pi^+$ decays}",
    note = "",
    eprint = "2603.10860",
    archivePrefix = "arXiv",
    primaryClass = "hep-ex",
    reportNumber = "LHCb-PAPER-2025-074, CERN-EP-2026-028",
    month = "3",
    year = "2026"
}

@article{Atwood:2000ck,
    author = "Atwood, David and Dunietz, Isard and Soni, Amarjit",
    title = "{Improved Methods for Observing CP Violation in $B^\pm\to KD$~and Measuring the CKM Phase $\gamma$}",
    eprint = "hep-ph/0008090",
    archivePrefix = "arXiv",
    reportNumber = "AMES-HET-00-09, BNL-HET-00-31, FERMILAB-PUB-00-422-T",
    doi = "10.1103/PhysRevD.63.036005",
    journal = "Phys. Rev. D",
    volume = "63",
    pages = "036005",
    year = "2001"
}

@article{Belle:2021nyg,
    author = "Bloomfield, T. and others",
    collaboration = "Belle",
    title = "{Measurement of the branching fraction and $CP$ asymmetry for $B\to\bar{D}^{0} \pi$ decays}",
    eprint = "2111.12337",
    archivePrefix = "arXiv",
    primaryClass = "hep-ex",
    reportNumber = "Belle Preprint 2021-17, KEK Preprint 2021-15",
    doi = "10.1103/PhysRevD.105.072007",
    journal = "Phys. Rev. D",
    volume = "105",
    number = "7",
    pages = "072007",
    year = "2022"
}

@article{LHCb:2013jqb,
    author = "Aaij, R and others",
    collaboration = "LHCb",
    title = "{Observation of the suppressed ADS modes $B^\pm \to [\pi^\pm K^\mp \pi^+\pi^-]_D K^\pm$ and $B^\pm \to [\pi^\pm K^\mp \pi^+\pi^-]_D \pi^\pm$}",
    eprint = "1303.4646",
    archivePrefix = "arXiv",
    primaryClass = "hep-ex",
    reportNumber = "CERN-PH-EP-2013-038, LHCB-PAPER-2012-055",
    doi = "10.1016/j.physletb.2013.05.009",
    journal = "Phys. Lett. B",
    volume = "723",
    pages = "44--53",
    year = "2013"
}

@article{Belle:2006cuz,
    author = "Abe, Kazuo and others",
    collaboration = "Belle",
    title = "{Study of B+- ---{\ensuremath{>}} D(CP)K+- and D*(CP)K+- decays}",
    eprint = "hep-ex/0601032",
    archivePrefix = "arXiv",
    reportNumber = "BELLE-PREPRINT-2006-1, KEK-PREPRINT-2005-91",
    doi = "10.1103/PhysRevD.73.051106",
    journal = "Phys. Rev. D",
    volume = "73",
    pages = "051106",
    year = "2006"
}

@article{Brod:2014bfa,
    author = "Brod, Joachim and Lenz, Alexander and Tetlalmatzi-Xolocotzi, Gilberto and Wiebusch, Martin",
    title = "{New physics effects in tree-level decays and the precision in the determination of the quark mixing angle {\ensuremath{\gamma}}}",
    eprint = "1412.1446",
    archivePrefix = "arXiv",
    primaryClass = "hep-ph",
    doi = "10.1103/PhysRevD.92.033002",
    journal = "Phys. Rev. D",
    volume = "92",
    number = "3",
    pages = "033002",
    year = "2015"
}

@article{Buchalla:1995vs,
    author = "Buchalla, Gerhard and Buras, Andrzej J. and Lautenbacher, Markus E.",
    title = "{Weak Decays beyond Leading Logarithms}",
    eprint = "hep-ph/9512380",
    archivePrefix = "arXiv",
    reportNumber = "SLAC-PUB-7009, SLAC-PUB-95-7009, MPI-PH-95-104, TUM-T31-100-95, FERMILAB-PUB-95-305-T",
    doi = "10.1103/RevModPhys.68.1125",
    journal = "Rev. Mod. Phys.",
    volume = "68",
    pages = "1125--1144",
    year = "1996"
}

@article{Bobeth:2014rda,
    author = "Bobeth, Christoph and Haisch, Ulrich and Lenz, Alexander and Pecjak, Ben and Tetlalmatzi-Xolocotzi, Gilberto",
    title = "{On new physics in $\Delta\Gamma_{d}$}",
    eprint = "1404.2531",
    archivePrefix = "arXiv",
    primaryClass = "hep-ph",
    reportNumber = "FLAVOUR(267104)-ERC-66, IPPP-14-29, DCPT-14-58",
    doi = "10.1007/JHEP06(2014)040",
    journal = "J. High Energy Phys.",
    volume = "06",
    pages = "040",
    year = "2014"
}

@article{Lenz:2019lvd,
    author = "Lenz, Alexander and Tetlalmatzi-Xolocotzi, Gilberto",
    title = "{Model-independent bounds on new physics effects in non-leptonic tree-level decays of B-mesons}",
    eprint = "1912.07621",
    archivePrefix = "arXiv",
    primaryClass = "hep-ph",
    reportNumber = "IPPP/19/49, Nikhef-2019-054, SI-HEP-2019, P3H-19-044",
    doi = "10.1007/JHEP07(2020)177",
    journal = "J. High Energy Phys.",
    volume = "07",
    pages = "177",
    year = "2020"
}

@article{Atkinson:2024hqp,
    author = "Atkinson, Oliver and Englert, Christoph and Kirk, Matthew and Tetlalmatzi-Xolocotzi, Gilberto",
    title = "{Collider-flavour complementarity from the bottom to the top}",
    eprint = "2411.00940",
    archivePrefix = "arXiv",
    primaryClass = "hep-ph",
    reportNumber = "IPPP/24/72, SI-HEP-2024-24, P3H-24-086",
    doi = "10.1140/epjc/s10052-024-13739-w",
    journal = "Eur. Phys. J. C",
    volume = "85",
    number = "3",
    pages = "258",
    year = "2025"
}

@unpublished{Araz:2026zlu,
    author = "Araz, Jack Y. and Englert, Christoph and Kirk, Matthew and Tetlalmatzi-Xolocotzi, Gilberto",
    collaboration = "",
    title = "{New insights into the $b\rightarrow c \bar{u}q$ puzzle through Top-Bottom synergies}",
    note = "",
    eprint = "2604.25998",
    archivePrefix = "arXiv",
    primaryClass = "hep-ph",
    reportNumber = "IPPP/26/36, SI-HEP-2026-11, P3H-26-032",
    month = "4",
    year = "2026"
}

@article{Charles:2015gya,
    author = "Charles, J. and others",
    title = "{Current status of the Standard Model CKM fit and constraints on $\Delta F=2$ New Physics}",
    eprint = "1501.05013",
    archivePrefix = "arXiv",
    primaryClass = "hep-ph",
    reportNumber = "LPT-ORSAY-15-04",
    doi = "10.1103/PhysRevD.91.073007",
    journal = "Phys. Rev. D",
    volume = "91",
    number = "7",
    pages = "073007",
    year = "2015"
}

@misc{CKMfitterSummer2025,
  author       = "{CKMfitter Group}",
  title        = "{CKMfitter Global CKM Fit Results: Summer 2025}",
  howpublished = "\url{https://ckmfitter.in2p3.fr/}",
  year         = {2025},
  note         = "Numerical results from the CKMfitter global analysis, accessed July 2026"
}

\end{document}